\documentclass[a4paper,11pt]{article}
\pdfoutput=1 

\usepackage{jheppub} 

\usepackage[T1]{fontenc} 
\usepackage{tabu}
\usepackage{graphicx,amsmath,amssymb,multirow,array,bm,mathrsfs}
\usepackage{epstopdf}

\def\bp{\begin{pmatrix}}
\def\ep{\end{pmatrix}}
\def\ba{\begin{align}}
\def\ea{\end{align}}
\def\eps{\epsilon}

\def\l({\left(}
\def\r){\right)}
\def\be{\begin{equation}}
\def\ee{\end{equation}}

\newcommand{\prn}[1]{\left ( #1 \right )}
\newcommand{\brk}[1]{\left [ #1 \right ]}
\newcommand{\bigbr}[1]{\Bigl\{ #1 \Bigr\} }

\newcommand{\half}{\frac{1}{2}}

\newcommand{\form}[1]{\bm{#1}}

\newcommand{\hodge}{\star}

\newcommand{\fu}{\form{u}}
\newcommand{\acc}{a}
\newcommand{\fa}{\form{\acc}}
\newcommand{\fomega}{\form{\omega}}

\newcommand{\fA}{\form{A}}
\newcommand{\Ah}{\hat{A}}
\newcommand{\fAh}{\form{\hat{A}}}
\newcommand{\fF}{\form{F}}
\newcommand{\fFh}{\form{\hat{F}}}
\newcommand{\fB}{\form{B}}
\newcommand{\fBh}{\form{\hat{B}}}
\newcommand{\El}{E}
\newcommand{\fE}{\form{\El}}
\newcommand{\fEh}{\form{\hat{\El}}}

\newcommand{\JH}{J_{_H}}
\newcommand{\JHh}{\widehat{J}_{_H}}
\newcommand{\fJH}{\form{J}_{_H}}
\newcommand{\fJHh}{\form{\widehat{J}}_{_H}}

\newcommand{\JBZ}{J_{_{BZ}}}
\newcommand{\JBZh}{\hat{J}_{_{BZ}}}
\newcommand{\fJBZ}{\form{J}_{_{BZ}}}
\newcommand{\fJBZh}{\form{\widehat{J}}_{_{BZ}}}

\newcommand{\fsH}{\form{\sigma}_{_H}}

\newcommand{\fP}{{\form{\mathcal{P}}}}
\newcommand{\fPh}{{\widehat{\form{\mathcal{P}}}}}

\newcommand{\ICS}{{\form{I}}^{CS}_{2n+1}}
\newcommand{\hICS}{\widehat{\form{I}}^{CS}_{2n+1}}

\newcommand{\VP}{\form{\mathcal{T}}_{2n+1}}
\newcommand{\WCS}{\form{\mathrm{B}}_{2n}}

\newcommand{\JP}{J_\fP}
\newcommand{\fJP}{\form{J}_\fP}
\newcommand{\qP}{q_{_\fP}}
\newcommand{\fqP}{\form{q}_{_\fP}}

\newcommand{\adiff}{a}
\newcommand{\gdiff}{\gamma}

\definecolor{rust}{rgb}{0.8,0.2,0.2}
\definecolor{green}{rgb}{0.1,0.8,0.2}

\title{Effective actions for anomalous hydrodynamics}

\author[a]{Felix M. Haehl}
\author[b]{\!, R.\ Loganayagam}
\author[a]{\!, Mukund Rangamani}
\affiliation[\,a]{Centre for Particle Theory \& Department of Mathematical Sciences,\\
                     Science Laboratories, South Road, Durham DH1 3LE, UK.}
\affiliation[\,b]{Junior Fellow, Harvard Society of Fellows, Harvard University, Cambridge, MA 02138,USA.}
\emailAdd{f.m.haehl@durham.ac.uk}
\emailAdd{nayagam@gmail.com}
\emailAdd{mukund.rangamani@durham.ac.uk}

\abstract{We argue that an effective field theory of local fluid elements
 captures the constraints on hydrodynamic transport stemming from the presence 
 of quantum anomalies in the underlying microscopic theory. Focussing on global current anomalies
 for an arbitrary flavour group, we derive the anomalous constitutive relations 
 in arbitrary even dimensions. We demonstrate that our results agree with the
 constraints on anomaly governed transport derived hitherto using a local version of 
 the second law of thermodynamics. The construction crucially uses the anomaly inflow mechanism
 and involves a novel thermofield double construction. In particular, we show that the
 anomalous Ward identities necessitate   non-trivial interaction between the two parts of the Schwinger-Keldysh contour.}

\begin{document} 
	\begin{flushright} \small{DCPT-13/41} \end{flushright}
	
\maketitle
\flushbottom

\section{Introduction \& Summary}
\label{sec:intro}

The past few years have seen an increasing and successful effort in furthering our understanding of hydrodynamics,
which is perhaps one of the simplest effective field theories. The reasons for this are many-fold: theoretical results
from the study of holographic fluids via the fluid/gravity correspondence \cite{Hubeny:2011hd}, experimental results 
for strongly coupled systems in the hydrodynamic limit such as the quark-gluon plasma or cold atoms at unitarity, 
cf., \cite{Schafer:2009dj} and developments in condensed matter theory have driven much of the recent progress.  
Spurred by these and other developments we have now come to appreciate that the microscopic details of 
the underlying quantum dynamics leave behind indelible signatures in hydrodynamic transport.

The essential point, which forms the main focus of this investigation, is that quantum anomalies, which 
conventionally are discussed in the context of non-thermal observables, influence equilibrium thermodynamics 
and near-equilibrium hydrodynamical behaviour in an essential manner. The effects of pure flavour anomalies
were recently first noticed in holographic computations of charged fluids using the fluid/gravity correspondence \cite{Erdmenger:2008rm, Banerjee:2008th},
where their origins could be traced to gauge Chern-Simons terms in the dual supergravity theory.\footnote{There are earlier 
discussions of these terms in \cite{Vilenkin:1978hb,Vilenkin:1979ui}. A clear account of 
the historical development of anomalous transport can be found in the recent review 
\cite{Landsteiner:2012kd}. It should further be noted that even in the gauge/gravity context, 
there are signals in the holographic analysis of \cite{Bhattacharyya:2007vs} which point to 
anomalous effects in hydrodynamics.} Whilst surprising at first sight, it was quickly
realized in \cite{Son:2009tf} that consistency with a local version of the second law of 
thermodynamics in near-equilibrium situations necessitates that a part of the hydrodynamical 
transport is controlled directly by the quantum flavour anomaly. It is these terms which we 
refer to collectively as anomalous transport (we will be more specific below).

Following these early developments, much effort has been expended in understanding how
anomalies influence near-equilibrium thermal physics. First steps were taken 
in the derivation of Kubo formulae \cite{Kharzeev:2009pj, Amado:2011zx} to explain
the anomaly controlled phenomena of chiral magnetic and chiral vorticial conductivities
in four dimensions, whence it was realized that not only global flavour current anomalies
but also mixed flavour-Lorentz anomalies leave an imprint on fluid transport 
in flat spacetime \cite{Landsteiner:2011cp}. Together with independent analysis
of the macroscopic form of the second law demanding the existence of an entropy current 
with non-negative divergence generalizing \cite{Son:2009tf}, it was shown in
\cite{Loganayagam:2011mu,Kharzeev:2011ds} that, in any even spacetime dimension,
the part of hydrodynamic transport controlled by flavour anomalies could be 
completely determined assuming that such transport 
did not lead to entropy production.\footnote{This hypothesis of anomalous transport
being non-dissipative is independently supported by the fact that the Kubo formulae
determine these in terms of zero frequency correlators, supporting the notion 
that such processes are time-invariant and hence non-dissipative.} In addition 
accumulation of evidence  from holography \cite{Torabian:2009qk, Jensen:2010em, Kharzeev:2011ds}
and analysis of free field theories \cite{Loganayagam:2012pz} lead to a near-complete picture
of anomalous transport induced by flavour anomalies (see also \cite{ Neiman:2010zi,Jensen:2012jy} 
for related discussions and \cite{Nair:2011mk, Capasso:2013jva} for general symmetry based group-theoretic approach).

The current state of the art uses crucially the fact that anomalous transport
can be determined by equilibrium dynamics by virtue of their being non-dissipative. 
The logic exploited in \cite{Banerjee:2012iz, Jensen:2012jh} is that 
equilibrium configurations of a fluid in the presence of arbitrary time-independent
spatially varying sources (which thereby allow non-trivial stationary fluid flows) 
can be equivalently described by a partition function of the sources. By writing down
the most general partition function compatible with the symmetries in a systematic 
low energy gradient expansion, together with introduction of parity-odd terms to 
account for anomalies, it was shown in \cite{Banerjee:2012iz} that all 
known effects of global anomalies in hydrodynamics could be captured explicitly.
Furthermore, this was shown to be explicitly in agreement with
the general entropy analysis of \cite{Loganayagam:2011mu} in \cite{Banerjee:2012cr} 
(\cite{Jain:2012rh, Valle:2012em} use this formalism to 
explain anomalous transport in two dimensions with the latter focusing on the Lorentz
anomaly; cf., also \cite{Banerjee:2013qha,Banerjee:2013fqa}). The authors of \cite{Jensen:2012kj} argued 
that this framework needs to be extended further before 
contribution from gravitational anomalies can  be fully understood -- 
the key point which they exploit is the fact that in  equilibrium, characterized 
by a thermal partition function by considering the theory on a Euclidean time circle, 
one should be free to consider other cycles for thermal reduction, leading to 
further global constraints, which fix the contribution from the gravitational anomaly.
These results have been recently extended in \cite{Jensen:2013kka,Jensen:2013rga,Azeyanagi:2013xea}.

While these results are impressive, they have also  served to focus our attention 
on the aspects of hydrodynamics that we don't understand so well, viz., 
how does one think about this effective field theory in a standard effective action formulation?
As conventionally formulated, hydrodynamics is the universal low energy effective theory 
of any interacting quantum system, valid when the length scales of departure from equilibrium are 
large in units of the mean free path. We use thermodynamic variables such as local temperature
and chemical potential together with the (normalized) fluid velocity field to characterize 
the energy-momentum tensor $T^{\mu\nu}$ and charge current $J^\mu$. The constitutive relations
express the stress-energy tensor and the conserved currents in terms of fluid dynamical variables,
while the conservation equations (supplemented with sources if necessary) serve as equations of motion.

Note that this description does not correspond to the standard formulation of effective field theories. 
There is no low energy Lagrangian, and the stress-energy tensor and conserved currents are not 
constructed from more fundamental degrees of freedom of a true effective field theory 
(the equilibrium partition functions of \cite{Banerjee:2012iz,Jensen:2012jh} are functionals 
of background sources). It is therefore interesting to examine the extent to which an 
effective field theory of hydrodynamics can be formulated in terms of appropriate degrees of freedom.
The naive argument against effective actions is the fact that hydrodynamics is a dissipative theory. 
However, this does not preclude the existence of an effective action for non-dissipative fluids. 
Moreover, in light of the above discussion, one should expect to capture effects of
quantum anomalies in such a framework.

There is indeed a natural framework for discussing the dynamics of fluids using intrinsic variables
which has been used over the years to describe ideal fluid dynamics \cite{Taub:1954zz,Carter:1973fk,
Carter:1987qr,Brown:1992kc,Leutwyler:1996er,Jackiw:2004nm,Dubovsky:2005xd}. This framework was re-examined
recently in \cite{Dubovsky:2011sj} who suggested that it could be useful beyond the ideal fluid level;
by studying the corrections to the ideal fluid effective action  \cite{Bhattacharya:2012zx} argued 
that the transport of non-dissipative neutral fluids could be systematically understood in this framework.
Moreover, this effective action approach has proven useful to understand aspects of anomalous transport
in two dimensions \cite{Dubovsky:2011sk}, parity-odd Hall viscosity in three dimensions \cite{Nicolis:2011ey,
Haehl:2013kra} and more recently even been argued to be useful in understanding aspects 
of dissipation \cite{Nickel:2010pr,Endlich:2012vt, Grozdanov:2013dba}.

The basic idea \cite{Dubovsky:2011sj,Bhattacharya:2012zx} is to use as fields the local fluid element variable $\phi^I$;
we have $d-1$ degrees of freedom in these fields characterizing the dynamics of a $d$-dimensional hydrodynamic system.
Since these fluid element labels are arbitrary we need to allow arbitrary diffeomorphisms in the configuration space 
of $\phi^I$ (which is assumed to admit a manifold structure and called ${\cal M}_\phi$), so long as they preserve
the total volume of the fluid. As a result we have a theory which enjoys a large symmetry under the group of 
volume-preserving diffeomorphisms $\text{Sdiff}({\cal M}_\phi)$ of the configuration space ${\cal M}_\phi$. As shown 
in the works above, from these coarse-grained building blocks, a complete set of fluid dynamical variables 
can be constructed and one can systematically derive the constitutive relations for the conserved currents. By virtue
of the $\text{Sdiff}({\cal M}_\phi)$ symmetry, it transpires that the theory admits an exactly 
conserved vector field, which is identified with the canonical entropy current of fluid dynamics. This
current is conserved off-shell, and as a result the class of fluids one naturally describes in this framework
are non-dissipative. Generalizations to incorporate conserved charges are  straightforward: we introduce fields for
the local charge label and enlarge the symmetry to allow local reshuffling of the charges.

The effective action approach has hitherto been used to study neutral fluids \cite{Bhattacharya:2012zx}
and charged parity-odd fluids in $2+1$ dimensions \cite{Haehl:2013kra}. From these analyses one learns
that the family of non-dissipative fluids derivable from an effective action is restricted when compared
to those which one would obtain by writing down constitutive relations for conserved charges together 
with demanding the existence of a conserved entropy current \cite{Jensen:2011xb}. The latter construction is a variant
of conventional approaches to hydrodynamics, generalizing earlier analyses of \cite{Loganayagam:2008is,
Romatschke:2009kr,Bhattacharya:2011tra,Bhattacharyya:2012nq} (to restrict oneself to non-dissipative fluids).
While the statement seems natural, in that the existence of an effective action is a more stringent condition
than simply the lack of entropy production, the precise details of the effective actions described in 
the aforementioned references leave something to be desired. Before rushing to conclude 
that hydrodynamic constitutive relations obey constraints beyond the local form of the second law 
of thermodynamics (for generic fluids), we should ensure that the effective action approach 
can be reliably applied to obtain sensible results. 

The natural context to test the effective action formalism is to ascertain whether the formalism 
can tackle to derive the hydrodynamic transport driven by anomalies. The first step in this direction
was taken in  \cite{Dubovsky:2011sk} who show that the effective action captures the physics of 
anomalous fluid transport in two dimensions. However, they already noted that there are some puzzles
extending their analysis to higher dimensions; naively there does not seem to be a way to respecting 
the symmetries of the effective action formalism compatible with incorporating the anomaly. 

In the present work we demonstrate that it is indeed possible to write down an effective action 
which reproduces  the known anomalous transport data. To be specific we will restrict attention
to global current anomalies, though it seems quite plausible that our analysis can be generalized 
to Lorentz anomalies as well. The logic we employ is quite simple: using a combination of 
the symmetries of the effective action and the inflow mechanism for anomalies, we first argue that
we can construct an independent anomalous part of the effective action which reproduces
the known anomalous constitutive relation. In doing so we will allow some additional structures 
which deviate from the considerations of \cite{Dubovsky:2011sk} which will explain some of 
the difficulties that were encountered earlier.

In fact, the result for the anomalous part of the effective action is succinctly summarized
as\footnote{We will use bold-face symbols to denote differential forms. Sources are denoted 
by the usual one-forms for the vector potential, while we will find it convenient to present
results for currents in terms of the Hodge dual $d-1$ form.}
\begin{equation}\label{sanom1}
 \begin{split}
  S_{anom} = \int_{\mathcal{M}_{2n+1}} \VP[\fA, \fAh] = \int_{\mathcal{M}_{2n+1}} \frac{\fu}{2\fomega} \wedge \prn{\fP-\fPh} .
 \end{split}
\end{equation}
Here $\VP[\fA_1, \fA_2]$ is a transgression form -- it is essentially 
the difference of two Chern-Simons forms for the gauge potentials $\fA_1$ and $\fA_2$ 
up to the removal of an exact piece which ensures that the transgression form is covariant (we review this below).
In the effective action  $\fA$ is the background gauge potential, and $\fAh= \fA + \mu\, \fu$
is the {\it hydrodynamic shadow} gauge field, introduced in \cite{Jensen:2013kka}. In fact, all the
dependence on the fluid element variables has been packaged into the hydrodynamic velocity 1-form $\fu$ and
the chemical potential $\mu$ appearing in $\fAh$.\footnote{In the equation above, we have also provided 
an explicit expression for this transgression form 
in terms of the hydrodynamic velocity 1-form $\fu$, the vorticity 2-form $\fomega$ and the 
anomaly polynomial $\fP$ which is a $2n+2$ form.The symbol $\fPh$ denotes the anomaly polynomial
evaluated over the shadow gauge field. We use the notation previously described in \cite{Loganayagam:2011mu, Jensen:2012kj, Jensen:2013kka}; see also footnote \ref{f:formdiv}  and Appendix~\ref{appendix:conventions} for our conventions on differential forms.}

The subscript of $\mathcal{M}_{2n+1}$ denotes that the integral is over a manifold which is one dimension 
higher than the manifold in which the fluid lives. Here, we have found it useful to employ the 
anomaly inflow mechanism \cite{Callan:1984sa} (we refer the reader to the appendices of \cite{Jensen:2013kka}
for a recent review), i.e., viewing our anomalous field theory in $d=2n$ dimensions as the boundary dynamics
of a bulk topological theory (in $2n+1$ dimensions), to guide us in writing covariant effective actions.
So the physical part of the effective action in $2n$ dimensions is the boundary term arising from the transgression form,
with the Chern-Simons pieces ensuring cancellation of the anomaly contributions between bulk and the boundary.

We arrive at this action by simply following our nose in demanding that we have a gauge covariant action
preserving the basic symmetries (a generalization of the volume preserving diffeomorphisms to include
the charge degree of freedom) of our set-up.
 
The action \eqref{sanom1} passes several consistency checks:
\begin{enumerate}
\item the currents obtained from it are 
precisely the anomaly induced transport currents derived in \cite{Loganayagam:2011mu} using the 
entropy current analysis.\footnote{It was already remarked in \cite{Loganayagam:2011mu} that 
a generating function of the form $\frac{\fu}{2\fomega} \wedge \prn{\fP-\fPh}$ can be used 
to generate the anomaly-induced currents.}
\item It reduces quite simply to the anomalous equilibrium partition 
function derived in \cite{Banerjee:2012iz,Jensen:2013kka} upon restricting to stationary flows. 
\end{enumerate}

While the anomalous part of the effective action is a useful starting point, one would like to
have a complete action, which has the correct dynamical structure. A curious property of the anomalous 
part of the current $J^\alpha_{anom}$ is that its divergence is the difference of the anomalies of the background 
gauge field and the hydrodynamic shadow gauge field. For example in $d=4$ dimensions we find explicitly 
\begin{align}
  \nabla_\alpha J^\alpha_{anom} &=  \frac{3 c_{_A}}{4} \epsilon^{\alpha\beta\rho\sigma}F_{\alpha\beta}F_{\rho\sigma}
     - \frac{3 c_{_A}}{4} \epsilon^{\alpha\beta\rho\sigma}\hat{F}_{\alpha\beta}\hat{F}_{\rho\sigma} \,,\label{Summary1}\\
   \nabla_\beta T^{\alpha\beta}_{anom} &= F^\alpha{}_\beta J^\beta_{anom} - \mu \, u^\alpha\left( 
     \frac{3 c_{_A}}{4} \, \epsilon^{\beta\gamma\rho\sigma} \hat{F}_{\beta\gamma} \hat{F}_{\rho\sigma} \right) . \label{Summary2}
\end{align}
Here, $T^{\alpha\beta}_{anom}$ is the anomalous part of the energy momentum tensor. This structure of the 
anomalous current conservation was already noted in \cite{Loganayagam:2011mu}. However,
were we to naively write down a complete non-dissipative hydrodynamic action as $S_f = S_{non-anom} + S_{anom}$ 
then clearly the currents derived from $S_f$ would continue to obey an equation like \eqref{Summary1}, which of 
course are not the (anomalous) conservation equations of hydrodynamics. It therefore appears that while 
the effective action is capable of generating the correct off-shell currents accounting accurately for the 
details of anomalous transport, it does not capture the correct hydrodynamic Ward identities. 

Fortunately, there is a simple cure to this malaise. We argue that the correct (anomalous) Ward identities
should be recovered in a non-equilibrium system from a Schwinger-Keldysh construction involving two copies
of the dynamical degrees of freedom. Utilizing this formalism we argue that the hydrodynamic effective action,
not only consists of the two copies of the naive effective action, but also demands the existence of 
an interesting cross-term between the two to ensure that one has the desired anomaly. The anomalous piece
of the effective action can be succinctly summarized again in terms of transgression forms:
\begin{align}
 S_{SK} ^{anom}=   \int_{\mathcal{M}_{2n+1}} \left(\VP[\fA_R, \fAh_R]-\VP[\fA_L, \fAh_L]+\VP[\fAh_R, \fAh_L]\right) .
 \label{sanomSK}
\end{align}

The first two terms are the standard terms we would encounter in a thermofield double construction, with $\fA_R$
living on the forward part of the Schwinger-Keldysh contour and $\fA_L$ the source on the reversed part of the contour.
The new term is the third one which consists of a cross-term involving just the shadow fields on either piece of the contour.
Incorporating this term, we find a very simple effective action which gives the correct anomalous currents with
the expected anomalous conservation equations of hydrodynamics. The necessity for this term is easy to intuit 
once we realize that its presence is necessary to ensure that the correct amount of current inflows into 
each part of the contour.\footnote{Abelian anomalies in  two dimensions are special. Indeed in this case
one can write down an action which does not involve the shadow fields \cite{Dubovsky:2011sk}. This happens
in part  because the contribution from the cross-term turns out to be irrelevant in the hydrodynamic limit as we explain in 
Appendix \ref{appendix:2Danom}.} 

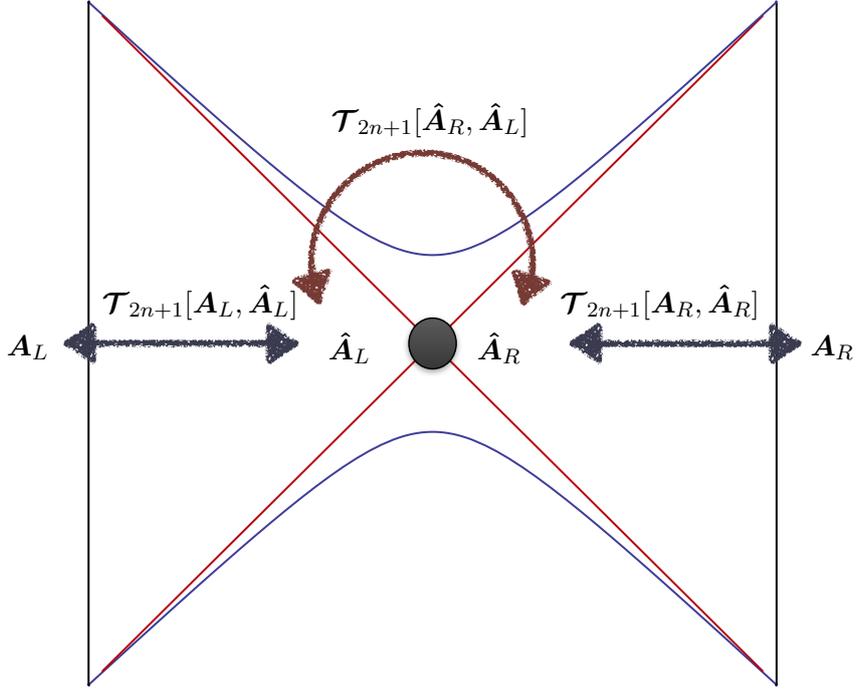
\begin{figure}
\centerline{\includegraphics[width=10cm]{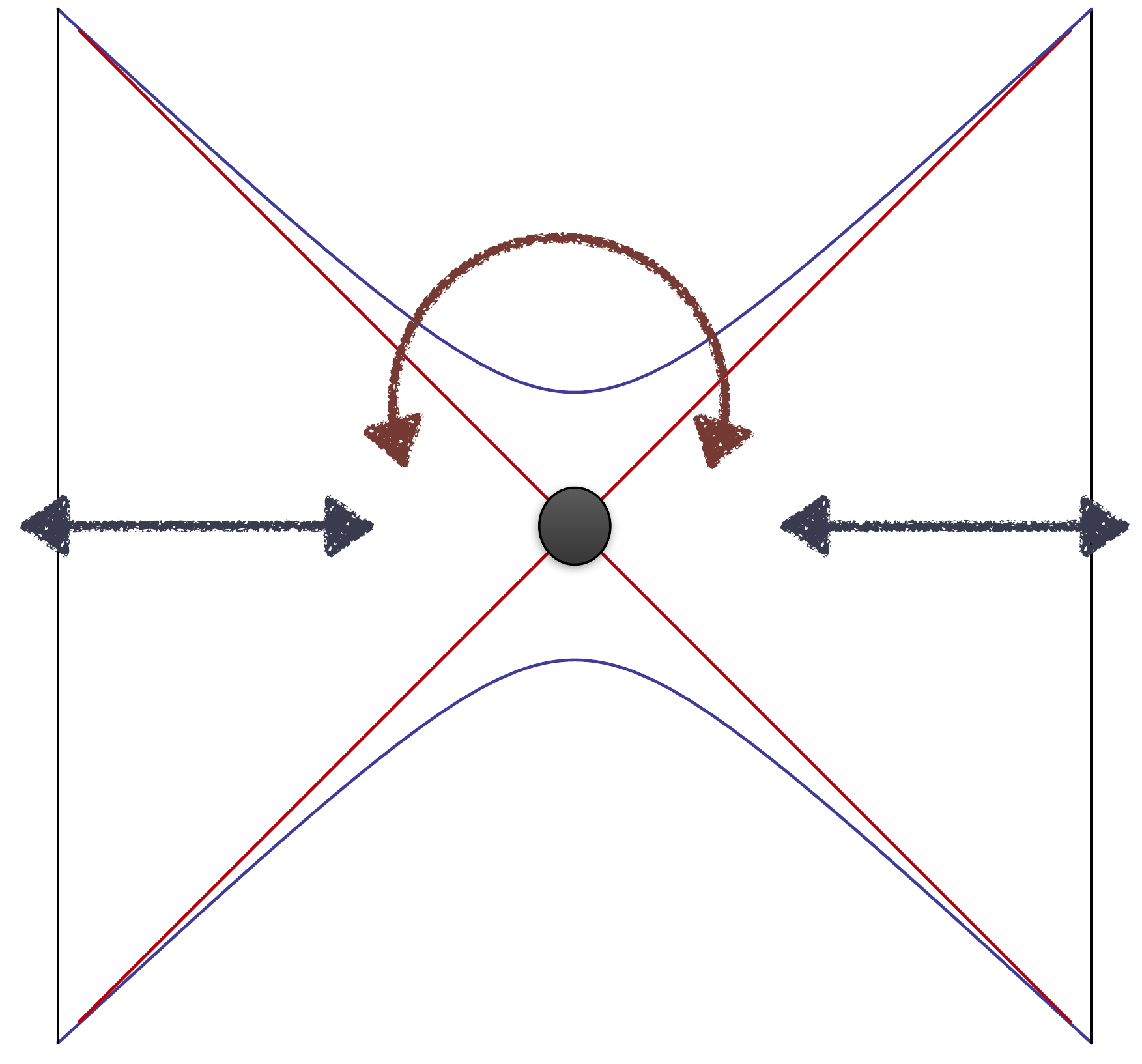}}
\setlength{\unitlength}{0.1\columnwidth}
\begin{picture}(0.3,0.4)(0,0)
\put(8.5,3.5){\makebox(0,0){$\fA_R$}}
\put(5.6,3.5){\makebox(0,0){$\fAh_R$}}
\put(7,3.9){\makebox(0,0){$\VP[\fA_R, \fAh_R]$}}
\put(3,3.9){\makebox(0,0){$\VP[\fA_L, \fAh_L]$}}
\put(5,5.5){\makebox(0,0){$\VP[\fAh_R, \fAh_L]$}}
\put(1.5,3.5){\makebox(0,0){$\fA_L$}}
\put(4.3,3.5){\makebox(0,0){$\fAh_L$}}
\end{picture}
\caption{The heuristic picture of anomalous effective actions suggested by the Schwinger-Keldysh thermofield double construction, illustrated holographically using the eternal black hole in AdS.
}
\label{fig:gluing}
\end{figure}

Heuristically, one can motivate the above action, by thinking about the hydrodynamic theory in the holographic
language using the fluid/gravity correspondence. Since we are restricting to adiabatic pieces of transport, 
we can consider a stationary (planar) black hole in AdS in its maximally extended Kruskal spacetime. The 
two sources $\fA_R$ and $\fA_L$ are the background fields we turn on on the right and left boundaries. The
hydrodynamic shadow gauge field can roughly be viewed as living on the horizon of the black hole (it secretly is 
a proxy for the horizon gauge field as we will describe later). The individual terms on 
the left and right given by the transgression form in \eqref{sanom1} can be viewed as the effective
action obtained by working with the Goldstone mode which is the Wilson line interpolating between
the boundary and the horizon as described in \cite{Nickel:2010pr} (see also \cite{Faulkner:2010jy} 
for a derivation from a holographic renormalization group perspective). While this individual 
decoupled left-right construction suffices for non-anomalous pieces of the transport, to 
correctly incorporate the failure of gauge invariance of the effective due to the anomaly, 
we need to demand a particular gluing condition across the horizon. This ensures that there
is a the desired amount of anomaly inflow, but in the process it couples the left and right
sectors through the horizon degrees of freedom. It is this gluing condition that leads 
to the third term in the effective action \eqref{sanomSK}, see Fig.~\ref{fig:gluing} for an illustration. While we find this heuristic picture compelling,
we should add that we do not derive it from first principles, but rather motivate its inclusion in 
the Schwinger-Keldysh formalism based on symmetry arguments. The fact that it reproduces much 
of the known results of anomaly driven transport and its rather simple structure (which as 
we will see is not obvious a-priori in our construction) leads us to believe that we are 
on the right track; it nevertheless would be nice to have an independent argument in favour 
of our construction.

This paper is organized as follows. To keep the paper self-contained we start 
with some basic preparatory material in \S\ref{s:prelim}. First we give an 
overview of anomalous transport in hydrodynamics in \S\ref{sec:hanom} focussing on
the results we require for our analysis. Following this in \S\ref{sec:0orderFormalism} we review
the effective action description of non-anomalous, charged, perfect fluids. Armed  
with this basic background material, we begin in \S\ref{sec:4dAnomaly} with a simple
and straightforward derivation of an effective action in $(3+1)$ dimensions which
captures the triangle anomaly for a single $U(1)$ current. We will go through the computation
explicitly to exhibit the tight cancellations to ensure the symmetries we want and then establish
consistency with the previous derivations. Our next step will be to show that although the currents
are taking the expected form, the effective action
does not give the correct anomalous dynamics -- the conservation equations are different from 
the expected ones. In \S\ref{sec:AllDimensions} we show that our method can be easily generalized
to all even dimensions and to arbitrary number of abelian currents.  In \S\ref{sec:NonAbelianFormalism}
we develop a more elegant formalism which expresses our result in a way that makes all the symmetries manifest.
In this language, anomalies of non-abelian gauge symmetries are also easily treated. In section 
\S\ref{sec:Schwinger-Keldysh} we present the final result of our paper; a careful treatment of the problem in
the Schwinger-Keldysh formalism. This will result in a simple expression for the effective action 
which gives the correct covariant currents including the correct anomalous dynamics. We conclude 
with a discussion and highlight some open issues in \S\ref{sec:Conclusion}. 

The appendices contain technical details explicating the calculations reported in the main text. In
Appendices \ref{appendix:pieces} and \ref{sec:relothers} we provide some details of the four dimensional construction,
while in Appendix~\ref{appendix:2Danom} we take a close look at the two dimensional story to contrast
with previous analysis. In Appendix \ref{appendix:transg} we give some useful formulae for variations of 
transgression forms, which we then use in Appendix \ref{sec:stensorA} to derive the currents and equations of motion.
Appendix \ref{sec:SK-review} gives a quick overview of the Schwinger-Keldysh construction, 
while Appendix \ref{appendix:conventions} lays out our conventions for differential forms.

\section{Preliminaries: review of some background material}
\label{s:prelim}
We collect some relevant data about anomalous transport and the general framework of
the non-dissipative hydrodynamic effective actions, which we use in the later sections. 
\subsection{Anomalies in hydrodynamics: brief review}
\label{sec:hanom}

In \S\ref{sec:intro} we have given a brief overview of the historical development 
of the subject of anomaly induced transport in hydrodynamics. While our primary interest
is in deriving an effective action for describing the anomaly contributions in hydrodynamics, 
it is worth reviewing the statements in the literature to set the proper stage for our discussion 
(and furthermore clarify what we mean by anomalous transport in precise terms). In general
the conserved currents of a hydrodynamical system take the form:
\begin{align}
T_{\alpha \beta}  &= \varepsilon \, u_\alpha \, u_\beta + p \, P_{\alpha \beta} 
+ \, q_{\alpha}\, u_{\beta} + \, u_{\alpha}\, q_{\beta} + \Pi_{\alpha \beta} \,,
\qquad \Pi_{\alpha\beta}\,u^\beta = 0 \,,\nonumber \\
J^\alpha & = \rho\, u^\alpha + \nu^\alpha\,,
\label{consrel}
\end{align}
with $P_{\alpha\beta}=g_{\alpha\beta}+u_\alpha u_\beta$ being the projector onto the space orthogonal 
to the velocity field, $P_{\alpha\beta} \, u^\beta =0$. The quantities $\{\varepsilon, P, \rho\} $ are thermodynamic parameters. Often these are viewed as functions of 
the local temperature and chemical potential. By a suitable change of thermodynamic state functions (Legendre transform)
we will view them as functions of the entropy density $s$ and the chemical potential $\mu$ instead. 

In writing the expression for the currents above we have made no a-priori choice of fluid frame, preferring to
keep the gradient terms captured in the heat current $q_\alpha$, the transverse part of energy-momentum $\Pi_{\alpha \beta}$
and the charge currents $\nu_\alpha$ arbitrary for the moment. The constitutive relations for the fluid 
express them in gradients of the fluid velocity and thermodynamic parameters, $\nabla_\alpha u_\beta$
and $\nabla_\alpha \left(\mu/T\right)$ respectively. Typically this is done on-shell in that we use
the ideal fluid equations of motion to eliminate the gradients of the temperature in favour of the derivatives above. 

The hydrodynamical equations of motion are the conservation of the energy-momentum and charge current 
in the absence of sources. However, if either the flavour symmetry associated with the charge or the Lorentz
symmetry is anomalous and we furthermore have background electromagnetic and gravitational fields turned 
on then the conservation equations have sources and read:
\begin{equation}
\nabla_\alpha J^\alpha = P_{A}(A, \Gamma) \,, \qquad \nabla_\beta T^{\beta\nu} = F^{\nu\alpha} \,J_\alpha + Q_A(A,\Gamma)
\label{cons1}
\end{equation}	
where the charge source $P_{A}(A, \Gamma)$ and the energy-momentum source $Q_A(A,\Gamma)$ are functionals of
background electromagnetic $(A)$ and gravitational $(\Gamma)$ connections. In what follows, we will focus 
on pure flavour anomalies\footnote{In
addition in certain circumstances we could encounter diffeomorphism anomalies and  Weyl anomalies (the latter in scale invariant theories). In two dimensions these already contribute to the thermodynamics and are well understood, cf., \cite{Jensen:2012kj} for a clear discussion of these contributions. The effects of Weyl anomalies in hydrodynamics is less clear beyond two dimensions (for some preliminary statements see \cite{Eling:2013bj}) though as we describe later it might be possible to 
use the action formalism to decipher these.}, i.e., the  case in which the functionals $P_{A}(A, \Gamma)$ 
and $Q_A(A,\Gamma)$ are independent of the background gravitational connection $\Gamma$.

We now turn to the allowed forms of these functionals in field theory.\footnote{We refer the reader
to \cite{Weinberg:1996fk,Harvey:2005it} for the essential machinery behind the classification of
possible anomalies  in a field theory.} Field theoretical considerations imply that possible anomalies
of a QFT in $d=2n$ spacetime dimensions are characterized by a Chern-Simons $(2n+1)$-form $\ICS[ \fA ]$ in one dimension higher.
In turn, this Chern-Simons form defines an anomaly polynomial $\fP[ \fF ]=d\ICS[ \fA ]$ which 
is a $(2n+2)$-form defined in two dimensions higher than the original field theory. As  the notation signifies, 
the anomaly polynomial $\fP$ is a gauge-invariant functional of the background field strength $\fF$.

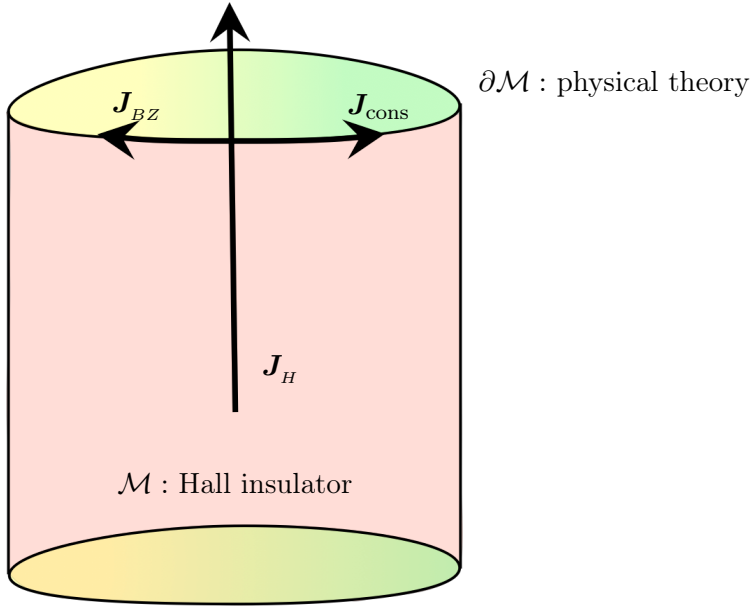
\begin{figure}
\centerline{\includegraphics[width=6cm]{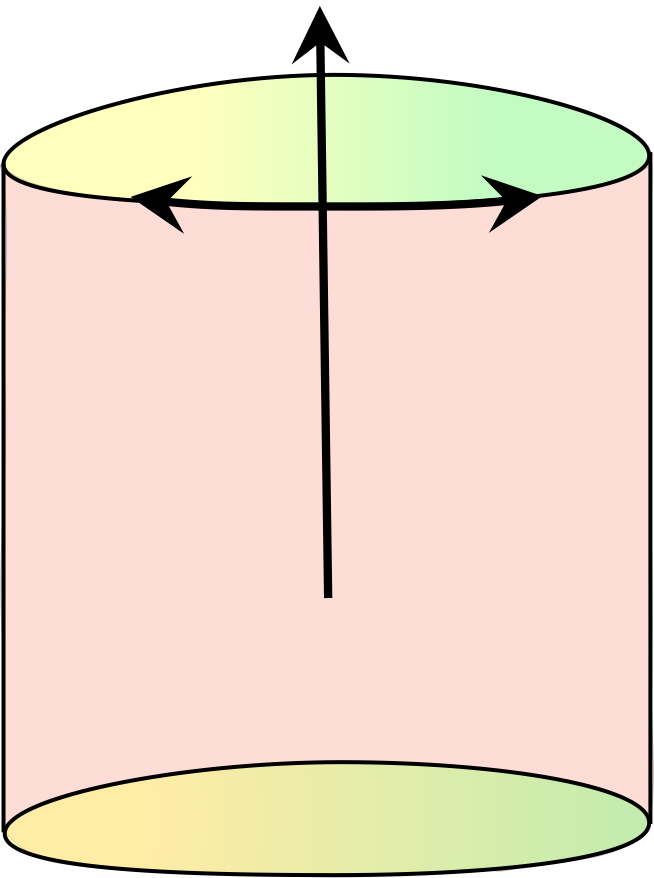}}
\setlength{\unitlength}{0.1\columnwidth}
\begin{picture}(0.3,0.4)(0,0)
\put(5.4,2.5){\makebox(0,0){$\fJH$}}
\put(4.15,4.8){\makebox(0,0){$\fJBZ$}}
\put(6.25,4.8){\makebox(0,0){$\form{J}_\text{cons}$}}
\put(8.3,5){\makebox(0,0){$\partial{\mathcal M}:$ physical theory}}
\put(5,1.5){\makebox(0,0){${\mathcal M}:$ Hall insulator}}
\end{picture}
\caption{Illustration of the anomaly inflow mechanism. The bulk theory
in ${\mathcal M}_{2n+1}$ is the Hall insulator theory, and on the boundary
we have the physical theory with the anomaly. The Hall current $\fJH$
propagates in the bulk and its inflow shows up as anomaly in the boundary theory.
Coupling to the Hall insulator corrects the physical current $\form{J}_\text{cons}$ by a Bardeen-Zumino
contribution $\fJBZ$. The consistent boundary current $\form{J}_\text{cons}$ together with the
Bardeen-Zumino term $\fJBZ$ gives the total current $\form{J}_\text{cov}$ which transforms covariantly.
}
\label{fig:inflow}
\end{figure}

To clarify the physical interpretation of these forms, we adopt an inflow picture of anomalies whereby 
one imagines placing the field theory under question at the boundary of an appropriate Hall insulator
in one dimension higher.\footnote{We refer the reader to the appendices of \cite{Jensen:2013kka} for 
a recent review of anomaly inflow along with the explicit form of anomalies that follows from this picture.}
The anomaly in the QFT is then  simply understood as a flow of a conserved charge from the Hall bulk to the boundary, see Fig.~\ref{fig:inflow}. 
These charge currents in the Hall insulator  are captured by a generating function $\ICS[\fA]$
which is the Chern-Simons $(2n+1)$-form introduced above. More explicitly, let the variation of
this Chern-Simons term  be characterized by 
\begin{equation}
 \delta \ICS = \delta \fA \wedge\star_{2n+1}\fJH   + d\brk{ \delta \fA \wedge \star\fJBZ } \,,
\end{equation}
or equivalently\footnote{We use lower case Greek indices for the QFT (boundary) directions and 
lower case Latin indices for the bulk Hall insulator theory. The direction normal to the boundary 
will often be denoted by $\perp$.} 
\begin{equation}
\begin{split}
\delta \int_{\mathcal{M}_{2n+1}} \ICS =  
\int_{\mathcal{M}_{2n+1}} \sqrt{-g_{2n+1}}\ \JH^a \delta A_a 
+ \int_{\partial\mathcal{M}_{2n+1}} \sqrt{-g_{2n}}\ J_{_{BZ}}^\alpha \delta A_\alpha \,,
\end{split}
\end{equation}
where we will call the bulk part of the charge current $\JH^a$ as the Hall current  and 
we will call the current $\JBZ^\mu$ induced along the boundary of a Hall insulator as
the Bardeen-Zumino current. The explicit form of these
currents is given by\footnote{See \cite{Jensen:2013kka}. We give a short derivation of this and related 
results in Appendix~\ref{appendix:transg}.}
\begin{equation}
 \begin{split}
 \fP &\equiv d\ICS\ \quad ,\quad
 \star_{2n+1} \fJH \equiv \frac{\partial \fP}{\partial \fF} \quad ,\quad
  \star \fJBZ \equiv \frac{\partial \ICS}{\partial \fF}. 
\end{split}
\end{equation}
Thus, the total boundary current in this picture is then the sum of two contributions: 
\begin{itemize}
\item the usual charge current of the boundary QFT obtained from varying the QFT path-integral
as in \eqref{ConRel02}. This current is neither gauge covariant, nor is conserved \cite{Bardeen:1984pm}.
It however satisfies the Wess-Zumino consistency condition obtained by  demanding commutativity of an explicit 
gauge transformation against variation of the background gauge potential. 
For this reason, this current is often called the consistent current.
\item Bardeen-Zumino current that arises from the bulk Chern-Simons term.
\end{itemize}
This total charge current in the boundary  transforms covariantly and is hence called the covariant current 
of the field theory under question. The divergence of this covariant current is non-zero in an 
anomalous field theory: in fact, using the inflow picture, the amount of boundary covariant charge
that is produced is equal to the charge injected by the bulk Hall currents, viz.,
\begin{align}
\nabla_\alpha J^\alpha_\text{cov} = \JH^\perp \,.
\label{cons2}
\end{align}
where $\perp$ denotes the direction of the outward normal to the boundary. 

As an example, let us begin by considering an abelian flavour symmetry that is anomalous in $d=2n$ spacetime 
dimensions. The most general  Chern-Simons $(2n+1)$-form made of a single abelian field is
given by $\ICS = c_{_A} \fA \wedge \fF^n$ where $c_{_A}$ is the anomaly coefficient of interest.
Then we define the anomaly polynomial $(2n+2)$-form $\fP\equiv d\ICS=  c_{_A} \fF^{n+1}$. The Hodge dual
of the Hall current $\JH^a$  in $(2n+1)$ dimensional bulk and the Hodge dual of 
Bardeen-Zumino current $\JBZ^\mu$  in the $2n$ dimensional boundary are given by 
\begin{equation}
 \begin{split}
 \star_{2n+1} \fJH &\equiv \frac{\partial \fP}{\partial \fF} = (n+1)\ c_{_A} \fF^n \,,\\
  \star \fJBZ &\equiv \frac{\partial \ICS}{\partial \fF} = n\ c_{_A}\fA\wedge\fF^n \,.\\
 \end{split}
\end{equation} 
In components, we have
\begin{equation}
 \begin{split}
J_{_H }^a &=   \frac{(n+1)\ c_{_A}}{2^n}\ \epsilon^{ap_1p_2\cdots p_{2n-1}p_{2n}}\
F_{p_1p_2}\cdots F_{p_{2n-1}p_{2n}} \,,\\
 J_{_H }^\perp &=   \frac{(n+1)\ c_{_A}}{2^n}
\eps^{\alpha_1\beta_1\cdots \alpha_n\beta_n} \, F_{\alpha_1\beta_1} \cdots F_{\alpha_n\beta_n} \,,\\
J_{_{BZ}}^\alpha &= \frac{n\ c_{_A}}{2^{n-1}}\ \epsilon^{\alpha\beta\delta_1\delta_2\cdots \delta_{2n-3}\delta_{2n-2}
}\ A_\beta\ F_{\delta_1\delta_2}\cdots F_{\delta_{2n-3}\delta_{2n-2}} \,.\\
 \end{split}
\end{equation}
The covariant currents of the field theory under question are then obtained by first computing 
the consistent current via \eqref{ConRel02} and then shifting it by the Bardeen-Zumino contribution above:
\begin{equation}
\begin{split}
J^\gamma_\text{cov}  &= J^\gamma_\text{cons} + J_{_{BZ}}^\gamma \\
&= J^\gamma_\text{cons} + \frac{n\ c_{_A}}{2^{n-1}}\ \epsilon^{\gamma\beta\delta_1\delta_2\ldots \delta_{2n-3}\delta_{2n-2}
}\ A_\beta\ F_{\delta_1\delta_2}\cdots F_{\delta_{2n-3}\delta_{2n-2}}\,.
\label{}
\end{split}
\end{equation}	
The anomalous current and energy-momentum conservation equations relevant for 
hydrodynamics are then given by the  behaviour of the covariant current, viz.,  
\begin{align}
\nabla_\alpha T^{\alpha\beta} = F^{\beta\alpha} \left(J_\text{cov}\right)_\alpha \,,\qquad 
\nabla_\alpha J^\alpha_\text{cov} &= J_{_H }^\perp =   \frac{(n+1)\ c_{_A}}{2^n}
\eps^{\alpha_1\beta_1\cdots \alpha_n\beta_n} \, F_{\alpha_1\beta_1} \cdots F_{\alpha_n\beta_n} \,.
\label{hydroan}
\end{align}
While the covariant current is what appears in hydrodynamics, it is the consistent current that is natural
from an effective action viewpoint. As a result we will keep track of both of these and also 
derive the relevant Bardeen-Zumino term to translate between the two currents via an explicit 
$(2n+1)$ dimensional Chern-Simons action.

The statement that anomalies influence hydrodynamical transport entails that 
there are contributions to $\{q_\alpha, \nu_\alpha , \Pi_{\alpha \beta}\}$ which are determined
explicitly by the quantum anomaly. In particular, it is convenient to view the contributions to these currents as:
\begin{equation}
q^\alpha = q^\alpha_{anom} + q^\alpha_{diss} \,, \qquad
\nu^\alpha = J^\alpha_{anom} + \nu^\alpha_{diss} \,, \qquad 
\Pi^{\alpha\beta} = \Pi^{\alpha\beta}_{anom} + \Pi^{\alpha\beta}_{diss} \,.
\label{}
\end{equation}	
where we use the subscripts to denote the contribution to the hydrodynamical transport to indicate 
their origins from the anomaly ({\em anom}) and the conventional dissipative contributions ({\em diss}).\footnote{It is typically possible to set $q^\alpha_{diss} =0$ without loss of generality and this 
is indeed exploited in various constructions.} The anomalous transport terms come in two varieties:
(i) contributions to the current which are functionals of background sources and (ii) contributions 
involving intrinsic fluid dynamical data (e.g., gradients of velocity field). An example of the former
is a term in the charge current proportional to the magnetic field (called the chiral magnetic effect in four dimensions),
while contributions involving the fluid vorticity (chiral vorticial effect) exemplify the latter. 
The key point to note is that terms involving intrinsic fluid variables remain non-vanishing even 
in the absence of external sources, which means that in hydrodynamics one can infer the presence of 
an anomaly without explicitly turning on background electromagnetic (or gravitational) fields.

While the conserved currents form the basic content of hydrodynamics, there is another object of interest, 
viz., the entropy current $J_\text{S}^\alpha$. Following the earlier decomposition we can write:
\begin{equation}
J_\text{S}^\alpha = s\, u^\alpha + J_{\text{S},\,anom}^\alpha+ J_{\text{S},\,diss}^\alpha
\label{}
\end{equation}	
explicitly demarcating the contributions of the perfect fluid, the anomaly and the  dissipative effects. The entropy current is required to satisfy a local form of the second law, $\nabla_\alpha J_\text{S}^\alpha \geq  0$; as has been discussed extensively in the literature this constrains transport in a non-trivial manner.

 The analysis of \cite{Loganayagam:2011mu} (generalizing the original argument of \cite{Son:2009tf}) 
 shows that by invoking adiabaticity of the anomalous contributions, one can reduce the requirement 
 of the second law of thermodynamics of local entropy increase to only involve the dissipative contributions. 
 This leads to an equation that should be satisfied by the {\em covariant} anomalous currents:
 \begin{equation}
\left(\nabla_\alpha + {\mathfrak a}_\alpha\right) q^\alpha_{anom} - J^\alpha_{anom}\, E_\alpha 
= T\, \nabla^\alpha J_{\text{S},\,anom}^\alpha + \mu \left(\nabla_\alpha  J^\alpha_{anom} - \JH^\perp \right)
 \label{adiabatc}
 \end{equation}	
 with $\JH^\perp $ being the covariant anomaly given in \eqref{cons2}, 
 while the electric field and fluid acceleration defined as $E_\alpha = F_{\alpha\beta} \,
 u^\beta$ and $\mathfrak{a}_\alpha = u^\beta\,\nabla_\beta u_\alpha$ respectively (see Table \ref{table:4Dclassification}).
 It was shown that this equation has a consistent solution, which furthermore could be  obtained 
 from a Gibbs potential constructed from the anomaly polynomial. It was later demonstrated that this
 solution is also obtained from a free energy by considering equilibrium configurations on backgrounds
 with arbitrary spatial dependence (with slow variations as appropriate to hydrodynamics) \cite{Banerjee:2012cr, Jensen:2013kka}. 
 For the moment we refrain from writing down the particularities of the solution; the reader will find all
 the relevant details in \S\ref{sec:4dAnomaly} and \S\ref{sec:AllDimensions}.
 
\subsection{An effective field theory for {\em non-dissipative} fluids}
\label{sec:0orderFormalism}

We now quickly review the essential ingredients of the effective action approach. Following \cite{Dubovsky:2011sj} (see also \cite{Bhattacharya:2012zx, Haehl:2013kra} for further comments) the basic degrees of freedom for a fluid in $d$ dimensions are $d-1$ fundamental fields $\{\phi^I\}$ which give the position of local fluid elements in physical space (coordinates $x^\alpha$). Additional fields characterize the charge label of a single fluid element; for a global $U(1)$ charge we can take a phase field $\psi$ to capture this degree of freedom (the non-abelian generalization is discussed later). The $U(1)$ particle number symmetry is implemented as a translation in field space
\begin{align}
\psi \longrightarrow \psi + c \,.
\label{cShiftSymm}
\end{align}
We assume that this configuration space has a manifold structure and denote it by ${\cal M}_{\phi,\psi}$. 

Noting that the labels of individual fluid elements is an arbitrary choice, we demand reparametetrization invariance under arbitrary diffeomorphisms of ${\cal M}_\phi \subset {\cal M}_{\phi,\psi}$ subject to the condition that the total volume of the fluid remains fixed, i.e.,
\begin{equation}
\phi^I \to \xi^I(\phi) \ , \qquad \text{Jacobian}(\xi,\phi) = 1 \,,
\label{sdiffphi}
\end{equation}	
leaves the effective action invariant. This leads us to considering a theory with volume-preserving diffeomorphisms in configuration space. Furthermore, the phase field itself can be made to depend on the co-moving position within the fluid because  charge conservation holds \textit{locally} (at least for the non-dissipative fluids that we are dealing with). Therefore, the particle number symmetry \eqref{cShiftSymm} can be enlarged to a symmetry of the theory under the following transformation:
\begin{align}
\psi \longrightarrow \psi + \mathfrak{f}(\phi^I) \, , \label{shiftSymm}
\end{align}
where $\mathfrak{f}(\phi^I)$ is an arbitrary function of the $\phi^I$.
Following \cite{Dubovsky:2011sj}, we will call this the \textit{chemical shift symmetry}. 
In effect the theory is invariant under the combined symmetry group generated by \eqref{sdiffphi}
and \eqref{shiftSymm}; for brevity we refer to this complete symmetry group as $\widetilde{\text{Sdiff}({\cal M}_{\phi,\psi})}$.

There are some important consequences of the symmetry $\widetilde{\text{Sdiff}({\cal M}_{\phi,\psi})}$ which were explained in \cite{Dubovsky:2011sj,Bhattacharya:2012zx}:
\begin{itemize}
\item An effective field theory with this symmetry naturally ensures that the dynamical  equations of motion of an effective Lagrangian contain no more data than the conservation of the energy-momentum tensor and charge current.
\item The fields $\phi^I$ and $\psi$ are Goldstone modes; thus canonical assignment of dimensions is $\left[d\phi^I \right] = [d\psi] =0$. The effective action is then built out of the gradients of $\phi^I$ and $\psi$ in a systematic low energy expansion.
\item An important consequence of volume preserving diffeomorphisms which we  denote as $\text{Sdiff}({\cal M}_\phi)
\subset \widetilde{\text{Sdiff}({\cal M}_{\phi,\psi})}$ , is that the vector field
\begin{align}
 J_\text{S}^\beta = \frac{1}{(d-1)!}\; \eps^{\beta \alpha_1 ... \alpha_{d-1}}\; \eps_{I_1...I_{d-1}} \; \prod_{j=1}^{d-1}
    \partial_{\alpha_j} \phi^{I_j} \, .
\end{align}
is trivially conserved $\nabla_\alpha J_\text{S}^\alpha = 0$ and $[J_\text{S}^\alpha] =0$. We interpret this object as the entropy current of our fluid and the fact that it is divergence-free amounts to saying that the formalism deals with non-dissipative fluids.
\end{itemize}

To construct the effective action we also introduce background sources; the metric $g_{\alpha \beta}$ on the physical spacetime and background gauge fields $A_\alpha$ which couple to the charge. The abelian gauge transformations are implemented via 
\begin{align}
\psi \longrightarrow \psi + \Lambda(x) \, , \qquad A_\alpha \longrightarrow A_\alpha - \partial_\alpha \Lambda(x) \,,
  \label{gaugeTrf}
\end{align}
for an arbitrary scalar parameter $\Lambda(x)$. This in particular means that the gauge covariant derivative of the field $\psi$ is that appropriate to a phase field
\begin{equation}
D_\alpha \psi  = \nabla_\alpha \psi + A_\alpha \,.
\label{Dcov}
\end{equation}	

Armed with these basic variables we are now in a position to describe the hydrodynamic fields relevant for charged fluid dynamics. We define a velocity field $u^\alpha$ along the direction of $J_\text{S}^\alpha$ which is normalized to $u^\alpha u_\alpha = -1$ and the entropy density via the norm of $J_\text{S}^\alpha$. We also introduce a chemical potential $\mu$ that couples of the conserved $U(1)$ charge. To wit,
\begin{align}
J_\text{S}^\alpha = s\, u^\alpha \, , \qquad s = \sqrt{-\, g_{\alpha\beta}\, J_\text{S}^\alpha\, J_\text{S}^\beta}  \, , \qquad\mu = u^\alpha\, D_\alpha \psi \,.
\label{Jsall}
\end{align}
The chemical potential $\mu$ is invariant under gauge transformations and the chemical shift symmetry \eqref{shiftSymm}, since 
 $J^\alpha_\text{S}$ is co-moving with the fluid elements (i.e., $J_\text{S}^\alpha \partial_\alpha \phi^I=0$). 
 
The leading order action invariant under $\text{Sdiff}({\cal M}_\phi)$ \eqref{sdiffphi}, the chemical shift symmetry (\ref{shiftSymm}) and given our assignment of scaling dimensions $[d\phi^I] = [d\psi] =0$ is just \cite{Dubovsky:2011sj}
\begin{align}
 S_0 = \int \sqrt{-g}\,  f(s,\mu) \, .
\label{S0def}
\end{align}
Abbreviating $f_{,s}\equiv \tfrac{\partial}{\partial s} f$ and $f_{,\mu} \equiv \tfrac{\partial}{\partial \mu}f$, we find the following stress tensor and charge current (which as always are defined by varying the background sources):
\begin{align}
T^{\alpha\beta}_{(0)} &= \frac{2}{\sqrt{-g}} \frac{\delta S_0}{\delta g_{\alpha\beta}}  = \left( f-s  f_{,s} \right) g^{\alpha\beta} + \left(
                    \mu\, f_{,\mu}- s \,f_{,s}\right) u^\alpha u^\beta   \,,\label{ConRel01} \\
J^\alpha_{(0)} &= \frac{1}{\sqrt{-g}} \frac{\delta S_0}{\delta A_\alpha} = f_{,\mu}\, u^\alpha \,. \label{ConRel02}
\end{align}
We thus have a post-facto justification of our identification of the thermodynamic parameters in this framework, obtaining the energy density $\varepsilon =(\mu\,  f_{,\mu}-f)$, the pressure $P=(f-s f_{,s})$ and the charge  density $\rho=f_{,\mu}$. Finally  since the entropy current is given by \eqref{Jsall} at every order in the gradient expansion, the hydrodynamic  constitutive relations are obtained in the  {\em entropy frame}.

\section{An effective action for the triangle anomaly}
\label{sec:4dAnomaly}

We have now assembled all the ingredients to address the question of deriving an effective action for anomalous hydrodynamics. To illustrate the procedure we will first explain how the global current anomaly in four spacetime dimensions manifests itself in hydrodynamic transport. 
This turns out to be the prototypical case and generalizations to other even dimensions turn out to be quite straightforward (and are explained in \S\ref{sec:AllDimensions}). As noted in  \S\ref{sec:intro} \cite{Dubovsky:2011sk} derive the anomalous hydrodynamical transport equations in $1+1$ dimensions using the effective action approach; our approach has some similarities, but deviates from them in an essential manner, which we will indicate where appropriate.

 \subsection{Preliminaries: a useful basis of fields}
 \label{sec:4dvars}
 
While we will use the effective action formulated in terms of the fluid element fields $\{\phi^I,\psi\}$ it is useful to construct a basis of independent fluid data relevant to our analysis. First of all, we start with the fluid velocity $u^\alpha$ whose gradient can be decomposed in a standard form
\begin{align}
\nabla_\alpha u_\beta = \sigma_{\alpha\beta} + w_{\alpha\beta} - u_\alpha\,{\mathfrak a}_\beta + \frac{1}{d-1}\, \Theta\, P_{\alpha\beta} \,,
\label{}
\end{align}
into the symmetric-traceless shear $\sigma_{\alpha\beta}$, antisymmetric vorticity tensor $w_{\alpha\beta}$, transverse vector acceleration ${\mathfrak a}_\alpha$, and scalar expansion $\Theta$; explicit definitions for these are given in Table \ref{table:4Dclassification}. An object that plays a key role in the anomalous transport is the vorticity vector
\begin{align}
\omega^\alpha = \frac{1}{2} \, \eps^{\alpha\nu\rho\sigma} \;u_\nu \,\nabla_{\rho}u_\sigma \,.
\label{vorticity}
\end{align}
Note that this vector can be expressed in terms of the vorticity tensor and the acceleration. In addition to the fluid velocity we also have the thermodynamic parameters temperature $T$ and chemical potential $\mu$. We view the former as a function of the entropy density $s$ when necessary. 

The background fields are the metric $g_{\alpha\beta}$ which for the most part plays a passive role and the electromagnetic potential $A_\alpha$. The electric and magnetic fields are defined by projection of the gauge field strength with respect to the velocity field, i.e.,
\begin{equation}
E^\alpha = F^{\alpha\beta}\, u_\beta \,, \qquad
\ B^\alpha = \frac{1}{2}\, \eps^{\alpha\nu\rho\sigma}\; u_\nu F_{\rho\sigma} \,,\qquad F_{\alpha\beta} \equiv 2\, \nabla_{[\alpha} A_{\beta]}\,.
\label{}
\end{equation}	
We also find it convenient following \cite{Banerjee:2012cr, Jensen:2012kj} to introduce a transverse gauge field ${\hat A}_\alpha$: 
\begin{align}
{\hat A}_\alpha &= A_\alpha + \mu\, u_\alpha  \,,\qquad {\hat F}_{\alpha\beta} \equiv 2\, \nabla_{[\alpha} {\hat A}_{\beta]} \,.
\label{transhat}
\end{align}
In particular, it will be useful to note that the transverse electric and magnetic fields are given by
\begin{align}
   \hat E^\alpha &\equiv \hat F^{\alpha\beta}\,u_\beta =E^\alpha - \mu \, {\mathfrak a}^\alpha - P^{\alpha\beta} \nabla_\beta \mu \, ,  \label{I3}\\
   \hat B^\alpha &\equiv \tfrac{1}{2}\,\eps^{\alpha\beta\rho\sigma}\; u_\beta \, \hat F_{\rho\sigma} = B^\alpha + 2\,\mu\,\omega^\alpha \, ,
\end{align}
with the transverse field $\hat E^\alpha = 0$ in equilibrium.\footnote{In deriving this identity we had occasion to  exploit the ideal fluid equation of motion
$P_{\mu\lambda} \nabla_\nu T^{\nu\lambda} = P_{\mu\lambda} F^{\lambda \nu} J_\nu$
and the Gibbs-Duhem relation $dP = s \,dT + q \, d\mu$.}

In Table \ref{table:4Dclassification} we summarize the basic hydrodynamic fields we use, listing in addition to their transformation properties under the spatial $SO(3)$ symmetry, the objects which typically are taken to be on-shell independent after imposition of the ideal fluid conservation equations. We will use some of the these variables viewed as functions of $\{\phi^I,\psi\}$ in addition to other tensors built from $\psi$ as the basic building blocks of our effective action.

  \begin{table}[h!]
 \centering
{\tabulinesep=1.5mm
\begin{tabu}{|c||c|c|}
 \hline
     & Data & On-shell independent
      \\ 
 \hline \hline
 Scalars & $\Theta \equiv \nabla_\alpha u^\alpha$ & \\ 
         & $u^\alpha\nabla_\alpha T$ &  $\Theta$   \\
         & $u^\alpha\nabla_\alpha \mu$  & \\
\hline
		 & ${\mathfrak a}^\alpha \equiv u^\beta\nabla_\beta u^\alpha$ &  ${\mathfrak a}^\alpha$ \\
Vectors  & $E^\alpha \equiv F^{\alpha\beta}u_\beta$ & $E^\alpha$ \\
       & $ P^{\alpha\beta}\nabla_\beta T$ &  ${\hat E}^\alpha \equiv E^\alpha- \mu\, {\mathfrak a}^\alpha- P^{\alpha\beta}\nabla_\beta \mu $ \\
         & $P^{\alpha\beta}\nabla_\beta (\tfrac{\mu}{T})$ &   \\
      \hline
 Pseudo- & $\omega^\alpha \equiv \tfrac{1}{2} \eps^{\alpha\nu\rho\sigma}\; u_\nu \nabla_\rho 
           u_\sigma$  & $\omega^\alpha$ \\
 Vectors & $B^\alpha \equiv \tfrac{1}{2} \eps^{\alpha\nu\rho\sigma} \;u_\nu F_{\rho\sigma}$ 
           &  $B^\alpha$ \\
      \hline
 Tensors & $\sigma^{\alpha\beta} \equiv P^{\alpha\rho} P^{\beta\sigma} \left( \nabla_{(\rho} u_{\sigma)} - \tfrac{\Theta}{3}  \,
           P_{\rho\sigma}\right)$ & $\sigma^{\alpha\beta}$ \\
 \hline
\end{tabu}}
 \caption{Complete basis of $(3+1)$-dimensional fluid and background (gauge field) data at first order in derivatives which is useful for our construction. We have used the ideal fluid equations to arrive at the on-shell independent data, listed here for convenience.}
 \label{table:4Dclassification}
 \end{table}

\subsection{The reveal: anomalous action summarized}
\label{sec:reveal}

Before going into details, let us summarize the basic logic and the main result of our construction. 
As described earlier, the effective action built out of the fluid element variables is required 
to respect $\widetilde{\text{Sdiff}({\cal M}_{\phi,\psi})}$ symmetry described by the transformations \eqref{sdiffphi} 
and \eqref{shiftSymm}. We also want to ensure that the effective action correctly reproduces the 
abelian flavour anomaly of the current. These are a-priori the only requirements on the effective action 
we wish to construct. 

The anomaly inflow mechanism \cite{Callan:1984sa} reviewed in \S\ref{sec:hanom} guarantees that 
\begin{align}
\int_{{\cal M}_5} \, \ICS + \int_{{\cal M}_4} {\cal L}_{eff} 
\label{}
\end{align}
is invariant under local gauge transformations $\fA\to \fA - d\Lambda$. In the case of
four dimensional $U(1)^3$ triangle anomalies it is easy to argue that the Chern-Simons term of
relevance is simply $\ICS = c_{_A}\, \fA \wedge \fF \wedge \fF$. The corresponding Hall and
Bardeen-Zumino currents are given by $\star_{2n+1} \fJH = 3 \,c_{_A} \fF \wedge \fF$ and
$\star \fJBZ = 2 \,c_{_A}\, \fA \wedge \fF $.

Since this Chern-Simons form guarantees the correct anomaly, we start with this term
and proceed to determine the form of  $S_{eff} = \int_{{\cal M}_4}\, {\cal L}_{eff}$ built from the fields ${\phi^I,\psi}$ that
(i) ensures that the total action is gauge invariant and (ii) respects the $\widetilde{\text{Sdiff}({\cal M}_{\phi,\psi})}$
symmetry. It transpires that the volume preserving diffeomorphisms $\text{Sdiff}({\cal M}_\phi)$ are easily 
taken care of by working directly with the fluid velocity and the entropy density (since $J_\text{S}^\alpha$ is invariant under \eqref{sdiffphi}). The non-trivial constraint comes from \eqref{shiftSymm}, which demands that in addition to the Chern-Simons term involving the background gauge field above, we also include a Chern-Simons term  written in terms of the transverse field $\fAh$. In particular, as already anticipated in \cite{Dubovsky:2011sk} there is no action that respects the chemical shift symmetry and reproduces the anomaly.

To proceed, we take inspiration from the analysis of \cite{Loganayagam:2011mu}. As described in \S\ref{sec:hanom} the adiabaticity argument leads to an equation for the anomalous hydrodynamic currents \eqref{adiabatc}. A solution to this equation is given as $J_{\text{S},\,anom}^\alpha = 0$ and an anomalous current $J^\alpha_{anom}$ 
which turns out to satisfy $d\star\form{J}_{anom} = 3 \,c_{_A} (\fF\wedge \fF - \fFh \wedge \fFh)$ or
\begin{align}
 \nabla_\alpha J^\alpha_{anom}= \frac{3\,c_{_A}}{4}\, \eps^{\alpha\beta\gamma\delta}
\left(F_{\alpha\beta} \, F_{\gamma\delta} - {\hat F}_{\alpha\beta}\, {\hat F}_{\gamma\delta} \right) = - 6\, c_{_A} \left[ E_\nu B^\nu - \hat E_\nu \hat B^\nu \right] \,. 
\label{AnConservation}
\end{align}
Since our effective action formalism naturally conserves the entropy current, we ask whether
it is possible to engineer a situation wherein we reproduce the solution to \eqref{adiabatc} directly.

The main claim then which we justify in the rest of the section is that the action
\begin{equation}
 \begin{split}
  S_{anom} &= -\,c_{_A} \int ({\bf D}\psi \wedge \fA \wedge \fF 
  - {\bf \hat D} \psi\wedge \fAh\wedge \fFh)
+ c_{_A} \int_{\mathcal{M}_5} (\fA\wedge \fF\wedge \fF - \fAh\wedge \fFh \wedge \fFh )\\
                 &\qquad \qquad \qquad -\; c_{_A}\int \sqrt{-g} \;  D_\alpha \psi \left( \mu^2 \,(2\omega^\alpha) +2\mu \, B^\alpha \right)\\
  &= -\, c_{_A} \int ({\bf D\psi} \wedge \fA \wedge \fF 
  - {\bf \hat D}\psi \wedge \fAh\wedge \fFh)
   - c_{_A}\int{\bf D}\psi \wedge \frac{\fu}{2\fomega} \wedge \prn{\hat{\fB}^2-\fB^2}   \\
  &\qquad \qquad \qquad+\; c_{_A} \int_{\mathcal{M}_5} (\fA\wedge \fF\wedge \fF - \fAh\wedge \fFh \wedge \fFh )
\label{Sa}  
 \end{split}
\end{equation}
satisfies all the requirements listed above and reproduces precisely the anomalous transport currents
derived earlier using the entropy considerations (see Appendix~\ref{appendix:conventions} for some conventions
concerning integration of differential forms and Hodge duals).\footnote{To retain compact expressions we perform some formal manipulations with differential forms as in \cite{Loganayagam:2011mu, Jensen:2012kj,Jensen:2013kka}. Divisions by a differential form implicitly indicates that the numerator when expanded out always has a factor which cancels the form we divide by; see the first step in the manipulation of \eqref{Bconst} for an illustration. \label{f:formdiv}}  In particular, from the  action $S_{anom}$ in 
\eqref{Sa} we argue that one obtains the currents
\begin{align}
q^\alpha_{anom} &=  -4\,c_{_A}\,\mu^3 \, \omega^{\alpha}
  -3\,c_{_A}\,\mu^2 \, B^{\alpha} \, , 
  \qquad  \Pi^{\alpha\beta}_{anom} = 0 \,,\label{T_sFrame0}\\                   
J^\alpha_{anom} &= -6\,c_{_A}\,\mu^2 \, \omega^\alpha -6\,c_{_A}\,\mu \, B^\alpha  \,,
\label{J_sFrame0}
\end{align}
which are precisely those derived by solving \eqref{adiabatc} in \cite{Loganayagam:2011mu} .

In the action (\ref{Sa}) we use the gauge invariant one-form ${\bf (D\psi)} = d\psi + \fA$ and
hatted quantities are constructed from the transverse gauge field introduced in \eqref{transhat}. 

Note that the anomalous pieces of $q^\alpha_{anom}$, $\Pi^{\alpha\beta}_{anom}$, and $J^\alpha_{anom}$ provide an off-shell
solution to anomalous transport for similar reasons as in \cite{Loganayagam:2011mu}. These terms by themselves 
don't satisfy traditional hydrodynamic equations of motion and furthermore are derived without ever referring to the lower order equations of motion. Indeed, we never need to invoke the dynamics of fluid transport equations to derive these contributions. We will return to on-shell data once we justify the construction above and describe some interesting conundrums from our ignorance of their presence.

\subsection{Derivation of the anomalous action}
\label{sec:dervive4d}

We will now derive the action (\ref{Sa}) in a constructive way, exploiting as remarked above, the fact that the anomalous current derived from the adiabaticity argument is known from \cite{Loganayagam:2011mu}.\footnote{Even in the absence of this result we could have taken the known Landau frame result derived in \cite{Son:2009tf} and rotated it to the entropy frame. We describe the connection between various frame choices in Appendix~\ref{sec:frames}.}   The anomalous current (in the entropy frame) $J^\alpha_{anom}$ is required to satisfy \eqref{AnConservation}; this provides a valuable clue. A natural way to start, is to ignore the symmetries of the system momentarily and ask if we can write down an action which reproduces the anomalous variation to give the r.h.s.\ of \eqref{AnConservation}. 

This is rather easy to do. Consider, the parity-odd topological term
\begin{align}
  S_{wzI} =   \aleph \, \int_{{\cal M}_4} \left[{\bf D\psi} \wedge \fA \wedge \fF  - {\bf \hat D} \psi\wedge \fAh\wedge  \fFh\right] 
\label{S_anom_ng}
\end{align}
with constant $\aleph$ tuned such that the correct anomalous variation is produced. We have already been inspired by the symmetric form (\ref{AnConservation}) of the divergence of the anomaly piece of the current,
to introduce just one constant parameter.\footnote{We could, of course, not refer to the knowledge about this conservation equation and indeed introduce a second constant $\hat \aleph$ at this stage, and determine it later from demanding consistency. This would make the calculations slightly more complicated with the same end result.}

Now under a gauge transformation (\ref{gaugeTrf}), this action is clearly not invariant. It transforms as
\begin{align}
 \delta S_{wzI} &= - \frac{\aleph}{2}\; \int \sqrt{-g} \;\eps^{\alpha\beta\rho\sigma}
    \left[\, D_\alpha \psi  \, \partial_\beta \Lambda \, F_{\rho\sigma} - \,\hat D_\alpha \psi  \, \partial_\beta \Lambda \, \hat F_{\rho\sigma}  \right] \notag \\
    &= -\frac{\aleph}{4}\; \int \sqrt{-g} \;  \Lambda\,\eps^{\alpha\beta\rho\sigma}
    \left[ F_{\alpha\beta}F_{\rho\sigma} - \hat F_{\alpha\beta} \hat F_{\rho\sigma} \right]\, . \label{gaugeVar}
\end{align}
Of course, due or the presence of an anomaly we cannot have a gauge-invariant action in $3+1$ dimensions. However, exploiting the inflow mechanism \cite{Callan:1984sa}, all we need is a five dimensional Chern-Simons action which will serve to cancel the anomalous variation in \eqref{gaugeVar} above. This is simply
\begin{align}
  S_\text{CS} = -\aleph \int_{\mathcal{M}_5} \left[ \fA\wedge \fF\wedge \fF - \fAh\wedge \fFh \wedge \fFh\right] \label{Scs}
\end{align}
Under a gauge transformation, the Chern-Simons term gives a boundary term on the four dimensional spacetime which cancels the variation (\ref{gaugeVar}). So $S_{wzI}+S_\text{CS}$ is a gauge invariant action.

Having dealt with the gauge invariance we need to address the second requirement, viz., demanding that the terms in the effective action respect the chemical shift symmetry \eqref{shiftSymm}. It is a simple exercise to check that $S_{wzI}$ by itself is not invariant under this symmetry ($S_\text{CS}$  is blind to transformations of the $\psi$ field); we need to augment the action with other terms. However, since we have taken the trouble to ensure gauge invariance, we now should restrict attention to terms that are explicitly gauge invariant. 

Thus we need to determine the most general parity-odd, gauge invariant term which we can build from one-derivative data that can appear in the action. The focus on one-derivative data is because the action $S_{wzI}$ is itself built out of the first order gradients and we know from earlier analysis that the four dimensional global anomaly is manifested at this order. 

This set of requirements is in fact quite constraining. There are precisely two one derivative parity-odd vector fields at our disposal: the vorticity $\omega^\alpha$ and the magnetic field $B^\alpha$.\footnote{In the counting of derivatives the electromagnetic potential has mass dimension zero, $[\fA] = 0$. Also note that the other obvious parity-odd vector $\hat B_\alpha$ is related to the two we use via \eqref{I3}, so does not comprise independent data.} The only scalars at hand are the entropy density $s$, the chemical potential $\mu$ and the norm $\chi$ of the transverse part of $D_\alpha \psi$:
\begin{align}
\chi\equiv \zeta_\alpha \zeta^\alpha\,, \qquad \text{with} \;\; \zeta_\alpha \equiv P_\alpha^\beta \,D_\beta\psi \,.
\label{}
\end{align}
All of these are parity-even and cannot appear in the action with gradients. The only other potential term of interest at this order is the parity-odd vector
$\eps^{\alpha\nu\rho\sigma}\; u_\nu\, \nabla_\rho \zeta_\sigma$. The three vectors in question can contract against the zero-derivative object $D_\alpha\psi$. However, not all three scalars thus constructed are independent. Indeed
\begin{align}
  &\nabla_\lambda D_\sigma \psi - \nabla_\sigma D_\lambda \psi = F_{\lambda \sigma} \;\;   \Rightarrow \;\; \eps^{\mu\nu\lambda\sigma}u_\nu (\nabla_\lambda D_\sigma \psi - \nabla_\sigma D_\lambda \psi - F_{\lambda \sigma}) 
     = 0 \notag \\
\;\;  \Rightarrow \;\; &\eps^{\alpha\nu\rho\sigma} D_\alpha\psi\, u_\nu \nabla_\rho \zeta_\sigma = 
    (2\,\mu \,\omega^\alpha + B^\alpha) D_\alpha\psi \notag  \,.
\end{align}
As a result the only combination of terms that is allowed at this order is
\begin{align}
  S_{wzII} = \int \sqrt{-g} \; \left[ \eta_\omega(s,\mu,\chi) \,\omega^\alpha + \eta_B(s,\mu,\chi) \,B^\alpha \right] D_\alpha \psi \; ,
    \label{S_ansatz}
\end{align} 
where we have allowed arbitrary functions $\eta_\omega$, $\eta_B$ of the three scalars.
Note that at zeroth order we refrained from introducing an explicit dependence on $\chi$: this was disallowed by the chemical shift symmetry \eqref{shiftSymm}. Now a dependence on $\chi$ is allowed because we do not require the action (\ref{S_ansatz}) to by itself be invariant under the chemical shift symmetry. Rather, only the sum $S_{wzI} + S_{wzII}$ 
needs to respect that symmetry.

We now need to determine the coefficients $\eta_\omega$ and $\eta_B$ such that the full action
\begin{align}
  S_{anom} =  S_{wzI} + S_{wzII} + S_\text{CS} 
\end{align}
is invariant under the chemical shift symmetry. This uniquely fixes the coefficient functions $\eta_\omega$ and $\eta_B$ in terms of the constant $\aleph$ and demanding that we have the correct $U(1)^3$ anomaly leads to \eqref{Sa}.

This is actually rather straightforward.  Under $\psi \rightarrow \psi + \mathfrak{f}(\phi^I)$, the variation is
\begin{align}
  \delta (S_{wzI} + S_{wzII}) &= \int \sqrt{-g} \left[ \left( \eta_\omega \,\omega^\alpha + \eta_B \, B^\alpha
      \right) \partial_\alpha \mathfrak{f} + \frac{\aleph}{2}\, \eps^{\alpha\beta\rho\sigma} \left\{(\partial_\alpha \mathfrak{f}) A_\beta F_{\rho\sigma} 
       - (\partial_\alpha \mathfrak{f}) \hat A_\beta \hat F_{\rho\sigma} \right\}\right]\notag\\
    &= \int \sqrt{-g} \left[-\nabla_\alpha(\eta_\omega \,\omega^\alpha + \eta_B\, B^\alpha) - \frac{\aleph}{4} \eps^{\alpha\beta\rho\sigma}
      \left( F_{\alpha\beta} F_{\rho\sigma}-\hat F_{\alpha\beta} \hat F_{\rho\sigma} \right)\right] \mathfrak{f}
 \label{ChemShift}
\end{align}
which implies that in order to get an invariant action we need to make the choice
\begin{equation}
\eta_\omega \, \omega^\alpha + \eta_B\, B^\alpha = \frac{1}{3}\, J_{anom}^\alpha  + {\cal J}^\alpha 
\,, \qquad \aleph = -c_{_A}
\label{etassol}
\end{equation}	
where ${\cal J}^\alpha $ is an exactly conserved gauge and chemical shift-invariant vector $\nabla_\alpha {\cal J}^\alpha = 0$.

The divergence of the current  $J^\alpha_{anom}$  in Eq. \eqref{J_sFrame0} has been calculated using:
\begin{align}
  \nabla_\alpha \left( \mu^2 \,\omega^\alpha + \mu\, B^\alpha \right) &= \left[(2\,\mu \,\omega^\alpha+B^\alpha) \,\nabla_\alpha \mu 
   +  \mu\, \eps^{\alpha\beta\rho\sigma} \, \nabla_\alpha u_\beta \nabla_\rho A_\sigma
   + \frac{\mu^2}{2} \eps^{\alpha\beta\rho\sigma}\, \nabla_\alpha u_\beta \nabla_\rho u_\sigma \right] \notag \\
   & =   -\frac{1}{8} \eps^{\alpha\beta\rho\sigma} (F_{\alpha\beta} F_{\rho\sigma} -\hat F_{\alpha\beta} \hat F_{\rho\sigma})
=  \left[ B_\alpha E^\alpha - \hat B_\alpha \hat E^\alpha\right]   
   \, , \label{CalcDiv}
\end{align}
where we applied (\ref{I3}), (\ref{I1}) and (\ref{I2}) on the first, second and third term of the first line, respectively. Therefore, demanding invariance under the chemical shift symmetry amounts to requiring (anomalous) current conservation. 

It is instructive to derive this result slightly differently to see the contribution from various terms. Consider the energy-momentum tensor and charge current derived from \eqref{Sa}. Since we work in entropy frame, we expect contributions to the heat current and the charge current proportional to the parity-odd vectors $\omega^\alpha$ and $B^\alpha$. Specifically, 
\begin{align}
q^{\alpha}_{anom} 
                 &=  {\mathfrak q}_\omega \, \omega^{\alpha} + {\mathfrak q}_B \, B^{\alpha}  \, , \qquad \Pi^{\alpha\beta}_{anom} = 0\,, 
\label{T_sFrame}\\
J^\alpha_{anom} & =  {\mathfrak j} _\omega \; \omega^\alpha + {\mathfrak j}_B \; B^\alpha \label{J_sFrame}
\end{align}
with coefficients $ {\mathfrak q}_\omega$, $ {\mathfrak q}_B$, ${\mathfrak j} _\omega$, ${\mathfrak j} _B$ which are a-priori functions of $s$, $\mu$ and $\chi$.  As we will see, demanding that we obtain the correct form of the conserved currents 
suffices to determine uniquely $\eta_\omega$, $\eta_B$ and reproduce the solution \eqref{etassol}.

Carrying out the variations with respect to the background sources, from the full anomaly action $S_{anom}$ we find the stress-energy tensor and current at first order to be:\footnote{For details on the calculation and the different contributions from distinct parts of the action, see Appendix~\ref{appendix:pieces}.}
\begin{align}
\Pi^{\alpha\beta}_{anom}
  &= -\left[ s\left( \eta_{\omega,s}\,  \omega^\nu + \eta_{B,s} \, B^\nu \right) D_\nu \psi\right] P^{\alpha\beta} \notag \\
  &\quad- \left[2\left(\eta_{\omega,\chi}\,\omega^\nu + \eta_{B,\chi} \, B^\nu\right) D_\nu\psi \right] \zeta^\alpha \zeta^\beta \,\label{Pianom}\\
2\,q^\alpha_{anom}  &= 4\, \mu \, \eta_\omega \,\omega^{\alpha}  +(\eta_\omega +2\,\mu\,\eta_B) B^{\alpha}
   +4\, \mu \,(\eta_{\omega,\chi}\, \omega^\nu+\eta_{B,\chi}\,B^\nu)\,D_\nu\psi\; \zeta^{\alpha} \notag \\
     &\quad + \left[ \left(\mu\, \eta_{\omega,\mu} - 2 \, \eta_\omega\right)\omega^\nu\,  	D_\nu\psi + \left(\mu\,\eta_{B,\mu} -  \eta_B\right) B^\nu\, D_\nu\psi\right] u^\alpha  \notag \\
	    &\quad
   +\;\eta_{\omega,\chi} \, \eps^{\alpha\nu\rho\sigma}\, D_\nu\psi \,u_\rho \nabla_\sigma \chi
   -\left(\eta_{\omega,\mu}-4\,\aleph\, \mu\right) \, \eps^{\alpha\nu\rho\sigma}\, D_\nu \psi \,u_\rho \hat E_\sigma\, , \label{qanom}\\
J^\alpha_{anom} 
  &= \left[(\eta_{\omega,\mu}-2\, \eta_B)\,\omega^\nu D_\nu \psi + \left(\eta_{B,\mu} - 2 \,\aleph\right) B^\nu \,D_\nu \psi\right] u^\alpha \notag \\
  &\quad+\left[ 2(\eta_{\omega,\chi}\, \omega^\nu + \eta_{B,\chi} \,B^\nu )D_\nu \psi\right] \zeta^\alpha\notag\\
  &\quad+  (\eta_\omega + 2\,\mu \,\eta_B) \omega^\alpha + \left(2\, \eta_B + 2\,\aleph\, \mu\right) B^\alpha
  +\eta_{B,\chi} \,\eps^{\alpha\nu\rho\sigma} \,D_\nu\psi\, u_\rho \,\nabla_\sigma\chi
   \notag\\
  &\quad 
   +\eps^{\alpha\nu\rho\sigma}\, D_\nu \psi\, u_\rho\, \left(2\,\aleph\,(\hat E_\sigma-E_\sigma) + \eta_{B,\mu} \, \nabla_\sigma \mu 
   + \eta_B \, {\mathfrak a}_\sigma \right) \, . \label{Janom}   
\end{align}

We must now require that
(\ref{Pianom}, \ref{qanom}, \ref{Janom}) take the forms (\ref{T_sFrame}, \ref{J_sFrame}) and solve for the coefficient functions $\eta_\omega$ and $\eta_B$
in terms of the thus far undetermined constant $\aleph$. This means that we need to solve the following equations:
\begin{eqnarray}
 \begin{aligned}
   \eta_{\omega,s} = \eta_{B,s} &=0 \,,\\
   \mu\, \eta_{B,\mu}-\eta_B &= 0\,, \\
   \eta_{\omega,\mu} - 4\, \aleph\,\mu &= 0 \,,
 \end{aligned}
\quad  \quad \begin{aligned}
   \mu \, \eta_{\omega,\mu}-2\,\eta_\omega &= 0  \,, \\
   \eta_{\omega,\chi} = \eta_{B,\chi} &= 0 \,,\\
   \eta_{\omega,\mu} -2\,\eta_B &= 0 \,,
 \end{aligned}\\
  2\,\aleph\,\hat E_\sigma-2\,\aleph \,E_\sigma + \eta_{B,\mu} \, \nabla_\sigma \mu 
   + \eta_B \, {\mathfrak a}_\sigma = 0 \;\,\notag
\end{eqnarray}
This system of equations has a rather simple solution: 
\begin{align}
  \eta_\omega = 2\, \aleph\, \mu^2\, , \qquad \eta_B = 2\, \aleph \, \mu \; . 
\end{align}
The expressions (\ref{Pianom}, \ref{qanom}, \ref{Janom}) thus take the forms (\ref{T_sFrame}, \ref{J_sFrame}) with
\begin{align}
{\mathfrak q}_\omega = 4\,\aleph \,\mu^3  \, , \qquad {\mathfrak q}_B = 3\, \aleph\,\mu^2  \, ,
 \qquad 
{\mathfrak j}_\omega = 6 \,\aleph \,\mu^2 \,, \qquad {\mathfrak j}_B = 6 \,\aleph \,\mu  \; . \label{Coeffs}
\end{align}
Finally, we see that comparison with \eqref{AnConservation} fixes $\aleph$ as quoted in \eqref{etassol}.

This confirms that there exists an effective action with the proposed symmetries (gauge invariance and chemical shift symmetry) which is indeed able to capture the constitutive results for the anomalous parts of the currents. The stress tensor and the charge  current which are defined by Eqs.\ (\ref{T_sFrame}, \ref{J_sFrame}) and (\ref{Coeffs}) coincide with the result of the entropy current analysis of \cite{Loganayagam:2011mu} (in entropy frame). The reader is invited to consult Appendix~\ref{sec:relothers} for a comparison with earlier discussions in the literature -- we map our answer to the Landau frame and demonstrate that the shadow terms vanish in equilibrium there.

\subsection{Currents \& dynamics}
\label{s:curdyn}

So far we have established that using the fluid element variables, we can construct an action that reproduces the solution \eqref{T_sFrame0} and \eqref{J_sFrame0} to \eqref{adiabatc}. As described in \cite{Loganayagam:2011mu} the latter equation arises as an off-shell constraint on anomalous fluid dynamics. Note that in deriving it a crucial assumption has been made, viz., the local form of the second law of thermodynamics is unchanged by the presence of quantum anomalies. More specifically, the constraints from the second law on dissipative parts of transport are independent of the theory suffering from anomalies.

A-priori it would appear that we are pretty much done. We have a parity-odd first order action $S_{anom}$ in \eqref{Sa} which together with the zeroth order action \eqref{S0def} should serve to determine the dynamics of the system. Ideally, we would like to see that the dynamics of the system is given by \eqref{hydroan}. Unfortunately this is not true as we now describe.

As we remarked in \S\ref{sec:0orderFormalism} there are the dynamical Euler-Lagrange equations of the system $S_0 + S_{anom}$ obtained by varying the fields $\phi^I$ and $\psi$. In the absence of anomalies these are isomorphic to the conservation of energy-momentum and charge. However, we need to be careful in the presence of anomalies \cite{Dubovsky:2011sk}. 

Consider first the charge current dynamics; we have the equation of motion of the $\psi$ field 
\begin{equation}
\nabla_\alpha J_\text{N}^\alpha = 0 \,,\qquad  J_\text{N}^\alpha \equiv \frac{\delta (S_0+S_{anom})}{\delta \nabla_\alpha \psi} \,,
\label{}
\end{equation}	
where the subscript N denotes the fact that here we are calculating a Noether current. It is a simple matter to check that 
\begin{align}
J_\text{N}^\alpha &= \rho\, u^\alpha + \frac{1}{3}\, J_{anom}^\alpha  - \frac{c_{_A}}{2}\, \eps^{\alpha\beta\gamma\delta}\left(A_\beta\, F_{\gamma\delta} - {\hat A}_\beta \, {\hat F}_{\gamma\delta}\right) \notag \\
&= \left(J_\text{cov}\right)^\alpha - \frac{2}{3}\, J_{anom}^\alpha   - \frac{c_{_A}}{2}\, \eps^{\alpha\beta\gamma\delta}\left(A_\beta\, F_{\gamma\delta} - {\hat A}_\beta \, {\hat F}_{\gamma\delta}\right)
\label{}
\end{align}	
where we have written the total covariant current arising from the action $S_0+S_{anom}$. It is then easy to see
using \eqref{AnConservation} that the conservation of the Noether current associated with $\psi$-translations implies that 
\begin{equation}
\nabla_\alpha J_\text{cov}^\alpha = \frac{3\, c_{_A}}{4}\, \eps^{\alpha\beta\gamma\delta}
\left(F_{\alpha\beta} \, F_{\gamma\delta} - {\hat F}_{\alpha\beta}\, {\hat F}_{\gamma\delta} \right)
\label{}
\end{equation}	
or $d\star \form{J}_\text{cov} =3\, c_{_A} \left(\fF\wedge \fF - \fFh \wedge \fFh \right) $.
So the covariant current does not reproduce the correct anomaly, but rather the difference of the anomaly in the gauge field $\fA$ and its hatted counterpart $\fAh$. It is easy to check that the consistent current 
\begin{equation}
J_\text{cons}^\alpha = \frac{\delta (S_0+S_{anom})}{\delta A_\alpha} = J_\text{cov}^\alpha - c_{_A}\, \eps^{\alpha\beta\gamma\delta}\left(A_\beta\, F_{\gamma\delta} - {\hat A}_\beta \, {\hat F}_{\gamma\delta}\right)
\label{}
\end{equation}	
and again gives the difference of the consistent anomaly for the gauge field and its hatted counterpart. A similar problem arises in case of the stress tensor conservation equation as indicated in \eqref{Summary2}; we demonstrate 
this explicitly in Appendix~\ref{sec:stensorA}.
We conclude that these equations of motion do not have the form of conservation equations of anomalous hydrodynamics. Note, however, that the corrections as compared to traditional hydrodynamic equations of motion only involve transverse gauge fields and therefore vanish in equilibrium.  We will resolve this issue in \S\ref{sec:Schwinger-Keldysh} invoking a formalism suitable for non-equilibrium physics.

\section{Global anomalies in all even dimensions}
\label{sec:AllDimensions}

Thus far we have focussed on the triangle anomaly in $3+1$ dimensions. The extension of our analysis to include global anomalies in all even dimensions is relatively simple. Aided by the key observation which is the generalization of \eqref{AnConservation}, we show that the effective action can be written down exploiting the analysis of \cite{Loganayagam:2011mu}. We sketch the general structure below, demonstrating that the physical requirements are met.  We contrast our analysis in $d=2$ with the earlier discussion of \cite{Dubovsky:2011sk}.

\subsection{Deriving the anomalous effective action}
\label{sec:GeneralResult}

There is a straightforward way to generalize the procedure of the previous section
to all even dimensions: in $d=2n$ dimensions the non-invariant action which is supposed to eventually give the anomaly becomes
\begin{align}
  S_{wzI} = - c_{_A} \, \int_{{\cal M}_{2n}} \left[{\bf D}\psi \wedge \fA \wedge \fF^{n-1} - {\bf\hat  D} \psi \wedge \fAh\wedge \fFh^{n-1} \right]\, ,
  \label{SanomAllg}
\end{align} 
where $\fF^{n-1}$ denotes the wedge product of $n-1$ copies of $\fF$. Similarly, the Chern-Simons action which restores gauge invariance will be
\begin{align}
 S_\text{CS} = c_{_A} \int \left[\fA \wedge \fF^n - \fAh\wedge \fFh^n \right] \,.
 \label{eq:SCS2}
\end{align}
These two actions must be supplemented by the most general gauge invariant (but not chemical shift-invariant)
action $S_{wzII}$ that can be written down in $d$ dimensions. Each term in this action is accompanied by an a-priori undetermined 
function of all possible independent scalar data. By requiring the stress-energy tensor and charge current to take the general
form (\ref{T_sFrame}, \ref{J_sFrame}), one determines these coefficient functions. After having determined these free
parameters, the complete action $S_{wzI}  + S_\text{CS} + S_{wzII}$ should be invariant under
the chemical shift symmetry. 

In fact, we can carry out this idea explicitly in arbitrary dimensions. We have seen in the four-dimensional analysis that the integrand of $S_{wzII}$ was proportional to $\star\form{J}_{anom} \wedge {\bf D}\psi$, where $\star\form{J}_{anom}$ is the $(2n-1)$-form current Hodge dual to $\form{J}_{anom}$. We thus use as an ansatz that $S_{wzII}$ in $d$-dimensions should have an integrand proportional to the $d$-dimensional analog of $\star\form{J}_{anom}\wedge {\bf D}\psi$. From \cite{Loganayagam:2011mu} we know what $\star\form{J}_{anom}$ looks like in arbitrary dimensions. Using that knowledge, we claim that the solution in $d=2n$ dimensions reads explicitly
\begin{align}
  S_{wzII} = -c_{_A} \int \frac{\fB^n- \fBh^n}{2{\boldsymbol \omega}} \wedge \fu \wedge {\bf D}\psi \,, \label{SPsi}
\end{align}
where we have already fixed the numerical pre-factor relative to $S_{wzI}$ with the benefit of hindsight. For convenience we use only differential forms here. The magnetic 2-form $\fB$ and the vorticity 2-form ${\boldsymbol \omega}$ are defined as
\begin{align}
  \fB \equiv \fF - \fu\wedge \fE \,, \qquad 2{\boldsymbol \omega}\equiv d\fu + \fu \wedge \fa \,,
\end{align}
where $\fa$ is the acceleration 1-form. Note that dividing by a 2-form in (\ref{SPsi}) is a well defined procedure because every term in the numerator contains at least one factor of $2{\boldsymbol \omega}$. Also, every term in the numerator is a product of 2-forms, so their ordering is arbitrary and there is no problem with sign conventions when performing the division. 

The above Eqs.\ (\ref{SanomAllg}-\ref{SPsi}) are the main result of this section and they solve the problem of anomalies in the effective action approach in all even dimensions. 
One can check that the above three contributions to the anomaly action give the following 
anomalous contributions to the charge current:
\begin{align}
\star\form{J}_{anom} &= \star \left(\form{J}_{wzI} + \form{J}_{wzII} + \form{J}_\text{CS} \right)
               =-(n+1) c_{_A} \, \frac{\fBh^n - \fB^n}{2{\boldsymbol\omega}} \wedge \fu \,,\label{J_2n_sol}\\
\star \form{ q}_{anom} &=- c_{_A} \frac{\fB^{n+1} - \fBh^{n+1} + (n+1)\,(2\fomega \mu)\wedge \fBh^n}{(2\fomega)^2} \wedge \fu \,.
  \label{Pi_2n_sol}
\end{align}
such that the gauge invariant current and stress tensor are given by what one would expect from the analysis in \cite{Loganayagam:2011mu}.
Here $\form{q}_{anom}$ is the $(2n-1)$-form Hodge dual of $q^\alpha_{anom}$ that can be derived in a similar fashion as $\form{J}_{anom}$.

We can easily check that the combined action $S_{wzI}+S_{wzII}$ is invariant under the chemical shifts $\psi \rightarrow \psi + \mathfrak{f}(\phi^I)$. The variations are, respectively:
\begin{align}
 \delta S_{wzI} &=-c_{_A} \int \prn{ d\mathfrak{f}\wedge \fA \wedge \fF^{n-1}
    - d\mathfrak{f} \wedge \fAh\wedge \fFh^{n-1} } \notag \\
    &= c_{_A} \int  \left( \fF^n - \fFh^n \right) \cdot \mathfrak{f} \,,\\
 \delta S_{wzII} &=- c_{_A} \int \frac{\fB^n-\fBh^n}{2{\boldsymbol \omega}} \wedge \fu \wedge d\mathfrak{f} \notag\\
    &=n c_{_A} \int \left(\form{E}\wedge \fB^{n-1}-\form{{\hat E}}\wedge \fBh^{n-1}\right) \wedge \fu \cdot \mathfrak{f} \notag\\
    &= -c_{_A} \int \left( \fF^n - \fFh^n \right) \cdot \mathfrak{f} \,,
\end{align}
where we used integration by parts in both calculations. In the last step we used
\begin{align}
  \fF^n = n \cdot \fB^{n-1} \wedge \fu \wedge \fE 
\end{align}
and similarly for the transverse gauge field (cf., \cite{Loganayagam:2011mu}).

Finally, in Appendix~\ref{sec:stensorA} we show that the conservation equations satisfied by this off-shell solution (in the sense of traditional hydrodynamics) are 
\begin{align}
 \nabla_\alpha J^\alpha_{anom}  &= \JH^\perp -\JHh^\perp
      \,, \qquad
  \nabla_\beta T^{\alpha\beta}_{anom} = F^\alpha{}_\beta J^\beta_{anom} - \mu\ u^\alpha  \, \JHh^\perp\,.
\label{AllDimConsEq}
\end{align}
where $\star_{2n+1} \fJH =(n+1) c_{_A} \fF^n $ and $\star_{2n+1} \fJHh =(n+1) c_{_A} \fFh^n $.
Again these equations encode hydrodynamical dynamics up to unwanted terms that are built from the hatted gauge connection. 
The resolution of this problem is postponed to \S\ref{sec:Schwinger-Keldysh}.

\subsection{Consistency checks}
\label{sec:Consistency}

One can immediately check that the action (\ref{SPsi}) in $d=4$ reduces to our result in \S\ref{sec:4dAnomaly}. In order to do so, we just need to note that
\begin{align}
 B^\alpha = \frac{1}{2} \eps^{\alpha\beta\gamma\delta} u_\beta B_{\gamma\delta} \,, \qquad
 \omega^\alpha = \frac{1}{2} \eps^{\alpha\beta\gamma\delta} u_\beta \omega_{\gamma\delta} \,,
\end{align}
and obtain in $d=4$:
\begin{align}
  S_{wzII}  = -c_{_A} \int \frac{\fB^2-\fBh^2}{2{\boldsymbol \omega}} \wedge \fu \wedge {\bf D}\psi 
          = -2 \,c_{_A} \int \sqrt{-g}\, (\mu\, B^\alpha + \mu^2 \, \omega^\alpha ) D_\alpha \psi \,.
\end{align}
 
As a further check of this proposal, we demonstrate that we can reproduce the anomaly in $d=2$ using our techniques. This is of a certain interest because the two-dimensional problem has been solved in \cite{Dubovsky:2011sk} in a slightly different manner. As we show in Appendix~\ref{appendix:2Danom},
following the logic we employ in the 4-dimensional case \S\ref{sec:4dAnomaly} equally well works in two dimensions and we obtain the result of \cite{Dubovsky:2011sk}.
The main difference from their analysis is that we allow for the presence of the transverse fields built from $\fAh$. As a result of the previous discussion for arbitrary dimensions, we propose the following action for the anomalous transport in $d=2$ (note that now we are using the fact that the anomaly appears at zero derivative order)
\begin{equation}\begin{split}
  S_{anom}^\text{(2d)} &= -c_{_A} \int_{{\cal M}_2 } ({\bf D\psi} \wedge \fA - {\bf \hat D} \wedge \fAh)
   -c_{_A}  \int \sqrt{-g}\, \mu \, \eps^{\alpha\beta} \,D_\alpha \psi\, u_\beta \\
   &\qquad + c_{_A}  \int_{\mathcal{M}_3} (\fA\wedge \fF- \fAh\wedge \fFh)
\end{split}\end{equation}
This actions  gives rise to the following stress tensor and current:
\begin{align}
 q_{anom}^{\alpha} = -2\, c_{_A} \,\mu^2 \, \eps^{\alpha\rho}\, u_\rho  \, , 
 \qquad J_{anom}^\alpha = -2 \,c_{_A} \,\mu \,\eps^{\alpha\rho} u_\rho \, .  \label{qJ2d}
\end{align}
We see again that the anomalous first order piece of the current satisfies an analog of Eq.\ (\ref{AnConservation}):
\begin{align}
   \nabla_\alpha J^\alpha_{anom} = c_{_A}  \, \eps^{\alpha\beta} (F_{\alpha\beta}-\hat F_{\alpha\beta}) \, .
  \label{Jcons2d}
\end{align}
Eq.\ (\ref{qJ2d}) is precisely the result obtained in \cite{Dubovsky:2011sk}. Once again only the dynamics
described by Eq.\ (\ref{Jcons2d}) is  spoiled by
the presence of $\fFh$. Note that this however doesn't affect equilibrium results. It is easy to see that we reproduce the equilibrium partition function of \cite{Banerjee:2012cr} in all dimensions.

In order to be ultimately able to repair this unfortunate state of affairs, the next section introduces some formalism which 
enable us to write our results so far in a much more efficient way.

\section{Transgression forms and non-abelian effective actions }
\label{sec:NonAbelianFormalism}

Thus far we have used the basic framework of the non-dissipative effective actions to capture the anomalous contribution. While the computation is straightforward, one can present the result in a much more elegant fashion. We will now show that the
effective action which encodes the gauge anomaly of non-dissipative hydrodynamics can be written as a transgression form. Along the way we will see a simple way to generalize our construction to non-abelian global symmetries. In order to do so we take inspiration from  holographic ideas \cite{Nickel:2010pr}, viewing the charge field $\psi$ as a Wilson line interpolating between the boundary and the horizon. This will then naturally suggest a more general picture a la Schwinger-Keldysh which we will eventually use to fix the Ward identities. 

\subsection{Deconstructing anomalous liquids}
\label{sec:non-abelian}

Let us imagine for a moment that we are working with a holographic liquid such as
those that  naturally arise in the fluid/gravity context. The gauge field source $\fA=A_\mu\, dx^\mu$
then lives on the boundary of some asymptotically AdS bulk spacetime and should be the
boundary value of a bulk gauge field $A_a\, dx^a$. Here $\{x^\mu\}$ are the boundary coordinates
and $\fA$ is valued under the  boundary  global symmetry group $G_\partial$.\footnote{For non-abelian
symmetry algebras, we will work in an anti-hermitian basis. We follow the conventions described in 
\cite{Jensen:2013kka} for non-abelian gauge fields, currents etc.}
The bulk geometry dual to the fluid is generically a black hole spacetime \cite{Hubeny:2011hd}. 
While for a general fluid flow such a black hole is inhomogenous and dynamical, to discuss 
anomalies we are only interested in adiabatic flows. So we imagine that we have a stationary
inhomogeneous black hole geometry. On the spatial sections of the horizon we introduce coordinates
$\{\phi^I\}$: these will be conflated with the fluid element fields we used to construct our 
effective action.\footnote{For practical purposes it suffices to think about the 
stretched horizon and the fields $\phi^I$ living on spatial sections of a timelike hypersurface straddling the true event horizon.} For completeness let us also record that the temporal direction will be denoted by affine parameter $v$ along the horizon (which goes along for the ride by virtue of adiabaticity).

To describe the charge sector, following the analysis of \cite{Nickel:2010pr} we introduce a Goldstone mode $c$ which generalizes the phase $\psi$ that was used earlier to discuss abelian currents. We assume that there is an independent gauge symmetry on the horizon, with gauge group $G_\text{h}$ and a horizon gauge field $\fA_\text{h}$. Clearly, since we are visualizing the horizon degrees of freedom as corresponding to the effective action variables, we should demand that the chemical shift symmetry is realized as a gauge transformation of this horizon gauge field. Note  that the in subscripts $G_\partial$ and $G_\text{h}$ are only used to refer to the locations where the symmetries act; we are interested in fluids carrying $G$-valued charges.

With the basic fields in place let us turn to the symmetries: the gauge potentials transform under the appropriate symmetries as usual, being adjoint-valued in the Lie algebras. The charge field $c$ should interpolate between the boundary and the horizon; it therefore must transform as a bi-fundamental of $G_\partial \times G_\text{h}$. For group elements  $(g(x^\mu),f(\phi^I,v)) \in G_\partial \times G_\text{h}$, the various fields transform as:
\begin{align}
c &\to g^{-1} \, c \, f \,,\qquad
\fA \to g^{-1} \fA \, g + g^{-1} \, dg  \,\qquad
\fA_\text{h} \to f^{-1} \fA_\text{h} \, f + f^{-1} \, df \,.
\end{align}

What we achieved by this interpretation is simply to gauge the chemical shift symmetry \eqref{shiftSymm} (more precisely its non-abelian generalization). The horizon gauge transformations are the chemical shifts of the $c$ field and $\fA_\text{h}$  gauges this symmetry. From the transformation rules we see that the natural gauge-covariant derivative is just
\begin{align}
 {\bf \mathcal{D}}c  = dc + \fA c - c \fA_\text{h} \qquad\text{such that}\qquad \mathcal{D}c \to g^{-1} \mathcal{D}c \, f \,.
 \label{Dcdef}
\end{align}
We are now in a position to define the non-abelian gauge covariant chemical potential as 
\begin{align}
\mu = i_u \left[  {\bf \mathcal{D}} c \, c^{-1} \right]\qquad \text{such that} \qquad \mu \to  g^{-1} \mu \,g \,,
\label{munadef}
\end{align}
where $i_u$ denotes contractions, i.e., $i_u(\fA) = u^\mu A_\mu$. It is easy to check that this definition is gauge-covariant and chemical shift invariant.

We now make the following claim: the hydrodynamic shadow gauge field which appeared somewhat mysterious in the previous analysis, 
has a natural interpretation in terms of the horizon gauge field. In fact, 
\begin{align}
 \fAh= c\, \fA_\text{h} \, c^{-1} - dc \, c^{-1} \,, \label{eq:AHat}
\end{align}
This definition of $\fAh$ is basically motivated from the requirement that the 
abelian identity $\mu = i_u(\fA-\fAh)$ generalize to the non-abelian case. As  such this leaves us the freedom to shift $\fAh$ by a transverse piece relative to the definition \eqref{eq:AHat}; we gauge fix this extra freedom and simply work with \eqref{eq:AHat}. 

To motivate our identification of the hydrodynamic shadow field with the horizon gauge field, let us recall that the fluid/gravity correspondence posits that the horizon ought to be regular \cite{Bhattacharyya:2008xc}. Since we are turning on boundary values of bulk gauge fields on the horizon in $\fA_\text{h}$, demanding horizon regularity is tantamount to asking the chemical potential to vanish there, i.e.,
\begin{equation}
i_u (\fA_\text{h}) = 0 \,.
\label{}
\end{equation}	
Computing then the chemical potential difference between the boundary and the horizon using 
\eqref{munadef} we can show that \eqref{eq:AHat} follows.
By viewing the gauge symmetry and the chemical shift symmetries as 
two manifestations of the same symmetry on the boundary and at the horizon respectively, in this deconstruction language we see that the shadow gauge field $\fAh$ is
just the boundary manifestation of the horizon gauge symmetry or equivalently the chemical shift symmetry (up to dressing by a Wilson line).  

Operationally, the identification (\ref{eq:AHat}) is useful because it defines $\fAh$ as a gauge field 
which transforms in the same way as $\fA$ under gauge transformations: 
\begin{align}
\fAh\to g^{-1}\fAh\, g + g^{-1} dg \,.
\end{align}
We will now proceed to show how the results of our previous analysis can be recast in terms of a simple transgression form using this picture.

\subsection{Anomalous effective action as a transgression form}
\label{sec:transgression}
Having formulated a non-abelian setup for effective field theories of non-dissipative hydrodynamics we 
can now proceed to generalize the action that was responsible for
the anomaly. It turns out that such a generalization can easily be found if we look at 
our previous results from the point of view of transgression forms.

Given a non-abelian Chern-Simons term $\ICS$ (and the corresponding anomaly polynomial $\fP=d\ICS$ ),
the transgression form for two gauge fields $\fA_1$ and $\fA_2$
is defined as \cite{Nakahara:2003fk,Mora:2006ka,Jensen:2013kka}
\begin{align}
\VP \brk{\fA_t} &\equiv \int_0^1 dt \brk{\frac{d \fA_t}{dt}  \cdot \prn{\frac{\partial \fP}{ \partial \fF} }_t
 }
 = \int_0^1 dt\ \frac{d \fA_t}{dt}  \cdot \prn{\star_{2n+1} \fJH  }_t \,, \label{eq:TransgrDef}
\end{align}
where  $\fA_t$ is an interpolating field
\begin{align}
 \fA_t &\equiv t\, \fA_1 + (1-t) \,\fA_2 \,,\qquad \fF_t = d\fA_t + \fA_t^2 \,,\qquad \frac{d \fA_t}{dt} = \fA_1 -\fA_2\equiv \Delta \fA.
\end{align}
Here, $\cdot$ indicates a trace over gauge group adjoint indices. 
By construction, the transgression form  transforms covariantly. In fact, 
it  can be written as the difference of two Chern-Simons forms plus an exact form (see Appendix~\ref{appendix:transg}): 
\begin{align}
 \VP[\fA_1,\fA_2]
 \equiv \ICS[\fA_1] - \ICS[\fA_2] - d\WCS[\fA_1,\fA_2] \,,
 \label{eq:Transgr}
\end{align}
where 

\begin{equation}
\begin{split}
\WCS \brk{\fA_t} &\equiv  \int_0^1 dt \brk{\frac{d \fA_t}{dt}  \cdot \prn{\frac{\partial \ICS}{ \partial \fF} }_t
 }
 =\int_0^1 dt\ \frac{d \fA_t}{dt}  \cdot \prn{\star \fJBZ }_t
\end{split} 
\end{equation}
The fact that the transgression form written as Eq.\ (\ref{eq:Transgr}) is gauge covariant is easily seen from 
the fact that it only depends on the {\it difference} of two gauge connections, such that their 
respective gauge non-invariance cancels out.

We will now prove the claim that our effective action $S_{anom}$ given by \eqref{Sa} in $d=4$ or 
its generalization discussed in \S\ref{sec:AllDimensions} is just the transgression form for 
the two gauge fields $\fA$, $\fAh$, i.e.,\
\begin{align}
 S_{anom}^{\text{(abelian)}} \equiv S_{wzI} + S_{wzII} + S_\text{CS} 
 = \left[ \int_{\mathcal{M}_{2n+1}} \VP[\fA,\fAh]\right]_\text{abelian} \, . 
  \label{eq:AbelianClaim}
\end{align}
It is easy to intuit this result: firstly $\fA$ and $\fAh$ transform similarly under gauge transformations
(which include both the boundary gauge transformations as well as the chemical shift transformations). This being the case,
their difference $\Delta \fA$ as well as the interpolating field strength $\fF_t$ transform as 
tensors such that their trace is gauge invariant.  All the symmetries that we imposed in
the construction of our abelian effective action (gauge invariance and chemical shift symmetry) are thus
\textit{manifestly} preserved by the transgression form.\footnote{The consistent use of differential
form language makes  sure that also that diffeomorphism invariance is manifest.}  So clearly 
the effective action must have a simple realization in terms of this object.

Let us verify our claim explicitly: it is trivial to see how the two Chern-Simons terms in 
Eq.\ \eqref{eq:Transgr} reproduce $S_\text{CS}$.
Let us therefore consider the remaining part $S_{wzI}+S_{wzII}$. It is useful to rewrite
\begin{align}
 S_{wzI} &= -c_{_A} \int_{\mathcal{M}_{2n}} \left[ {\bf D}\psi \wedge \fA \wedge \fF^n
     - {\bf \hat D}\psi\wedge \fAh\wedge \fFh^{n-1}\right]\notag\\
  &= -c_{_A} \int d\psi \wedge d\left( \frac{\fu}{2{\boldsymbol \omega}} \right) \wedge \left(\fA\fF^{n-1} - \fAh\fFh^{n-1} \right)  \label{eq:SA}
\end{align}
where we used the fact that $d \left( \tfrac{\fu}{2{\boldsymbol \omega}}\right)= 1$. We can write $S_{wzII}$ as 
\begin{align}
 S_{wzII} &= - c_{_A} \int_{\mathcal{M}_{2n}} \frac{\fB^n- \fBh^n}{2{\boldsymbol \omega}} \wedge \fu \wedge {\bf D}\psi   \notag\\
   &= -c_{_A} \left[\int_{\mathcal{M}_{2n}} \frac{\fu}{2{\boldsymbol\omega}} \left( \fA\fF^n - \fAh\fFh^n \right) - \int_{\mathcal{M}_{2n}}
     d\psi \wedge \frac{\fu}{2{\boldsymbol \omega}}  \wedge \left(\fF^{n} - \fFh^n\right) \right]\,, \label{eq:Spsi}
\end{align}
where we used $\fB^n\wedge \fu = \fF^n\wedge \fu$ and similarly $\fu \wedge {\bf D}\psi =\fu\wedge( d\psi + \fA) = \fu\wedge(d\psi+\fAh)$.
If we now add $S_{wzI}+S_{wzII}$, we see that by means of $\fF = d\fA$, the r.h.s.\ of 
Eq.\ (\ref{eq:SA}) together with the second integral in the r.h.s.\ of (\ref{eq:Spsi}) gives an integral over an exact form which vanishes. We conclude 
\begin{align}
S_{wzI} + S_{wzII} &= -c_{_A}\int_{\mathcal{M}_{2n}} \frac{\fu}{2{\boldsymbol\omega}} \left( \fA\fF^n - \fAh\fFh^n \right) \notag\\
  &=- \left[\int_{\mathcal{M}_{2n}} \WCS[\fA,\fAh] \right]_\text{abelian}\,, 
 \label{Bconst}
\end{align}
where we have used \eqref{eq:HatUnhTransgr} in the last step. This proves the claim (\ref{eq:AbelianClaim}) 
that $S_{anom}^\text{(abelian)}$ is just given by the integral of a transgression form.

\subsection{An effective action for non-abelian anomalous hydrodynamics}
\label{sec:non-abelian-action}

Motivated by the result of the previous section, it is now clear what the non-abelian generalization
of our anomalous effective action should be. We just write  down the transgression form 
with $\fA$ and $\fAh$ using the definitions \eqref{munadef} and \eqref{eq:AHat}. In short,
\begin{equation}\label{eq:NonAbelianClaim}
\begin{split}
 S_{anom} &= \int_{\mathcal{M}_{2n+1}} \VP[\fA,\fAh] =  \int_{\mathcal{M}_{2n+1}} \frac{\fu}{2\fomega}\wedge \prn{\fP-\fPh} \\
    &=   \int_{\mathcal{M}_{2n+1}} \left[\ICS [\fA]-  \ICS [\fAh]\right]
      -  \int_{\mathcal{M}_{2n}} \WCS[\fA,\fAh] \\
    &=   \int_{\mathcal{M}_{2n+1}} \left[\ICS-\hICS\right]
      -  \int_{\mathcal{M}_{2n}} \frac{\fu}{2\fomega}\wedge \prn{\ICS-\hICS} \,.
\end{split}
\end{equation}
Note that as far as this construction goes two Chern-Simons terms provide appropriate amounts of anomaly inflow
into the theory with action $\WCS$ on the boundary. Clearly, we have an anomalous contribution both
in the original gauge symmetry as well as in the shadow fields. This explains the appearance of transverse fields
in the anomalous conservation equations which are derived from these actions.

In Appendix~\ref{sec:stensorA}  we re-derive the anomalous conservation equations in this (non-abelian) formalism. 
In order to get a feeling for the non-abelian case, we quote the result for the anomalous part of the charge current in two-dimensional theories:
\begin{align}
 \star\form{J}_{2}^I = -2c_{_A}\, \mu^I\, \fu \, .
\end{align}
with $I$ being the adjoint index associated with the symmetry group $G$.  In the abelian case, this result reduces to the well-known expression \eqref{qJ2d} obtained before in \cite{Dubovsky:2011sk}.

To summarize, we have seen how it is useful and insightful to interpret the Lagrangian for the anomalous part of the charge current and stress tensor
as a transgression form. This has several advantages: first of all, the symmetries (diffeomorphism invariance, gauge invariance and chemical shift symmetry)
do not have to be imposed by hand, but are manifest by construction. Secondly, the transgression form is very easy to vary with respect to the gauge 
field, so the charge current is easily obtained from it.\footnote{Note that because the action satisfies the adiabaticity constraint by construction, 
we can efficiently get the stress tensor by plugging the charge current into the adiabaticity constraint and solving for the stress tensor.}
Furthermore, since the transgression formalism constructs covariant actions, we immediately are able to generalize the discussion to non-abelian symmetries. Finally, the fact that $S_{anom}$ is given by a transgression form, shines some light on the physical significance of the transverse gauge field as originating from a horizon field that treats the
chemical shift symmetry as another gauge symmetry.

\section{Schwinger-Keldysh formalism: covariant currents with correct dynamics}
\label{sec:Schwinger-Keldysh}

Our discussion thus far has involved constructing an effective action which respects
the symmetries of the underlying physical system formulated in terms of the fluid element variables. 
The general picture we developed in \S\ref{sec:NonAbelianFormalism} makes 
it clear that the transgression form is the natural action that is compatible with the symmetries. 
Phrased this way we have a clear problem with the Ward identities: the anomaly inflowing into 
the boundary is a combination of the (desired) global anomaly associated with the background gauge field
and the undesirable one associated with the hydrodynamic shadow field. As we have argued 
before this shadow contribution vanishes in equilibrium, rendering our picture correct in that limited context.

What does it take to ensure that there is an effective action valid outside equilibrium that allows us 
to obtain the correct anomalous currents along with the correct Ward identities?  We can turn this question
around and ask how to derive the Ward identities in general for anomalous systems in the non-equilibrium regime. 
This is clearly the remit of the real-time Schwinger-Keldysh (SK) formalism. In this section we will 
therefore examine the Ward identities  more carefully. After developing a framework for treating anomalous
systems in the Schwinger-Keldysh language we proceed to apply this prescription to hydrodynamics. At the end
of the day we will be able to argue that the correct Schwinger-Keldysh action constructed by doubling
the non-dissipative hydrodynamic system, necessarily has an additional term which ensures that 
the correct anomalous Ward identities are attained, whilst maintaining the form of the anomalous currents
derived hitherto. While we will not be able to rigorously justify each step of 
our argument, the final result is compelling in its simplicity to suggest that we are on the right track.

\subsection{Anomalies in the Schwinger-Keldysh formalism}
\label{sec:SK-anomalies}

Before we get into the discussion of anomalies let us recall some of the basic facts about the Schwinger-Keldysh (or in-in) formalism. The interested reader can find a more detailed description in Appendix~\ref{sec:SK-review}; for the present we will content ourselves with a brief reminder. 

To describe real-time dynamics of a quantum system with a Hilbert space ${\cal H}$ and Hamiltonian $H$, we begin by doubling the degrees of freedom. To wit, we consider an enlarged system with a Hilbert space ${\cal H}_R \otimes {\cal H}_L$, with the indices being used to refer to the two copies. Our choices of $L,R$ systems is inspired by the realization of the Schwinger-Keldysh formalism in the gravitational AdS/CFT context in terms of the eternal black hole with the two quantum systems being the right $(R)$ and left $(L)$ CFTs on the two boundaries as envisaged originally by Israel \cite{Israel:1976ur} and subsequently by Maldacena \cite{Maldacena:2001kr}. This formalism has of course proved to be useful in deriving the real-time prescription for computing correlation functions of single trace operators \cite{Herzog:2002pc}.   The dynamics is implemented in this framework by the difference Hamiltonian $H_R-H_L$. 

\begin{figure}[t!]
 \begin{center}
\includegraphics[width=.8\textwidth]{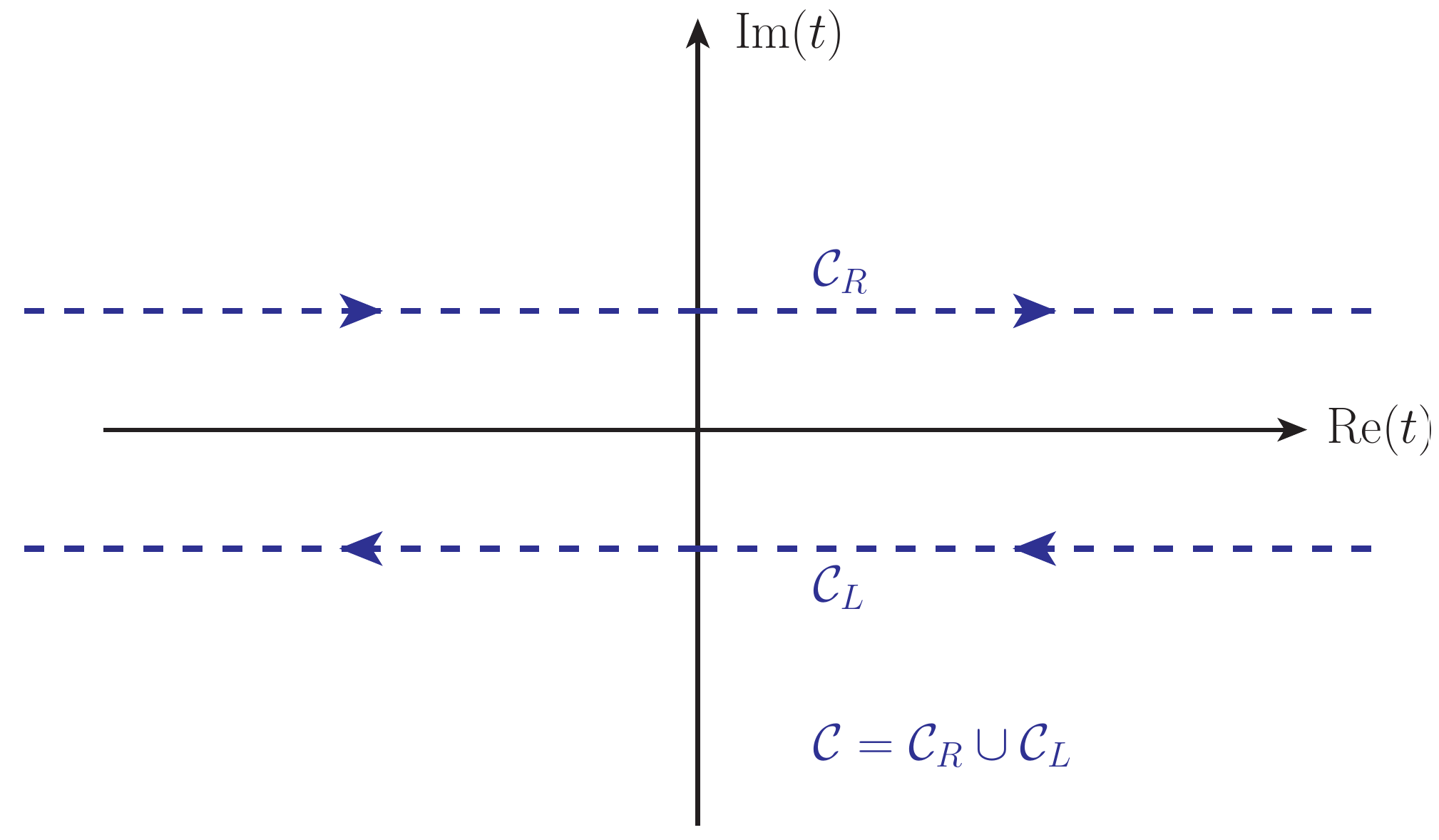}
\caption{Complex time Keldysh contour for systems out of equilibrium in Schwinger-Keldysh formalism.}
\label{fig:SK-contour1}
\end{center}
\end{figure}

However, for the purposes of discussing the effective action it is more useful to think about the complex Schwinger-Keldysh contour to describe the time-ordering prescription; see  Fig.~\ref{fig:SK-contour1}. We complexify the time coordinate and consider two anti-parallel contours ${\cal C}_R$ and ${\cal C}_L$ which refer to the two parts of the doubled system. Often the two contours are joined through a purely imaginary part (for equilibrium questions the imaginary time separation is set by the inverse temperature) to obtain the so called Keldysh-Baym contour, but this detail is not relevant for what we wish to discuss here. The time-ordering is left to right on the top ${\cal C}_R$ contour, and opposite on the bottom ${\cal C}_L$ contour. 

If our quantum system is described by an effective action $S_{eff}$ then in the Schwinger-Keldysh framework we consider $S_{eff,R} - S_{eff,L}$ to be the action inserted into the path integral, consistent with the evolution described earlier. The main question to address is when we couple the basic degrees of freedom to sources so as to be able to write down the generating function for computing correlators. We introduce of course independent sources ${\cal J}_R, {\cal J}_L$ for the two halves and thus would have to take 
\begin{equation}
{\cal S}_{SK} = S_{eff,R}( {\cal J}_R) -  S_{eff,L}( {\cal J}_L) 
\label{}
\end{equation}	
Since we have independent sources we can obtain generically a tensor of correlators with operator insertions on both contours. However, the main object of interest which enters any discussion of non-equilibrium dynamics is the causal retarded correlators. These one can argue are obtained by considering linear combinations of correlation functions with insertions on ${\cal C}_R$ and ${\cal C}_L$ respectively. In particular, defining ${\tt J }= \frac{1}{2} \left({\cal J}_R + {\cal J}_L\right)$ and ${\tt j } = {\cal J}_R - {\cal J}_L$ we can argue that the causal correlation functions of interest have a single variation with respect to the difference source ${\tt j}$ \cite{Chou:1984es}.

\subsubsection{General prescription for anomalous theories}
\label{sec:SK-anomalies-general}

We would like to use the Schwinger-Keldysh formalism described above to work out the effective action for a quantum field theory with a global anomaly. The most straightforward way to proceed is to use the anomaly inflow mechanism. Consider a physical anomalous theory in $2n$ dimensions with an effective action $S_{eff}[\fA; \ldots]$ with the background gauge source alone explicitly indicated and let the global current coupled to $\fA$ have an anomaly as indicated by  \eqref{hydroan}.

We take this anomalous theory and construct an explicit anomaly-free theory in one higher dimension by exploiting the inflow picture. Viewing the manifold on which the field theory lives as the co-dimension one boundary of a higher dimensional spacetime, $\mathcal{M}_{2n}=\partial \mathcal{M}_{2n+1}$, we offset the anomaly by introducing a bulk topological sector:
\begin{align}
 S =  \int_{\partial \mathcal{M}_{2n+1}} S_{eff}[\fA;\ldots]
      + \int_{\mathcal{M}_{2n+1}} \, I_\text{CS}[\fA] \,,
 \label{AnomThy}
\end{align}
where dots denote matter fields that do not play a role in this treatment. By construction $S$ respects the underlying gauge invariance, so in this enlarged theory we are in a position to use the standard Schwinger-Keldysh formalism.
 
From our discussion above it is then clear that for the anomaly free bulk+boundary theory, the Schwinger-Keldysh action in real time representation would thus be
\begin{align}
 S_{SK} = S_{eff}[\fA_R;\ldots] - S_{eff}[\fA_L;\ldots] + S_\text{CS}[\fA_R] - S_\text{CS}[\fA_L] \,,
  \label{S_SK-allg}
\end{align}
The left and right covariant currents are defined as before by varying with respect to the appropriate sources
\begin{align}
 \star\, \form{J}_{R,L} = \frac{\delta S_{SK}}{\delta \fA_{R,L}} \,.
\end{align}
These currents of course have the correct anomaly and satisfy \eqref{hydroan} with the appropriate $L,R$ subscripts. It also follows that the linear combination of currents $\form{J} = \frac{1}{2} \left(\form{J}_L + \form{J}_R\right)$ also has the desired anomaly. 

We assert that the ``hydrodynamical current'' is the one which enters the retarded correlation function and is given precisely by the linear combination $\form{J}$.  It can be obtained by writing the action in terms of the common and difference sources $\fA$ and ${\bf a}$ respectively defined as
\begin{align}
 \fA \equiv \frac{1}{2} \left( \fA_R + \fA_L \right) \,,\qquad
 {\bf a} \equiv \fA_R - \fA_L \,,
\end{align}
and varying with respect to the difference source. To be specific,
\begin{align}
\star \,\form{J}_\text{hydro} \equiv \frac{\delta S_{SK}}{\delta {\bf a}} \bigg{|}_{{\bf a}=0} 
     = \frac{1}{2} \left( \form{J}_R + \form{J}_L \right) \bigg{|}_{{\bf a}= 0} \,.
 \label{J_hydro_calc}
\end{align}
We have noted that this current satisfies the anomalous Ward identity \eqref{hydroan}  by construction.\footnote{This can easily be shown by a similar argument as in Appendix~\ref{sec:stensorA}. However,
the observation that one recovers the correct anomalous conservation equations is quite obvious from the fact that 
the structure of the Chern-Simons terms in the action (\ref{S_SK-allg}) is such that it gives the correct inflow.} Our discussion so far has been sufficiently general and has made no reference to the specific form of $S_{eff}$; the next task is to specify this discussion to the non-dissipative effective action derived in the previous sections.

\subsubsection{Construction of an action for hydrodynamics and anomalous current conservation}
\label{sec:SK-anomalies-hydro}

In our construction of an  effective action for anomalous hydrodynamics \eqref{eq:NonAbelianClaim}, we have seen that the anomaly always turned out to 
contain unexpected contributions from hydrodynamic shadow fields which don't generically vanish out of equilibrium. While working with a single copy of the effective action has provided us with no avenue out of this impasse we will now argue that the extra freedom inherent in the doubled Schwinger-Keldysh formalism possesses 
enough structure to cure this problem. What we do below is to identify an additional contribution necessary in the doubled theory to ensure that the anomaly inflow into the left and right theories is the correct amount.

Let us recall our basic goal: we wish to construct a 
Schwinger-Keldysh action $S_{SK}$ which contains in it an anomalous contribution
$\mathcal{L}_{SK}^{anom}[\fA_R, \, \fA_L]$. Our discussion of \S\ref{sec:SK-anomalies-general} 
indicates that in order to have the correct anomaly inflow the action should take the form
\begin{align}
 S_{SK} &=  S_{n-a}[\fA_R] - S_{n-a}[\fA_L] + S_{SK}^{anom} \,,\\
 S_{SK}^{anom} &=\int_{\mathcal{M}_{2n+1}} \ICS[\fA_R] - \ICS[\fA_L] 
          + \int_{\mathcal{M}_{2n}} \mathcal{L}_{SK}^{anom}[\fA_R,\fA_L] \,,
  \label{S_SK-ansatz}
\end{align}
where $S_{n-a}[\fA]$ is the part of the hydrodynamical effective action that is non-anomalous and 
$\ICS[\fA]$ has been defined in \S\ref{sec:transgression}.
By the identity (\ref{eq:Transgr}), we can rewrite the Chern-Simons bulk pieces of Eq.\ (\ref{S_SK-ansatz}) in terms of transgression forms:
\begin{align}
 S_{SK}^{anom} &=  \int_{\mathcal{M}_{2n+1}}\prn{\mathcal{T}_{2n+1}[\fA_R, \fAh_R] - \mathcal{T}_{2n+1}[\fA_L,\fAh_L]
                 + \ICS[\fAh_R] - \ICS[\fAh_L] }  \notag\\
               &\quad + \int_{\mathcal{M}_{2n}} \prn{ \WCS[\fA_R,\fAh_R] - \WCS[\fA_L
                 , \fAh_L]  + \mathcal{L}_{SK}^{anom} [\fA_R,\fA_L]} \,.
\label{step1}				 
\end{align}
While there is more than one way to write the difference of the left and right Chern-Simons forms in terms of a transgression form (one could simply have written ${\cal T}(\fA_R, \fA_L)$ for example), we are using the prior knowledge of \eqref{eq:NonAbelianClaim} to start assembling pieces that capture the anomaly in the fluid dynamical theory. 

The first two terms in \eqref{step1} are transgression forms and therefore gauge covariant; in fact they are nothing
but the anomalous effective action for the right and left currents, cf., \eqref{eq:NonAbelianClaim}. Thus
$\mathcal{L}_{SK}^{anom}$ must be such that together with the remaining terms we obtain a gauge invariant 
expression. But there is a unique gauge invariant object which contains the difference 
$\ICS[\fAh_R]-\ICS[\fAh_L]$ which is simply 
the transgression between the right and left hydrodynamic shadow fields, i.e., the
transgression form $\VP[\fAh_R, \fAh_L]$. Based on this observation, 
and the fact that ${\cal L}^{anom}_{SK}$ should also contain the $\fB$-terms between $\fA_{R,L}$ and 
their shadows, we  conclude that the unique gauge invariant action which is consistent with the
ansatz (\ref{S_SK-ansatz}) is simply
\begin{align}
 \mathcal{L}^{anom}_{SK}[\fA_R, \fA_L] = - \prn{ \WCS[\fA_R,\fAh_R] - \WCS[\fA_L
                 , \fAh_L] + \WCS[\fAh_R, \fAh_L] } \,,
\end{align}

This implies that the anomalous part of the Schwinger-Keldysh action is the sum of three transgression forms
\begin{align}
 S_{SK}^{anom} = \int_{\mathcal{M}_{2n+1}} \prn{ \mathcal{T}_{2n+1}[\fA_R, \fAh_R] - \mathcal{T}_{2n+1}[\fA_L,\fAh_L]
     + \mathcal{T}_{2n+1}[\fAh_R, \fAh_L] }\,.
  \label{S_SK-result}
\end{align}
This action clearly respects all symmetries of the theory and also has the desired amount of inflow into the left and right parts by construction. 

While the result was obtained by demanding that the Schwinger-Keldysh action obeys the correct symmetries (which in particular demands the appropriate amount of inflow) the final result can be interpreted in a simple manner. Recall that the  hydrodynamic shadow fields $\fAh_{L,R}$ are proxies for the gauge field on the horizon in the deconstruction picture \S\ref{sec:non-abelian}. Putting this together with the Schwinger-Keldysh formalism, we have a-priori independent terms for the left and right halves of the eternal black hole Kruskal geometry. However, since the geometry has a single bifurcation surface which connects the left and right black holes, one might ask if there isn't an `interaction' term which communicates across the bifurcation surface. In the absence of anomalies such a term is unnecessary and indeed from the geometric picture one expects the left and right theories to be decoupled. However, in an anomalous theory the anomaly inflow between the left and right parts must be carefully 
regulated to ensure gauge covariance of the action. The extra term $\mathcal{T}_{2n+1}[\fAh_R, \fAh_L]$ encountered in \eqref{S_SK-result} 
can therefore be interpreted as a gluing condition to join the $R$ and $L$ contours that are described by the first two transgression forms. Note that to infer its presence we really need to work with the doubled system: a single copy of the action is incapable of revealing this intricate structure.

One can argue for the presence of such a gluing term more generally. From the general form of the Schwinger-Keldysh action, Eq.\ (\ref{S_SK-allg}), one might naively expect that the doubled system has a  gauge symmetry $G_R\times G_L$ acting independently on the two branches. For example in the abelian case with $G= U(1)$ the action is 
$
 \fA_{R,L} \rightarrow \fA_{R,L} - d\Lambda_{R,L}$ or equivalently $ 
 \fA\rightarrow \fA - d\Lambda $, and $ {\bf a} \rightarrow {\bf a} - d\lambda 
$ 
with $\Lambda = \tfrac{1}{2}(\Lambda_R + \Lambda_L)$ and $\lambda = \Lambda_R - \Lambda_L$.
However, the existence of non-vanishing cross-contour correlation functions
in the Schwinger-Keldysh formalism demands that this symmetry be spontaneously broken to a single $G_\text{diag} \subset G_R \times G_L$ (i.e., \ $\lambda=0$ in the abelian case). 
The new term ${\bf\mathcal T}_{2n+1}[\fAh_R, \fAh_L]$ precisely captures the correct gauge invariant data for such a symmetry breaking pattern.

Let us now show that the third term in the action (\ref{S_SK-result}) indeed modifies the anomalous Ward identities in the correct way. In fact, it easy to see that this term does not change the anomalous current at all, but only the anomaly. 
To this end, let us start by considering the current conservation equation for an abelian $U(1)$ current (the non-abelian generalization is straightforward). 
The variation with respect to the source ${\bf a}$ of the new contribution to the action is given by (see Eq.~\eqref{eq:CrossTransVar})
\begin{align}
 \delta_{\bf a} \left(\int_{\mathcal{M}_{2n+1}} \VP[\fAh_R, \fAh_L]\right)
     \bigg{|}_{\text{linear in }{\bf a}}
     = \int_{\mathcal{M}_{2n+1}}  \delta a_b \, \tilde J^b_{(2n+1)} 
\end{align}
where we only keep track of terms linear in ${\bf a}$ and defined
\begin{align}
 \tilde J^{\;a}_{(2n+1)} &= P^a_b \JHh^b \nonumber \\
& \xrightarrow{abelian} P^a_b \frac{(n+1) c_{_A}}{2^n}
   \epsilon^{b\, l_1m_1\cdots l_nm_n}  \hat{F}_{l_1m_1}\cdots \hat{F}_{l_nm_n} \,.
\end{align}
This is a pure bulk current and therefore, by means of an analogous argument as in Appendix~\ref{sec:stensorA}, it contributes 
to the anomaly inflow for the current associated to variations with respect to ${\bf a}$ (i.e., $\form{J}_{hydro}$)
but it does not affect the particular form of $\form{J}_{hydro}$ itself. Just as in Appendix~\ref{sec:stensorA},
its contribution to the anomalous current conservation is $\tilde J^\perp_{(2n+1)}$.\footnote{Note that as we consider two copies of the gauge
field, we also need to consider two copies of the phase field $\psi$. This defines an $R$ and an $L$ chemical potential as well as
$\fAh= \fA + \tfrac{1}{2}(\mu_R + \mu_L) \fu$ and ${\bf \hat a} = {\bf a} + (\mu_R-\mu_L) \fu$. Strictly
speaking, a doubling of the Goldstone fields ${\phi^I}$ is also implied. We can, however suppress this fact because it
is not relevant for our discussion. In the present context a doubling of the metric is not necessary either; this will however
be discussed in \S\ref{sec:SK-anomalies-hydro2} in the context of stress tensor Ward identities.} 
We conclude that
\begin{align}
d\star \form{J}_{hydro} &\equiv d \star \left( \form{J}_{n-a}+ \form{J}_{anom}\right) 
=    \star\left( \form{J}_H^\perp - \form{\hat J}_H^\perp + \form{\tilde J}^\perp_{(2n+1)} \right) =\star \form{J}_H^\perp
\nonumber \\
&\xrightarrow{abelian} (n+1)\,c_{_A}  \left( \fF^n -\fFh^n\right) +  \star\form{\tilde J}^\perp_{(2n+1)} 
=  (n+1)c_{_A}  \fF^n \,, 
  \label{final-WI-J}
\end{align}
as desired. Here $\form{J}_{anom}$ gets contributions from the first two pieces of the action (\ref{S_SK-result}) alone, which is the result computed earlier in  
Eq.\ (\ref{J_hydro_calc}), and $\form{J}_{n-a}$ is due to the non-anomalous action. 
This shows that the Schwinger-Keldysh  formalism as advertised is indeed powerful enough to solve the problem of the ``hydrodynamic shadow anomaly''. By a natural
modification of the action in the doubled system, we have successfully recovered the correct hydrodynamical Ward identity for the current. In the next subsection we show how this reasoning can be applied to the stress tensor Ward identity, as well, thereby completing our derivation of an anomalous effective action in hydrodynamics.

\subsubsection{Anomalous stress tensor conservation}
\label{sec:SK-anomalies-hydro2}

We can now perform a similar analysis for the stress tensor Ward identity which is connected to the metric and diffeomorphism invariance. 
By analogous reasoning as for the $U(1)$ gauge field, we need to consider two copies of the metric $g_R$ and $g_L$ with independent diffeomorphism symmetry (as expected for a generally covariant effective action).
The linear combination $g = \frac{1}{2}\left(g_R + g_L\right)$ is the classical background metric and $\gamma =  g_R - g_L$ the difference metric which describes small fluctuations.\footnote{In the same sense that
${\bf a}$ being small in \S\ref{sec:SK-anomalies-hydro}.}

In order to get physical hydrodynamical quantities, one only
needs to keep track of $\gamma$ at linear order and in the end take the coincidence limit $\gamma=0$. As in the case of the 
global symmetry, the anomalous Schwinger-Keldysh action \eqref{S_SK-result} reduces to the ``classical'' result only in this limit. To be specific, the metric dependence of the 
individual terms in the third transgression form that connects the $R$ and $L$ branches is assumed to be as follows:
\begin{align}
\VP[\fAh_R, \fAh_L; g_R,g_L]
   = \ICS[\fAh_R; g_R] - \ICS[\fAh_L; g_L]
      - d\WCS[\fAh_R \text{ with } g_R , \fAh_L \text{ with } g_L] \,,
  \label{T-metric}
\end{align}
where the last term is meant to be such that right and left fields are contracted with   $g_R$  and $g_L$ respectively. It 
is then straightforward to show that variations of (\ref{T-metric}) with respect to  the difference
metric $\gdiff$ are given by (see Eq.~\eqref{eq:CrossTransVar})
\begin{align}
  \delta_\gamma\left(  \int_{\mathcal{M}_{2n+1}} \VP[\fAh_R, \fAh_L; g_R,g_L]
   \right)\bigg{|}_{\text{linear in }\gamma, \, {\bf a}=0} 
   = \int_{\mathcal{M}_{2n+1}} \sqrt{-G} \; \frac{1}{2}\,\delta \gdiff_{ab} \, \tilde T^{ab} \,,
\end{align}
with a tensor
\begin{align}
 \tilde T^{ab} &= \mu \left(P_{\ c}^a u^b  + P_{\ c}^b u^a  \right) \, \hat{J}_H^c
 \nonumber \\
 &\xrightarrow{abelian}\; \mu \left( P_{\ c}^a u^b  + P_{\ c}^b u^a  \right) \left[ \frac{(n+1)c_{_A}}{2^n}
    \epsilon^{c\, l_1 m_1 \cdots l_n m_n} \hat{F}_{l_1m_1}\cdots \hat{F}_{l_nm_n}  \right] \,,
\end{align}
whose indices are all contracted with the common metric  $\tfrac{1}{2}(g_L + g_R)$. 
The same kind of argument that we outline in Appendix~\ref{sec:stensorA} implies that this
pure bulk tensor leaves the anomalous stress tensor itself unchanged. However, the conservation 
equations get a contribution proportional to the  
component $\tilde T^{\perp\alpha}$ and read 
\begin{align}
 \nabla_\beta T^{\alpha\beta}_{hydro} \equiv \nabla_\beta\left( T^{\alpha\beta}_{n-a} + T^{\alpha\beta}_{anom}\right)
 & =  F^\alpha{}_\beta J^\beta_{hydro}  -\mu\, u^\alpha \, \hat{J}_H^\perp + \tilde T^{\perp \alpha}
   = F^\alpha{}_\beta J^\beta_{hydro} \,,
  \label{Final-WI-Pi}
\end{align}
which is the correct hydrodynamic anomaly without shadow field contributions. 
The upshot of this discussion is simply to affirm that we have an effective action which captures the physics of quantum dynamics in the presence of anomalies in the hydrodynamic limit.

\section{Conclusion}
\label{sec:Conclusion}

The main result of the paper is an effective action for non-dissipative fluids which captures the flavour anomaly contribution in the hydrodynamic limit.  The action is constructed exploiting the Schwinger-Keldysh framework, and involves a novel interaction term between the forward and reversed contours.

 In particular, in terms of the hydrodynamical variables $\phi^I_{L,R}$, $c_{L,R}$ (or $\psi_{L,R}$ in the abelian case) and the background source flavour gauge fields $\fA_{L,R}$ on the two contours,  the total effective action can be written as the sum of transgression forms. To wit, 
\begin{align}
 S_{SK} &= S_{n-a}[\phi_R, c_R, \fA_R] - S_{n-a}[\phi_L, c_L,\fA_L] \notag \\
      &\quad+\;\int_{\mathcal{M}_{2n+1}} 
     \mathcal{T}_{2n+1}[\fA_R, \fAh_R] - \mathcal{T}_{2n+1}[\fA_L,\fAh_L]
     + \mathcal{T}_{2n+1}[\fAh_R, \fAh_L] \,, \label{FinalResult}
\end{align}
where $S_{n-a}[\phi, c,\fA]$ is the anomaly free part of the effective action for non-dissipative fluids.  The transgression forms capture all the anomalous terms and are to be interpreted in the language of anomaly inflow.  Note that this piece of the effective action only depends on the effective field theory degrees of freedom $\phi^I$ and $c$ through the hydrodynamic velocity and chemical potential which enter into the definition of the shadow gauge potential $\fAh$.

The construction crucially employs the anomaly inflow mechanism. We have coupled our hydrodynamic effective field theory in $d=2n$ dimensions to a topological Hall insulator theory in $(2n+1)$-dimensions to ensure that the total Schwinger-Keldysh effective action is gauge invariant under the flavour symmetry. We have shown explicitly that the action \eqref{FinalResult} reproduces all of the anomaly induced hydrodynamic transport and the resulting currents obey the requisite Ward identities, cf., Eqs. (\ref{final-WI-J}) and (\ref{Final-WI-Pi}). 

The curious aspect our construction is the presence of the hydrodynamic shadow gauge field $\fAh =\fA+ \mu \,\fu$ which necessitates a non-trivial mixing between the two contours in the Schwinger-Keldysh construction.

The shadow gauge field which plays a critical role  needs a better understanding.  In \S\ref{sec:NonAbelianFormalism} we argued that in a holographic context  $\fAh$ should be related to a  ``horizon gauge field'' $\fA_\text{h}$ which gauges the chemical shift symmetry (\ref{shiftSymm}). 
This motivates our terminology of calling $\fAh$ a ``shadow'' of the gauge field $\fA$;  the field $\fAh$ lives at the boundary but it just reflects the horizon gauge symmetry, viz., the chemical shift symmetry. It captures the appropriate  coupling to the Goldstone mode (the Wilson line between horizon and boundary). Vis a vis, the cross-contour term in  \eqref{FinalResult} as remarked in \S\ref{sec:intro} one can give a heuristic argument by tracking the anomaly inflow between the two contours via the Hall insulator. We believe that this picture can be developed further using holographic embedding of anomalous fluid dynamics and in fact one should be able to derive \eqref{FinalResult}  using the techniques described in \cite{Son:2009tf, Faulkner:2010jy}.

We also note in passing that the shadow gauge field seems to appear whenever parity-odd transport phenomena is considered, cf.,  \cite{Haehl:2013kra} in the context of $(2+1)$-dimensions. It is interesting to speculate whether the inclusion of cross-contour terms is necessary to account for Hall viscosity in that context.
 
The present discussion has been confined to flavour anomalies in QFTs. Another source of anomaly induced transport 
is from Lorentz and mixed anomalies. It is interesting to ask whether the formalism developed herein 
can capture these contributions to hydrodynamics as well. We believe that a naive generalization of 
our construction to include transgression forms involving the background gravitational connection (and its shadow
counterpart) should be able to account for part of the transport. However, not all gravitational effects 
seem to be captured this way. In the language of \cite{Jensen:2012kj} the transcendental pieces of 
anomalous transport seem to be outside the remit of the effective action approach, in part because 
they necessitate non-trivial changes to the entropy current (and our effective action works primarily
in the entropy frame). It would be interesting to develop this argument further and ascertain where the bottle-necks are.

Finally, it is interesting to speculate that the formalism developed here could be used to go beyond hydrodynamics and provide insight into anomaly induced 
out-of-equilibrium dynamics.

\acknowledgments 

We would like to thank Jyotirmoy Bhattacharya and Sayantani Bhattacharyya for
collaboration on various aspects relating to effective actions for hydrodynamics
and for sharing their insights in the construction reported herein. It is also 
a pleasure to thank Veronika Hubeny, Sachin Jain, Kristen Jensen, Pavel Kovtun, Shiraz Minwalla, Amos Yarom for numerous 
enlightening discussions on topics related to hydrodynamics, entropy and effective actions.  We also would like to thank Jyotirmoy Bhattacharya, Sayantani Bhattacharyya, Kristan Jensen and Shiraz Minwalla for useful comments on a draft version of this paper.

FH and MR would like to thank Centro de Ciencias de Benasque and LMU Munich for their kind hospitality.  RL and MR would like to thank the workshop on 
Relativistic Hydrodynamics and the Gauge Gravity Duality held at Technion University, Israel. MR in addition acknowledges the hospitality of CERN and ITF, University of Amsterdam, KITP (Santa Barbara), Newton Institute (Cambridge)  during the course of this project. FH is supported by a Durham Doctoral Studentship and MR is supported in part by the  STFC Consolidated Grant ST/J000426/1.

\appendix

\section{Contributions to the anomalous constitutive relations in \texorpdfstring{$d=(3+1)$}{d=3+1}}
\label{appendix:pieces}

The complete action which consistently gives rise to the anomaly in 3+1 dimensions is presented in \eqref{Sa} which we decomposed as $S_{anom} = S_{wzI} + S_{wzII} + S_\text{CS}$. In this appendix, we list the contributions to the currents from each of these three terms. Our definition of the currents is the standard one where we vary the effective action with respect to the sources:
\begin{align}
T^{\alpha\beta}_{anom} = \frac{2}{\sqrt{-g}} \frac{\delta S_{anom}}{\delta g_{\alpha\beta}} \,, \qquad
J^\alpha_{anom} = \frac{1}{\sqrt{-g}} \frac{\delta S_{anom}}{\delta A_\alpha} \, .
\end{align}
We will make use of the following (off-shell) identities at various stages to simplify the terms
\begin{align}
  u^{[\alpha} \, B^{\beta]} &= \tfrac{1}{4} \eps^{\alpha\beta\rho\sigma} (F_{\rho\sigma} - 2\,  u_\rho \,E_\sigma) \,, \label{I1}\\
  u^{[\alpha} \hat B^{\beta]} &= \tfrac{1}{4} \eps^{\alpha\beta\rho\sigma} (\hat F_{\rho\sigma} - 2 \, u_\rho\, \hat E_\sigma) \,, \label{I1hat}\\
  u^{[\alpha} \omega^{\beta]} &= \tfrac{1}{4} \eps^{\alpha\beta\rho\sigma} ( \partial_\rho u_\sigma +u_\rho \, {\mathfrak a}_\sigma) \, ,\label{I2} 
\end{align}
along with Eq.\ (\ref{I3}). For convenience, let us first record the variations of various quantities with respect to the metric:
\begin{align}
  \delta_g u^\alpha &= \left(\frac{u^\alpha}{2}\right) \, u^\alpha u^\beta \delta g_{\alpha\beta}\,, \\
  \delta_g s &= -\left(\frac{s}{2} \right) \, P^{\alpha\beta} \delta g_{\alpha\beta} \, ,\\
  \delta_g \mu &= \left(\frac{\mu}{2}\right) \, u^\alpha u^\beta \delta g_{\alpha\beta} \,, \\
  \delta_g \chi &= \left( 2\mu \zeta^\alpha u^\beta -\zeta^\alpha \zeta^\beta \right) \delta g_{\alpha\beta} \,. 
\end{align}
The first two (4-dimensional) terms in the full action, $S_{wzI}+S_{wzII}$, yield the following contributions to the current:
\begin{align}
  J_{wzI}^\alpha 
      &= -2\, \aleph\left[ B^\nu D_\nu \psi +\hat B^\nu (\hat D_\nu \psi - \hat A_\nu) \right] u^\alpha  + 2\, \aleph\, \mu \, B^\alpha + \aleph\, \eps^{\alpha\nu\rho\sigma}
         A_\nu F_{\rho\sigma}  \notag \\
         &\qquad\qquad +\; \aleph\, \eps^{\alpha\nu\rho\sigma} (\hat D_\nu \psi- \hat A_\nu) \hat F_{\rho\sigma} 
         - 2\, \aleph \, \eps^{\alpha\nu\rho\sigma}\,  D_\nu \psi \, u_\rho \, E_\sigma \,,\label{J1} \\
J_{wzII}^{\alpha} &= \left[(\eta_{\omega,\mu}-2\,\eta_B)\omega^\nu D_\nu \psi + \eta_{B,\mu} B^\nu D_\nu \psi\right] u^\alpha \notag \\
	     &\quad+\left[ 2(\eta_{\omega,\chi} \omega^\nu + \eta_{B,\nu} B^\nu )D_\nu \psi\right] \zeta^\alpha\notag\\
	     &\quad+  (\eta_\omega + 2\mu \eta_B) \omega^\alpha + 2 \eta_B B^\alpha  \notag \\
	      &\quad+\eta_{B,\chi} \eps^{\alpha\nu\rho\sigma} D_\nu\psi u_\rho \nabla_\sigma\chi\notag\\ 
	      &\quad + \eta_{B,\mu} \,  \eps^{\alpha\nu\rho\sigma} D_\nu \psi\, u_\rho \nabla_\sigma \mu 
	         + \eta_B \, \eps^{\alpha\nu\rho\sigma} D_\nu \psi u_\rho {\mathfrak a}_\sigma\,, \label{Jpsi}
\end{align}
\begin{align}
  T_{wzI}^{\alpha\beta}
      &= -4\aleph\, \mu \hat B^\nu (\hat D_\nu \psi - \hat A_\nu) u^\alpha u^\beta
         + 2\aleph \, \mu \eps^{(\alpha\nu\rho\sigma}(\hat D_\nu \psi - \hat A_\nu) \hat F_{\rho\sigma}u^{\beta)}  \, ,   \label{T1} \\
 T_{wzII}^{\alpha\beta} &=  -\left[ s\left( \eta_{\omega,s} \omega^\nu + \eta_{B,s} B^\nu \right) D_\nu \psi\right] P^{\alpha\beta} \notag \\
  &\quad+ \left[ \left(\mu\eta_{\omega,\mu} - 2 \eta_\omega\right)\omega^\nu D_\nu\psi
            + \left(\mu\eta_{B,\mu} -  \eta_B\right) B^\nu D_\nu\psi\right] u^\alpha u^\beta \notag \\
  &\quad- \left[2\left(\eta_{\omega,\chi}\omega^\nu + \eta_{B,\chi} B^\nu\right) D_\nu\psi \right] \zeta^\alpha \zeta^\beta \notag\\
  &\quad+\big[4\mu \eta_\omega \omega^{(\alpha}  +(\eta_\omega +2\mu\eta_B) B^{(\alpha} 
   +4\mu (\eta_{\omega,\chi}\omega^\nu+\eta_{B,\chi}B^\nu) D_\nu\psi\,\zeta^{(\alpha} \notag \\
   &\qquad\;\;\,+\eta_{\omega,\chi} \eps^{(\alpha\nu\rho\sigma}D_\nu\psi u_\rho \nabla_\sigma \chi
   -\eta_{\omega,\mu} \, \eps^{(\alpha\nu\rho\sigma}D_\nu \psi \,u_\rho \hat E_\sigma
   \big]u^{\beta)} \,,  \label{Tpsi}
\end{align}

There are several gauge non-invariant pieces in the expressions (\ref{J1}, \ref{T1}). This is to be expected since the computation above picks out the analog of the consistent current for our theory. In particular, without the hatted term in $S_{wzI}$
we would precisely have been computing the consistent current for our theory. These are the currents that satisfy the Wess-Zumino consistency conditions (commutativity of variations of the gauge field with gauge transformations). However, their gauge variation is non-vanishing $\delta_A (S_{wzI} + S_{wzII}) \neq 0$.

To compute the gauge-covariant currents we would just add appropriate Bardeen-Zumino terms \cite{Bardeen:1984pm}; we implement this using the Chern-Simons action in the bulk spacetime explicitly.  Varying $S_\text{CS}$ with respect to either $g_{\alpha\beta}$ or $A_\alpha$ gives after an integration by parts
\begin{align}
  \delta S_\text{CS} 
                     &= -3\, \aleph\,  \int_{\mathcal{M}_5} \left[ \delta \fA\wedge \fF\wedge \fF - \delta \fAh\wedge{\bf  \hat F}\wedge \fFh\right]
                        - 2\, \aleph\, \int_{\partial \mathcal{M}_5} (\delta \fA \wedge \fA\wedge \fF-\delta\fAh \wedge\fAh \wedge \fFh) \, .
\end{align}
We see that the variations leave behind a boundary term that can be included in our definition of the current.  The contribution on the 4-dimensional physical space is therefore
\begin{align}
  \delta S_\text{CS}\big{|}_\text{4-dim} = -\aleph \int \sqrt{-g}\, \eps^{\alpha\beta\rho\sigma} (\delta A_\alpha\, A_\beta F_{\rho\sigma}
    - \delta \hat A_\alpha \,\hat A_\beta \hat F_{\rho\sigma})\,,
\end{align}
from which we obtain 
\begin{align}
T^{\alpha\beta}_\text{CS} &=-\left[4\aleph\, \mu\, \hat B^\nu \hat A_\nu \right] u^\alpha u^\beta
         + 2\aleph \, \mu \,\eps^{(\alpha\nu\rho\sigma} \hat A_\nu \hat F_{\rho\sigma}u^{\beta)}   \,,\\
J^\alpha_\text{CS} &= \left[-2\aleph\, \hat B^\nu \hat A_\nu\right] u^\alpha
                    -   \aleph\, \eps^{\alpha\nu\rho\sigma} (A_\nu F_{\rho\sigma} -\hat A_\nu \hat F_{\rho\sigma} )   \,,
\end{align}
which precisely cancels the gauge non-invariant pieces in (\ref{T1}, \ref{J1}). Using the identity (\ref{I1}), we can then simplify the 
resulting gauge invariant objects:
\begin{align}
 T_{wzI}^{\alpha\beta} + T_\text{CS}^{\alpha\beta}
   &= -\left[4\aleph\, \mu \,\hat B^\nu \hat D_\nu \psi  \right]u^\alpha u^\beta
         + 2\aleph \, \mu \eps^{(\alpha\nu\rho\sigma}\hat D_\nu \psi  \hat F_{\rho\sigma}u^{\beta)}  \notag\\
   &= 4\, \aleph \,\mu \, \eps^{(\alpha\nu\rho\sigma}  D_\nu \psi u_\rho \hat E_\sigma \, u^{\beta)} \, ,\label{2ofThemT} \\
  J_{wzI}^\alpha + J_\text{CS}^\alpha
   &= - 2\, \aleph\left[ B^\nu D_\nu \psi +\hat B^\nu \hat D_\nu \psi  \right] u^\alpha  + 2\,\aleph\, \mu \, B^\alpha 
         +\aleph\,\eps^{\alpha\nu\rho\sigma} \hat D_\nu \psi \hat F_{\rho\sigma}\notag\\
        &\qquad \qquad \,-\;2\aleph\, \eps^{\alpha\nu\rho\sigma} D_\nu \psi u_\rho E_\sigma\notag\\
   &= 2\,\aleph\, \eps^{\alpha\nu\rho\sigma} D_\nu \psi u_\rho (\hat E_\sigma-E_\sigma) - 2\,\aleph\, B^\nu D_\nu \psi \, u^\alpha 
        +2\,\aleph\, \mu \, B^\alpha \, . \label{2ofThemJ}
\end{align}
The full stress-energy tensor and charge current in \S\ref{sec:dervive4d} are now easily obtained by adding the contributions (\ref{2ofThemT}, \ref{2ofThemJ}) to
(\ref{Tpsi}, \ref{Jpsi}).

\section{Relation to other approaches}
\label{sec:relothers}

For completeness, we now show that our result for the anomalous transport agrees with other computations in the literature.  Firstly we illustrate that the standard result presented in \cite{Son:2009tf} is indeed recovered by rotating our answer from the entropy frame to the  Landau frame. Since this has been already established in \cite{Loganayagam:2011mu} we will be relatively brief. We also then demonstrate that our action when restricted to stationary fluid flows 
reproduces the anomalous part of the equilibrium partition described in \cite{Banerjee:2012iz}.

\subsection{The anomaly in Landau frame}
\label{sec:frames}

The precise form of the solution (\ref{T_sFrame0}, \ref{J_sFrame0}) to anomalous transport, is parameterized  by the fluid dynamical variables $u^\mu,\,s,\,\mu$. Within hydrodynamics these fields are a-priori ambiguous as they may be changed by a 
field redefinition. So the theory is only unambiguously defined once we eliminate the freedom of field redefinitions by supplementing the constitutive relations \eqref{consrel} with some additional constraints.  Eliminating the field redefinition ambiguity thus, is referred to as choosing a \emph{fluid frame}.
One choice of a fluid frame which is commonly used is the \textit{Landau frame}
where the field redefinition ambiguity is fixed by imposing $q_\alpha =  u_\alpha \Pi^{\alpha\beta} = u_\alpha \,\nu^\alpha= 0$.
The effective action formalism, on the other hand, automatically lands us in the  \textit{entropy frame} wherein 
imposes $J_\text{S}^\alpha = s\, u^\alpha$ as an exact statement to all orders in the derivative expansion. 

The result (\ref{T_sFrame0}, \ref{J_sFrame0}) is the first order realization of the anomalous part of the constitutive relations in entropy frame. In order to rotate this result to Landau frame, we need to make the anomalous heat current contribution in $T^{\alpha\beta}$ vanish. 
This is achieved by performing the field redefinition $u^\alpha \rightarrow u^\alpha - \tfrac{1}{2}\tfrac{q_{anom}^\alpha}{\varepsilon+P}$
and taking $c_{_A} = -\frac{C}{6}$  yields (see also  \cite{Loganayagam:2011mu})  
\begin{align}
 T^{\alpha\beta}_\text{(Landau)} &= (\varepsilon+P) u^\alpha u^\beta + P g^{\alpha\beta} + \ldots \,,\\
 J^\alpha_\text{(Landau)} &= \rho\, u^\alpha +C\,\mu^2 \left( 1 - \frac{2}{3} \frac{\rho\, \mu}{\varepsilon +P} \right) \omega^\alpha
     + C\mu \left( 1 - \frac{1}{2} \frac{\rho\, \mu}{\varepsilon+P} \right) B^\alpha + \ldots \,,\\
 J^\alpha_\text{s\,(Landau)} &= s\, u^\alpha - C\, \frac{s}{\varepsilon+P} \left( \frac{2}{3} \, \mu^3 \, \omega^\alpha
     + \frac{1}{2} \,\mu^2 \,B^\alpha \right)+ \ldots\,,
\label{ssresult}	 
\end{align}
where ellipses represent first and higher order corrections in gradients. 

The hydrodynamical constitutive relations \eqref{ssresult} is 
just the answer for anomalous transport derived in \cite{Son:2009tf} using the entropy current analysis. 
In deriving this expression use was made of the zeroth order fluid equations of motion, 
rendering the construction dependent on on-shell data. In particular, determining 
the contribution of the anomalous transport to higher orders, requires knowledge of 
 $q_{anom}^\alpha$ and equations of motion beyond the leading order in gradients. 
 Our construction is absolved of such complications since the entropy frame analysis
 as noted already in \cite{Loganayagam:2011mu} side-steps the issue by being explicitly off-shell.

\subsection{Recovering the equilibrium partition function}
\label{sec:equip}

In \cite{Banerjee:2012iz, Jensen:2012jh} it has been shown how the hydrodynamic constitutive relations are constrained by the requirement of the existence of a stationary flow on arbitrary spatially (slowly) varying backgrounds. Furthermore, the physics of such time-independent flows can be encapsulated in an equilibrium partition function of the sources. If we restrict the action which we proposed in  \S\ref{sec:4dAnomaly} to such stationary flows, one would expect to recover the anomalous part of the equilibrium partition function of \cite{Banerjee:2012iz}.\footnote{See also \cite{Jain:2012rh, Banerjee:2012cr} for related discussions of anomalous transport from the equilibrium partition function perspective.} We now demonstrate this explicitly for our action $S_{anom}$, providing yet another consistency check.

In order to write down the equilibrium solution, we pick an arbitrary background geometry and gauge field configuration which is time-independent. Essentially we write down the most general set of background sources with a timelike  Killing vector $\partial_t$. The metric and gauge field can be brought into a canonical form (choosing spatial coordinates $x^i$ ($i=1,2,3$))
\begin{align}
 ds^2 &= -e^{2\sigma(\vec{x})} (dt^2+a_i(\vec{x})dx^i)^2 + g_{ij}(\vec{x}) dx^i dx^j \,,\\
 A^\alpha &= (A^0(\vec{x}), A^i(\vec{x})) \,,
\end{align}
with Kaluza-Klein gauge field $a_i(\vec{x})$. Written in terms of the above parametrization, there exists a stationary (\emph{equilibrium}) solution to the prefect fluid equations of motion:
\begin{align}
 u^\alpha_\text{eq}(\vec{x}) &= e^{-\sigma(\vec{x})} \delta^\alpha_t \, , \qquad
 T_\text{eq}(\vec{x}) = T_0 \,e^{-\sigma(\vec{x})} \,, \qquad
 \mu_\text{eq}(\vec{x}) = A_0 \,e^{-\sigma(\vec{x})}  \,. \label{EquilConfig}
\end{align}

To compare with the equilibrium action (viewed as a functional of the background sources), we just need to evaluate $S_{anom}$ on (\ref{EquilConfig}). The relevant fields take the following values in equilibrium: 
\begin{align}
 \omega^\alpha_\text{eq} &= \left(0,\, \tfrac{1}{4} \,e^\sigma\,\eps^{ijk} f_{jk}\right) \,, \\
 B^\alpha_\text{eq} &= \left(0\,, \tfrac{1}{2} \eps^{ijk}(\tilde F_{jk}+A_0 f_{jk})\right) \,,\\
 (D_\alpha \psi)_\text{eq} &= A_\alpha = (A_0 \,,A_i) \,, \label{EquilD} \\
 (\hat D_\alpha\psi)_\text{eq} &= \hat A _\alpha = (A_0+\mu_{(0)},\, A_i)\label{EquilHatD} \,,
\end{align}
where $\tilde F_{jk}=2\,\partial_{[j} \tilde A_{k]}$ and $f_{jk}=2\,\partial_{[j}a_{k]}$. Here, objects with a tilde refer to the Kaluza-Klein gauge invariant combination
\begin{align}
  \tilde A_\alpha \equiv ( A_0+\mu_{(0)}, \, A_i- A_0 \, a_i ) \,.
\end{align} 
Furthermore, to derive (\ref{EquilD}, \ref{EquilHatD}), we are assuming that the phase $\psi \rightarrow 0$ in equilibrium. 

We now plug the above stationary solution into our anomaly action (\ref{Sa}). In order to obtain
a partition function as in \cite{Banerjee:2012iz}, we reduce the resulting 
on-shell action on the Euclidean thermal circle with period $\tfrac{1}{T_0}$. The part of 
the action $S_{wzI}$ and the four-dimensional boundary contribution of $S_\text{CS}$ vanish in equilibrium.
We thus obtain for the (Euclidean) equilibrium action (taking $c_{_A} = -\frac{C}{6}$ )
\begin{align}
 S_{anom}\bigg|_\text{eq} &= S_{wzII}\bigg|_\text{eq} = \frac{C}{3} \; \int \, d^4x\,\sqrt{-g}  \left( \frac{1}{2} A_0\, e^{-\sigma} \, \eps^{ijk} \tilde A_i \tilde F_{jk}
                   + \frac{1}{4} A_0^2\, e^{-\sigma} \, \eps^{ijk} \tilde A_i f_{jk} \right) \notag \\
                &= \frac{C}{3}\; \int \,d^3x \,\sqrt{g_3} \left( \frac{A_0}{2\,T_0} \eps^{ijk} \tilde A_i \tilde F_{jk} + \frac{A_0^2}{4\, T_0} 
                   \,\eps^{ijk} \tilde A_i \,f_{jk} \right) \,, \label{SPsiEquil}
\end{align} 
where we have done the time-circle reduction and denote the 3-dimensional metric determinant by $g_3$. The result (\ref{SPsiEquil}) is exactly what the authors of \cite{Banerjee:2012iz} find for their equilibrium partition function.

\section{The anomaly in \texorpdfstring{$d=(1+1)$}{d=1+1} derived from our formalism}
\label{appendix:2Danom}

In this appendix we re-derive the result of \cite{Dubovsky:2011sk} for the anomaly in two dimensions using the formalism that we developed in \S\ref{sec:AllDimensions}. Although our logic is quite close to the one used there, we will point out the reason why \cite{Dubovsky:2011sk} got the correct equations of motion even without using an appropriate Schwinger-Keldysh formalism (c.f.\ \S\ref{sec:Schwinger-Keldysh}).

We start by writing down the relevant pieces of the action according to our prescription:
\begin{align}
  S_{wzI} &= -c_{_A}  \int_{\mathcal{M}_2} {\bf D}\psi \wedge \fA - \hat {\bf D}\psi\wedge \fAh \label{2DS1} \, ,\\
  S_\text{CS} &= c_{_A} \int_{\mathcal{M}_3} \fA\wedge \fF - \hat \fA\wedge \fFh \,, \label{2DS2}\\
  S_{wzII} &= -c_{_A} \int_{\mathcal{M}_2} \sqrt{-g} \;\eta(s,\mu) \, \eps^{\alpha\beta} D_\alpha\psi u_\beta \, ,
\end{align}
where $\eta(s,\mu)$ has to be determined such that the full action is invariant under the chemical shift symmetry. 
Adding up these actions gives the action of \cite{Dubovsky:2011sk} up to two differences: (i) we include a Chern-Simons 
action $S_\text{CS}$ to account for the anomaly inflow and (ii) we have extra terms in both $S_{wzI}$ and $S_\text{CS}$
due to hatted connections. The former point is not too important. We simply use $S_\text{CS}$ to derive the Bardeen-Zumino 
current which \cite{Dubovsky:2011sk} just add by hand. The second point however will be shown to lead to different equations
of motion than what is derived there. 

From the two parts of the action $S_{wzI}+S_\text{CS}$, one finds the following contributions: 
\begin{align}
  T_{wzI}^{\alpha\beta} &= \left[c_{_A}\mu \, \eps^{\rho\sigma} \partial_\rho\psi u_\sigma\right]
       u^\alpha u^\beta -c_{_A} \mu \, \eps^{(\alpha\nu} \partial_\nu\psi u^{\beta)} \, ,\\
   J_{wzI}^\alpha &=- c_{_A} \, \eps^{\alpha\nu} [(A_\nu-D_\nu\psi)-(\hat A-\hat D_\nu \psi)]
       +[c_{_A} \, \eps^{\rho\sigma} (\hat D_\rho\psi-\hat A_\rho)u_\sigma ] u^\alpha \notag\\
       &= -c_{_A}\, \eps^{\alpha\nu} (A_\nu-\hat A_\nu) -[c_{_A} \, \eps^{\rho\sigma} \hat A_\rho u_\sigma] u^\alpha
          +c_{_A}\, \eps^{\alpha\nu} D_\nu \psi   \,,\\
  T_\text{CS}^{\alpha\beta} &= [c_{_A}\mu \, \eps^{\rho\sigma} \hat A_\rho u_\sigma] u^\alpha u^\beta
       -c_{_A}\mu \, \eps^{(\alpha\nu} \hat A_\nu u^{\beta)} \notag \\      
       &=-T_{wzI}^{\alpha\beta}  \,,\\
   J_\text{CS}^\alpha &= c_{_A}\, \eps^{\alpha\nu} (A_\nu-\hat A_\nu)+ [c_{_A} \, \eps^{\rho\sigma} \hat A_\rho u_\sigma]
   u^\alpha \notag\\
       &= - J_{wzI}^\alpha +c_{_A} \, \eps^{\alpha\nu} D_\nu \psi  \,,
\end{align}
where the identity 
\begin{align}
  u^{[\alpha} \tilde u^{\beta]} = \tfrac{1}{2} \eps^{\alpha\beta} \qquad \text{with } \tilde u^\alpha \equiv \eps^{\alpha\nu} u_\nu \label{2dIdent}
\end{align}
has been used. We see that the added currents 
$T_{wzI}^{\alpha\beta} +T_\text{CS}^{\alpha\beta} =0$ and 
$J_{wzI}^\alpha + J_\text{CS}^\alpha =c_{_A}\, \eps^{\alpha\nu} D_\nu \psi$ are much
simpler than in $d=4$ and in particular the contributions from hatted fields in the different actions cancel. 
In $d=2$ the hatted terms in the effective action do not have any effect on the form of the final stress-energy tensor and charge current. This is a first
hint that here, unlike in higher dimensions, they may be unnecessary. 

If we also include the contributions from $S_{wzII}$, we find for the full stress-energy tensor and charge current
\begin{align}
 T_{anom}^{\alpha\beta} &=  c_{_A}s\, \eta_{,s} \, (\tilde u^\nu D_\nu \psi )\, P^{\alpha\beta}
       +c_{_A}[\eta-\mu\, \eta_{,\mu}] (\tilde u^\nu D_\nu \psi) \, u^\alpha u^\beta 
       -2c_{_A}\mu \eta  \, \tilde u^{(\alpha} u^{\beta)} \,,\\
  J_{anom}^\alpha &= -c_{_A}[\eta_{,\mu}-1] (\tilde u^\nu D_\nu\psi) \, u^\alpha  -c_{_A}(\mu+\eta) \, \tilde u^\alpha \, .
\end{align}
Requiring these expressions to take the canoncial form (\ref{T_sFrame}) and (\ref{J_sFrame}), respectively, we need
to set $\eta=\mu$ and find the following gauge invariant data:
\begin{align}
T_{anom}^{\alpha \beta} = -2c_{_A}\mu^2 \, \tilde u^{(\alpha} u^{\beta)} \, , 
 \qquad J_{anom}^\alpha = -2c_{_A}\mu \tilde u^\alpha \, . \label{2dConstRel}
\end{align}
Naively calculating the divergence of just the anomalous piece of the current leads again to a contamination by hatted fields:
\begin{align}
\nabla_\alpha J^\alpha_{anom} = 2c_{_A} \,\nabla_\alpha \eps^{\alpha\beta} (A_\beta-\hat A_\beta)
    = c_{_A} \, \eps^{\alpha\beta} (F_{\alpha\beta}-\hat F_{\alpha\beta}) \, . \label{2Ddiv}
\end{align}
One can check that a Noether current argument as used in \cite{Dubovsky:2011sk} leads to the same equations of motion. 
We conclude that, although our prescription gives the same constitutive relations (\ref{2dConstRel}) as in \cite{Dubovsky:2011sk}, 
the equations of motion are different. 

In two dimensions, we thus encounter the same problem as in any higher dimension. So why was it nevertheless possible for 
\cite{Dubovsky:2011sk} to get the correct dynamics in $d=2$ without referring to a non-equilibrium Schwinger-Keldysh formalism
as we do in \S\ref{sec:Schwinger-Keldysh}? We will now argue that this is quite coincidental. 
Using the identity (\ref{I3}) which still holds in $d=2$, one can readily check that $S_{anom}$ as derived in this appendix
is invariant under the chemical shift $\psi \rightarrow \psi + \mathfrak{f}(\phi^I)$. The calculation is similar 
to the 4-dimensional case. The crucial difference is that the hatted parts of the action are already invariant under the chemical shift by themselves.
In particular,
\begin{align}
\delta_\psi \left[c_{_A}\int  {\bf \hat D}\psi \wedge \fAh\right] = c_{_A} \int \sqrt{-g}\, \eps^{\alpha\beta}
   \partial_\alpha \mathfrak{f} \,\hat A_\beta = 2c_{_A} \int \sqrt{-g} \, u^{[\alpha} \tilde u^{\beta]} 
   \partial_\alpha \mathfrak{f}\, \hat A_\beta = 0 \, .
\end{align}
Secretly, this is the reason why the authors of \cite{Dubovsky:2011sk} never had to talk about hatted fields in the action $S_{wzI}$ and 
therefore also never had to account for their gauge non-invariance by adding corresponding hatted Bardeen-Zumino terms; 
the invariance under chemical shift symmetry can be implemented in $d=2$ without introducing extra transverse contributions.
As we have shown above, adding these terms is of course allowed by symmetry and it does not change the constitutive relations. It does however  
change the equations of motion. While this change is optional (and therefore unnecessary) in two dimensions, it is unavoidable in higher dimensions. 

Let us now show consistency of this reasoning with the Schwinger-Keldysh picture that we advocate in \S\ref{sec:Schwinger-Keldysh}. We claim that treating the 2-dimensional case in Schwinger-Keldysh formalism is particularly simple because there are no cross-contour terms, i.e.\ the Schwinger-Keldysh action factorizes into two separate pieces which are just the $R$ and $L$ copies of the action found in \cite{Dubovsky:2011sk}. To be more precise, there is one piece in the Schwinger-Keldysh action which does not factorize but which is beyond hydrodynamics and does not affect the constitutive relations. To wit, we claim that for $d=2$
\begin{align}
 S_{SK}^{anom} &\equiv  \int_{\mathcal{M}_3} \mathcal{T}_3[\fA_R, \fAh_R] - \mathcal{T}_3[\fA_L,\fAh_L] + \mathcal{T}_3[\fAh_R,\fAh_L] \notag\\
  &= S_{DHN}[\fA_R] - S_{DHN}[\fA_L] + S_{cross}[\fA_R,\fA_L] \,, \label{2dSK_result}
\end{align}
where $S_{DHN}$ is the action of Dubovsky-Hui-Nicolis \cite{Dubovsky:2011sk} including a Chern-Simons term (this action contains no hatted gauge fields) and $S_{cross}$ is the remaining cross-contour piece,
\begin{align}
 S_{DHN}[\fA_{R,L}] &= c_{_A} \int_{\mathcal{M}_3} \fA_{R,L} \wedge \fF_{R,L} 
                 + \int_{\mathcal{M}_2} \left( \fA_{R,L}\wedge {\bf D}_{R,L} \psi_{R,L} + \mu_{R,L} \, \fu\wedge {\bf D}_{R,L} \psi_{R,L} \right) \,, \notag \\
 S_{cross} &= 2 c_{_A}\int_{\mathcal{M}_2} d(\psi_R-\psi_L) \wedge \fAh\,.
\end{align} 
The cross-contour piece is manifestly independent of ${\bf a}$, so it does not contribute to the current $\form{J}_{hydro}$ in Schwinger-Keldysh formalism. Similarly, one can check that it gives no contribution to the hydrodynamic stress tensor. This term is thus hydrodynamically irrelevant and $S_{SK}^{anom}$ in $d=2$ is physically equivalent to just two copies of $S_{DHNS}$. This explains why \cite{Dubovsky:2011sk} were able to treat anomalies in $d=2$ without referring to the Schwinger-Keldysh formalism; the extra data in the Schwinger-Keldysh action (which is important in higher dimensions) is irrelevant in $d=2$ where the Schwinger-Keldysh action simply factorizes.

We conclude by giving some details of the derivation of Eq.\ (\ref{2dSK_result}). The first two terms in the second line of (\ref{2dSK_result}) can be extracted from the full action in terms of transgression forms using the same arguments as in \S\ref{sec:transgression} . This leaves one with the following cross-contour piece:
\begin{align}
 S_{cross} &= c_{_A} \int {\bf \hat D}_R \psi_R \wedge \fAh_R - {\bf\hat D}_L \psi_L \wedge \fAh_L 
         - \fAh_R \wedge \fAh_L  \notag \\
         &= c_{_A} \int d\psi_R \wedge\left[ P_u(\fA_R) + \fu \, i_u(d\psi_R)\right] 
            - d\psi_L \wedge \left[ P_u(\fA_L) + \fu \, i_u (d\psi_L)\right] - \fAh_R \wedge \fAh_L \notag\\
         &= c_{_A}\int \sqrt{-g}\,\big[ -(u\cdot \partial\psi_R)(\tilde u\cdot D_R\psi_R) + (u\cdot \partial \psi_L) (\tilde u \cdot D_L\psi_L)  \notag \\
         &\qquad\qquad\qquad\quad\,  + (u\cdot \partial \psi_R) (\tilde u  \cdot A_R) - (u\cdot \partial \psi_L)(\tilde u\cdot A_L) - (u\cdot \partial(\psi_R+\psi_L)) (\tilde u \cdot a) \big] \notag \\
         &= c_{_A} \int \sqrt{-g} \, \left[ (u\cdot \partial\psi_R) \left[ (\tilde u \cdot \partial \psi_R) + 2 (\tilde u \cdot A)\right] - (u\cdot\partial \psi_L)\left[ (\tilde u \cdot \partial \psi_L) + 2 (\tilde u \cdot A)\right] \right] \notag\\
         &= c_{_A} \int \sqrt{-g}\, \left[- (u^\alpha\tilde  u^\beta - \tilde u^\alpha u^\beta) \partial_\alpha\psi_R\, \partial_\beta \psi_L\right] + 2 [u\cdot\partial(\psi_R-\psi_L)] \left[\tilde u\cdot {\bf \hat D} \Psi \right] \notag \\
         &=2 c_{_A} \int \sqrt{-g}\, \epsilon^{\alpha\beta} \partial_\alpha (\psi_R-\psi_L) {\hat A}_\beta \,,
\end{align}
where we defined the common phase field $\Psi \equiv \tfrac{1}{2}(\psi_R+ \psi_L)$ and the common covariant derivative ${\bf \hat D}\Psi = d\Psi + \fAh$, and we used the identity (\ref{2dIdent}) in the second and in the last step. 
$P_u$ is the generalized transverse projector which acts on $p$-forms ${\boldsymbol \alpha}$ as $P_u({\boldsymbol \alpha}) \equiv {\boldsymbol \alpha} + i_u({\boldsymbol \alpha}) \wedge  \fu$.

\section{Transgression forms and their variation}
\label{appendix:transg}

We collect various useful formulae for the variation of transgression forms which are useful to derive the Ward identities and currents described in the text.

\subsection{Variation of Chern-Simons term}
We begin by calculating the variation of $\ICS$
\begin{equation}
\begin{split}
\delta \ICS &=  \delta\fA \cdot \frac{\partial \ICS}{ \partial \fA} 
+ (D\delta \fA) \cdot \frac{\partial \ICS}{ \partial \fF}\\
&=  \delta\fA \cdot \brk{\frac{\partial \ICS}{ \partial \fA}+D\prn{\frac{\partial \ICS}{ \partial \fF}} } 
+ d\brk{ \delta\fA \cdot \frac{\partial \ICS}{ \partial \fF} 
   } \,.
\end{split} 
\end{equation} 

This can be simplified further by introducing $\fP=d\ICS$ where $\fP$ is the anomaly polynomial
associated with $\ICS$ made of wedge products of $\fF$. Using this, we get 
\begin{equation}
\begin{split}
0 &=d\delta \ICS -\delta d\ICS =d\delta \ICS -\delta \fP \\
&=  \delta\fF \cdot 
\brk{\frac{\partial \ICS}{ \partial \fA}+D\prn{\frac{\partial \ICS}{ \partial \fF}}-\frac{\partial \fP}{ \partial \fF} } 
 -\delta\fA \cdot D\brk{\frac{\partial \ICS}{ \partial \fA}+D\prn{\frac{\partial \ICS}{ \partial \fF}} } 
 \\
\end{split} 
\end{equation} 
which implies
\begin{equation}\label{eq:dPdFId}
\begin{split}
\frac{\partial \fP}{ \partial \fF} &= \frac{\partial \ICS}{ \partial \fA} 
+ D\prn{ \frac{\partial \ICS}{ \partial \fF} } \,.
\end{split} 
\end{equation}

Thus, we can write
\begin{equation}
\begin{split}
\delta \ICS &=  \delta\fA \cdot \frac{\partial \fP}{ \partial \fF} 
 + d\brk{ \delta\fA \cdot \frac{\partial \ICS}{ \partial \fF}  } \,.
\end{split} 
\end{equation}

We can think of $\int\ICS$ as the generating function describing the response of a Hall insulator
to probe electromagnetic fields. Then, by the above formula, the bulk Hall current is given by 
\begin{equation}
\star_{2n+1} \fJH = \frac{\partial \fP}{ \partial \fF} \,. 
\end{equation}
The normal component of this Hall current $\JH^\perp$ is the amount of charge that flows into the boundary 
in the inflow picture and hence, $\JH^\perp$ is also the covariant anomaly of the boundary theory.
It is convenient to define the (ac) Hall conductivity form $\fsH$  as
\begin{equation}
\fsH \equiv \frac{\partial }{ \partial \fF} \brk{\star_{2n+1} \fJH } = \frac{\partial^2 \fP}{\partial \fF \partial \fF} \,. 
\end{equation}

The boundary current contribution from $\int\ICS$ is  termed the Bardeen-Zumino current. It is given 
by
\begin{equation}
\star \fJBZ = \frac{\partial \ICS}{ \partial \fF} \,.
\end{equation}
This contribution, when added to the boundary charge current obtained by varying boundary action, covariatises
the boundary current.

In  terms of these currents we can write 
\begin{equation}
\begin{split}
\delta \ICS &=  \delta\fA \cdot \star_{2n+1} \fJH 
 + d\brk{\ \delta\fA \cdot \star \fJBZ\  } \,.
\end{split} 
\end{equation}

\subsection{Transgression forms}
Let us now consider a continuous set of connections $\fA_t$ parametrised by a 
parameter $t\in [0,1]$. One can think of this set as interpolating between the  connections 
$\fA_{t=0}$ and the connections $\fA_{t=1}$. The 
variational formula then gives
\begin{equation}
\begin{split}
\frac{d}{dt} (\ICS)_t &= \frac{d \fA_t}{dt}  \cdot \prn{\frac{\partial \fP}{ \partial \fF} }_t
 + d\brk{ \frac{d \fA_t}{dt}  \cdot \prn{\frac{\partial \ICS}{ \partial \fF} }_t } 
\end{split} 
\end{equation}
which can then be integrated to 
\begin{equation}
\begin{split}
 (\ICS)_{t=1}- (\ICS)_{t=0} &= \int_0^1 dt \brk{\frac{d \fA_t}{dt}  \cdot \prn{\frac{\partial \fP}{ \partial \fF} }_t
 }  + d\bigbr{ \int_0^1 dt \brk{\frac{d \fA_t}{dt}  \cdot \prn{\frac{\partial \ICS}{ \partial \fF} }_t
 } }\\
 &= \int_0^1 dt\ \frac{d \fA_t}{dt}  \cdot \prn{\star_{2n+1} \fJH  }_t \
   + \ d\bigbr{\ \int_0^1 dt\ \frac{d \fA_t}{dt}  \cdot \prn{\star \fJBZ }_t\
  } \,.
\end{split} 
\end{equation}
This is the basic transgression formula which can be used to write the difference between a Chern-Simons
form evaluated on two different connections. Note that the right hand side is a sum of a covariant term
(we remind the reader that expressions like $\frac{d \fA_t}{dt}$ depend on the difference of 
two connections and hence transform covariantly) and an exact term.

Introducing the transgression forms
\begin{equation}
\begin{split}
\VP \prn{\fA_t} &\equiv \int_0^1 dt \brk{\frac{d \fA_t}{dt}  \cdot \prn{\frac{\partial \fP}{ \partial \fF} }_t
 }
 = \int_0^1 dt\ \frac{d \fA_t}{dt}  \cdot \prn{\star_{2n+1} \fJH  }_t\\
\WCS \prn{\fA_t} &\equiv  \int_0^1 dt \brk{\frac{d \fA_t}{dt}  \cdot \prn{\frac{\partial \ICS}{ \partial \fF} }_t
 }
 =\int_0^1 dt\ \frac{d \fA_t}{dt}  \cdot \prn{\star \fJBZ }_t
\end{split} 
\end{equation}
we can write 
\begin{equation}
\begin{split}
 (\ICS)_{t=1}- (\ICS)_{t=0} &= \VP \prn{\fA_t}  + d\WCS \prn{\fA_t} \,.
\end{split} 
\end{equation}

It is common to take the linear interpolation 
\begin{equation}
\begin{split}
\fA_t = t\ \fA_{t=1}+ (1-t) \fA_{t=0} = \fA_{t=0} + t \Delta \fA
\end{split} 
\end{equation}
where we have defined $\frac{d \fA_t}{dt} = \Delta \fA = \fA_{t=1} -\fA_{t=0}$. 
The field strength is given by 
\begin{equation}
\begin{split}
\fF_t \equiv d\fA_t + \fA_t^2 = t\ \fF_{t=1}+ (1-t) \fF_{t=0} - t(1-t) (\Delta\fA)^2 \,.
\end{split} 
\end{equation}
Our 
expressions above can easily be specialised to this case.

\subsection{Variation of transgression forms}
Next, we would like to compute the variation of these transgression forms. A direct computation gives 
\begin{equation}
\begin{split}
\delta \VP \prn{\fA_t} &= \int_0^1 dt \brk{\frac{d \delta \fA_t}{dt}  \cdot \prn{\frac{\partial \fP}{ \partial \fF} }_t
 } + \int_0^1 dt \brk{\frac{d \fA_t}{dt}  \cdot \prn{\frac{\partial^2 \fP}{ \partial \fF \partial \fF} \cdot D \delta \fA }_t 
 } \\
 &= \int_0^1 dt \brk{\frac{d \delta \fA_t}{dt}  \cdot \prn{\frac{\partial \fP}{ \partial \fF} }_t
 }
+   \int_0^1 dt \brk{\frac{d \fF_t}{dt}  \cdot \prn{\frac{\partial^2 \fP}{ \partial \fF \partial \fF} 
 }_t \cdot \delta \fA_t 
 }   \\
&\qquad -d \bigbr{ \int_0^1 dt \brk{\frac{d \fA_t}{dt}  \cdot \prn{\frac{\partial^2 \fP}{ \partial \fF \partial \fF} 
 }_t \cdot \delta \fA_t 
 } } \\ 
 &= \int_0^1 dt \frac{d }{dt}\brk{ \delta \fA_t \cdot \prn{\frac{\partial \fP}{ \partial \fF} }_t
 }
 +d \bigbr{ \int_0^1 dt \brk{\delta \fA_t   \cdot \prn{\frac{\partial^2 \fP}{ \partial \fF \partial \fF} 
 }_t \cdot \frac{d \fA_t}{dt}
 } } \\  
 &=\prn{ \delta \fA \cdot \frac{\partial \fP}{ \partial \fF} }_{t=1}
- \prn{ \delta \fA \cdot \frac{\partial \fP}{ \partial \fF} }_{t=0}
 +d \bigbr{ \int_0^1 dt \brk{\delta \fA_t   \cdot \prn{\frac{\partial^2 \fP}{ \partial \fF \partial \fF} 
 }_t \cdot \frac{d \fA_t}{dt}
 } } \,. \\  
\end{split} 
\end{equation}

Next, we subtract from this expression the variation of the Chern-Simons terms
\begin{equation}
\begin{split}
 (\delta\ICS)_{t=1}- (\delta\ICS)_{t=0} &= \prn{ \delta \fA \cdot \frac{\partial \fP}{ \partial \fF} }_{t=1}
- \prn{ \delta \fA \cdot \frac{\partial \fP}{ \partial \fF} }_{t=0}\\
&\qquad  +d \bigbr{ \prn{ \delta \fA \cdot \frac{\partial \ICS}{ \partial \fF} }_{t=1}
- \prn{ \delta \fA \cdot \frac{\partial \ICS}{ \partial \fF} }_{t=0} } \\  
\end{split} 
\end{equation}
to get 
\begin{equation}
\begin{split}
-\delta \WCS \prn{\fA_t} 
  &= \int_0^1 dt \brk{\delta \fA_t   \cdot \prn{\frac{\partial^2 \fP}{ \partial \fF \partial \fF} 
 }_t \cdot \frac{d \fA_t}{dt}
 }+ \prn{ \delta \fA \cdot \frac{\partial \ICS}{ \partial \fF} }_{t=0}
 -\prn{ \delta \fA \cdot \frac{\partial \ICS}{ \partial \fF} }_{t=1}\\
 &\qquad + d\prn{\ldots} \,. \\  
\end{split} 
\end{equation}

We can write these variations in terms of Hall current, Hall conductivity and Bardeen-Zumino currents as  
\begin{equation}
\begin{split}
\delta \VP \prn{\fA_t} 
 &=\prn{ \delta \fA \cdot \star_{2n+1} \fJH   }_{t=1}
- \prn{ \delta \fA \cdot \star_{2n+1} \fJH }_{t=0}\\
&\qquad +d \bigbr{ \int_0^1 dt \brk{\delta \fA_t   \cdot \prn{\fsH}_t \cdot \frac{d \fA_t}{dt}
 } } \,,\\ 
-\delta \WCS \prn{\fA_t} 
  &= \int_0^1 dt \brk{\delta \fA_t   \cdot \prn{\fsH}_t \cdot \frac{d \fA_t}{dt}
 }+ \prn{ \delta \fA \cdot \star \fJBZ }_{t=0}
 -\prn{ \delta \fA \cdot \star \fJBZ }_{t=1}\\
 &\qquad + d\prn{\ldots} \,. \\  
\end{split} 
\end{equation}
We will now consider some examples which are useful study of anomaly-induced transport.

\subsection{Example I: transgression between \texorpdfstring{$A$}{A} and \texorpdfstring{$\Ah$}{Ahat} }
The first example we consider is the transgression with $\fA_t= \fA+(1-t) \mu\fu$
where $\mu$ is the chemical potential and $\fu$ is the velociy 1-form in hydrodynamics.
This is an interpolation from the hydrodynamic shadow field $\fAh\equiv \fA+ \mu\fu$
to $\fA$ with $\Delta \fA = -\mu \fu$. The corresponding field-strengths are given by 
\begin{equation}
\begin{split}
\fF &= d\fA + \fA^2  = \fB + \fu \wedge \fE \\
\fFh &= d\fAh + \fAh^2 = \fBh + \fu \wedge \fEh
 = \fB +2\fomega \mu + \fu \wedge \prn{\fE-D\mu-\fa \mu}
\end{split}
\end{equation}
where $\fa$  is the acceleration 1-form and $\fomega$ is the vorticity 2-form of the fluid.
$\fB$ and $\fE$ are the rest frame magnetic 2-form and electric 1-form respectively. 
The interpolating field-strength is $ \fF_t = t \fF + (1-t) \fFh$ since $(\Delta \fA)^2 = 0$ .
Further, we note that 
\begin{equation}
\begin{split}
\frac{d\fF_t}{dt}  &=  \fF - \fFh =  -2\fomega \mu + \fu \wedge \prn{D\mu+\fa \mu} \\
\frac{d\fA_t}{dt}  &=  \fA - \fAh =  - \mu \fu = \frac{\fu}{2\fomega} \wedge \frac{d\fF_t}{dt}
\end{split}
\end{equation}
from which it follows that
\begin{equation}\label{eq:HatUnhTransgr}
\begin{split}
\VP \prn{\fA,\fAh} &\equiv \int_0^1 dt \brk{\frac{d \fA_t}{dt}  \cdot \prn{\frac{\partial \fP}{ \partial \fF} }_t
 }
 = \frac{\fu}{2\fomega} \wedge\int_0^1 dt \brk{\frac{d \fF_t}{dt}  \cdot \prn{\frac{\partial \fP}{ \partial \fF} }_t
 } \\
 &= \frac{\fu}{2\fomega} \wedge \prn{\fP-\fPh} \,,\\
\WCS \prn{\fA,\fAh} &\equiv  \int_0^1 dt \brk{\frac{d \fA_t}{dt}  \cdot \prn{\frac{\partial \ICS}{ \partial \fF} }_t
 }
 = \frac{\fu}{2\fomega} \wedge\int_0^1 dt \brk{\frac{d \fF_t}{dt}  \cdot \prn{\frac{\partial \ICS}{ \partial \fF} }_t
 } \\
 &= \frac{\fu}{2\fomega} \wedge \prn{\ICS-\hICS} \,.\\
\end{split} 
\end{equation}

To compute the variation of these transgression forms , we need to evaluate
\begin{equation}
\begin{split}
 \int_0^1 dt \brk{\delta \fA_t   \cdot \prn{\frac{\partial^2 \fP}{ \partial \fF \partial \fF}}_t \cdot \frac{d \fA_t}{dt}
 } &= \delta \fA \cdot \int_0^1 dt \brk{ \prn{\frac{\partial^2 \fP}{ \partial \fF \partial \fF}}_t \cdot \frac{d \fA_t}{dt}
 }\\
 &\quad+  \delta \fu \wedge  \int_0^1 dt \brk{(1-t)\mu \cdot \prn{\frac{\partial^2 \fP}{ \partial \fF \partial \fF}}_t \cdot \frac{d \fA_t}{dt}
 }
\end{split}
\end{equation}
where we have used $\fu \wedge \frac{d \fA_t}{dt} =0$. We write 
\begin{equation}
\begin{split}
 \int_0^1 dt \brk{\delta \fA_t   \cdot \prn{\frac{\partial^2 \fP}{ \partial \fF \partial \fF}}_t \cdot \frac{d \fA_t}{dt}
 } &= \delta \fA \cdot \star \fJP+ \delta \fu \wedge \star\fqP \\
\qquad \text{where} \qquad
\star \fJP  &\equiv
 \int_0^1 dt \brk{ \prn{\frac{\partial^2 \fP}{ \partial \fF \partial \fF}}_t \cdot \frac{d \fA_t}{dt}
 }\\
\star\fqP &\equiv    \int_0^1 dt \int_t^1 ds\brk{\mu \cdot \prn{\frac{\partial^2 \fP}{ \partial \fF \partial \fF}}_t \cdot \frac{d \fA_t}{dt}
 }\\
 &= \int_0^1 ds \int_0^s dt\brk{\mu \cdot \prn{\frac{\partial^2 \fP}{ \partial \fF \partial \fF}}_t \cdot \frac{d \fA_t}{dt}
 } \,.
\end{split}
\end{equation}

These integrals can be easily computed: we get 
\begin{equation}\label{eq:JPdef}
\begin{split}
\star \fJP &= \int_0^1 dt \brk{ \prn{\frac{\partial^2 \fP}{ \partial \fF \partial \fF}}_t \cdot \frac{d \fA_t}{dt}
 } = \frac{\fu}{2\fomega} \wedge  \int_0^1 dt \brk{ \prn{\frac{\partial^2 \fP}{ \partial \fF \partial \fF}}_t 
 \cdot \frac{d \fF_t}{dt}
 }\\
 &= \frac{\fu}{2\fomega} \wedge \bigbr{\frac{\partial \fP}{ \partial \fF} -\frac{\partial \fPh}{ \partial \fFh} }\\
\end{split}
\end{equation}
and 
\begin{equation}\label{eq:qPdef}
\begin{split}
\star\fqP &= \int_0^1 ds \int_0^s dt\brk{\mu \cdot \prn{\frac{\partial^2 \fP}{ \partial \fF \partial \fF}}_t
\cdot \frac{d \fA_t}{dt}
 } \\
 &= -\frac{\fu}{(2\fomega)^2} \wedge 
 \int_0^1 ds \int_0^s dt\brk{\frac{d \fF_s}{ds} \cdot \prn{\frac{\partial^2 \fP}{ \partial \fF \partial \fF}}_t 
 \cdot \frac{d \fF_t}{dt} }\\
  &= -\frac{\fu}{(2\fomega)^2} \wedge 
 \int_0^1 ds\ \frac{d \fF_s}{ds} \cdot 
 \bigbr{ \prn{
    \frac{\partial \fP}{ \partial \fF } }_{t=s} 
- \frac{\partial \fPh}{ \partial \fFh }  }\\
  &= -\frac{\fu}{(2\fomega)^2} \wedge 
 \bigbr{ \fP -  \fPh
- \prn{\fF-\fFh}\cdot  \frac{\partial \fPh}{ \partial \fFh }  }\,.
\end{split}
\end{equation}

Thus, we have
\begin{equation}
\begin{split}
\delta \VP \prn{\fA,\fAh} 
 &=\delta \fA \cdot \star_{2n+1} \fJH 
- \delta \fAh \cdot \star_{2n+1} \fJHh
+d \bigbr{\delta \fA \cdot \star \fJP+ \delta \fu \wedge \star\fqP } \,,\\ 
-\delta \WCS \prn{\fA,\fAh}  
  &= \delta \fA \cdot \star \prn{\fJP-\fJBZ}+ \delta \fu \wedge \star\fqP +
  \delta \fAh \cdot \star \fJBZh  + d\prn{\ldots} \,. \\  
\end{split} 
\end{equation}

We now use
\begin{equation}
\begin{split}
\delta \Ah_a = P_a^b \delta A_b 
+ u_a u^b \delta \prn{c^{-1} \partial_b c}
 +  \half \mu \prn{ P_a^b u^c + u^b P_a^c} \delta g_{bc}\ , 
\\
\delta \Ah_\alpha = P_\alpha^\beta \delta A_\beta 
+ u_\alpha u^\beta \delta \prn{c^{-1} \partial_\beta c} 
+  \half\mu  \prn{ P_\alpha^\beta u^\rho + u^\beta P_\alpha^\rho} \delta g_{\beta\rho} 
\end{split} 
\end{equation}
to write 
\begin{equation}
\begin{split}
\delta \int_{\mathcal{M}_{2n+1}} \VP \prn{\fA,\fAh} 
 &=\int_{\mathcal{M}_{2n+1}} \sqrt{-g_{2n+1}} \Bigl\{  
  \frac{1}{2} \brk{ \mu \cdot \JHh^c \prn{P^a_c  u^b + u^a  P^b_c}} 
  \delta g_{ab} \Bigr.\\
  &\quad \Bigl. + \prn{ \JH^a+P^a_b \JHh^b }\cdot \delta A_a
  +\mu \cdot \JBZh^c u_c u^b \delta \prn{c^{-1} \partial_b c} \Bigr\}\\
&+\int_{\partial \mathcal{M}_{2n+1}}  \sqrt{-g_{2n}} \Bigl\{  
  \frac{1}{2} \brk{\qP^\alpha u^\beta
  + u^\alpha  \qP^\beta } 
  \delta g_{\alpha\beta}  + \JP^\alpha\cdot \delta A_\alpha \Bigr\} \\
  \end{split} 
\end{equation}
and
\begin{equation}
\begin{split}
-\delta \int_{\partial \mathcal{M}_{2n+1}}& \WCS \prn{\fA,\fAh} \\ 
  &= \int_{\partial \mathcal{M}_{2n+1}}  \sqrt{-g_{2n}} \Bigl\{  
  \frac{1}{2} \brk{ \prn{\qP^\alpha+\mu \cdot \JBZh^\rho P^\alpha_\rho } u^\beta
  + u^\alpha \prn{ \qP^\beta +\mu \cdot \JBZh^\rho  P^\beta_\rho}} 
  \delta g_{\alpha\beta} \Bigr.\\
  &\quad \Bigl. + \prn{ \JP^\alpha-\JBZ^\alpha+P^\alpha_\beta \JBZh^\beta }\cdot \delta A_\alpha
  +\mu \cdot \JBZh^\rho u_\rho u^\beta \delta \prn{c^{-1} \partial_\beta c} \Bigr\} \\
  &\quad + \text{Boundary terms} \\  
\end{split} 
\end{equation}
Hence, for a general non-abelian flavour symmetry, we have $J_{anom}^\mu = \JP^\mu$ and 
 $q_{anom}^\mu = \qP^\mu$ where  $\{ \JP^\mu,\qP^\mu \}$ are given by \eqref{eq:JPdef}
 and \eqref{eq:qPdef} respectively.

\subsection{Example II: right to left transgression form}
The second example we consider is the transgression with $\fA_t= t \fAh_R+(1-t) \fAh_L$
which is an interpolation from the hydrodynamic shadow field $\fAh_L$
in the left Schwinger-Keldysh contour to  $\fAh_R$ in the right Schwinger-Keldysh contour.
As we argued in the main text, this contribution to the 
Schwinger-Keldysh functional of an anomalous theory  is necessary to 
reproduce the correct conservation equations.

The corresponding field-strengths are given by 
\begin{equation}
\begin{split}
{\fFh}_R &= d\fAh_R + \fAh_R^2 = {\fBh}_R + \fu_R \wedge {\fEh}_R
 = \fB_R +2\fomega_R \mu_R + \fu_R \wedge \prn{\fE_R-D\mu_R-\fa_R \mu_R} \\
{\fFh}_L &= d\fAh_L + \fAh_L^2 = {\fBh}_L + \fu_L \wedge {\fEh}_L
 = \fB_L +2\fomega_L \mu_L + \fu_L \wedge \prn{\fE_L-D\mu_L-\fa_L \mu_L} \\ 
\end{split}
\end{equation}
where $\fa$  is the acceleration 1-form and $\fomega$ is the vorticity 2-form of the fluid.
$\fB$ and $\fE$ are the rest frame magnetic 2-form and electric 1-form respectively.
The subscripts $L$ and $R$ signify which set of fields $\{\phi,\psi\}$ and sources 
$\{g_{\mu\nu}, A_\mu \}$ are used in the construction of these fields.

The interpolating field-strength is 
\begin{equation}
\begin{split}
\fF_t &= t {\fFh}_R+ (1-t) {\fFh}_L - t(1-t) (\fAh_R-\fAh_L)^2\\
&= \frac{1}{2} \prn{{\fFh}_R+  {\fFh}_L} - \frac{1}{2}(1-2t)\prn{{\fFh}_R-  {\fFh}_L}
- t(1-t) (\fAh_R-\fAh_L)^2
\end{split}
\end{equation}
and $\frac{d\fA_t}{dt}  =  \fAh_R-\fAh_L$.
We are interested in studying the corresponding transgression forms 
\begin{equation}
\begin{split}
\VP^{\hat{R}\hat{L}}  &\equiv \int_0^1 dt \brk{\frac{d \fA_t}{dt}  \cdot \prn{\frac{\partial \fP}{ \partial \fF} }_t
 }
 \\
\WCS^{\hat{R}\hat{L}} &\equiv  \int_0^1 dt \brk{\frac{d \fA_t}{dt}  \cdot \prn{\frac{\partial \ICS}{ \partial \fF} }_t
 }
\end{split} 
\end{equation}
near the hydrodynamic limit, i.e., when the differences between the fields living on left and the right 
contours are taken to zero. While the transgression forms themselves vanish in this limit, as we 
will show below, their variations give a finite answer which solves the `shadow anomaly problem'.

We will begin by first looking at the boundary contribution:
\begin{equation}
\begin{split}
 \int_0^1 dt \brk{\delta \fA_t   \cdot \prn{\fsH}_t \cdot \frac{d \fA_t}{dt}
 } 
\end{split} 
\end{equation}
since $\frac{d \fA_t}{dt}=  \fAh_R-\fAh_L \to 0$ in the hydrodynamic limit, we conclude that 
there is no boundary contribution. This is consistent with the statement that $\{ J^\mu_{anom},q^\mu_{anom} \}$ calculated in 
hydrostatic equilibrium do not receive corrections when we turn on time-dependence.

The finite contributions are given by  
\begin{equation}
\begin{split}
\delta \VP^{\hat{R}\hat{L}}
 &= \delta \fAh_R \cdot \star_{2n+1} \prn{ \fJHh }_R 
-  \delta \fAh_L \cdot \star_{2n+1} \prn{\fJHh }_L  \,, \\
-\delta \WCS^{\hat{R}\hat{L}}
  &= -\delta \fAh_R \cdot \star \prn{\fJBZh }_R 
 +\delta \fAh_L \cdot \star \prn{\fJBZh }_L
+ d\prn{\ldots}  \,.
\end{split} 
\end{equation}  
We now use
\begin{equation}
\begin{split}
\delta \Ah_a = P_a^b \delta A_b 
 +  \half \mu \prn{ P_a^b u^c + u^b P_a^c} \delta g_{bc}\ , 
\\
\delta \Ah_\alpha = P_\alpha^\beta \delta A_\beta 
+  \half\mu  \prn{ P_\alpha^\beta u^\rho + u^\beta P_\alpha^\rho} \delta g_{\beta\rho} 
\end{split} 
\end{equation}
and then take the hydrodynamic limit, i.e., we set the fields on two Schwinger-Keldysh contours equal
keeping only linear terms in $A_R-A_L \equiv \adiff$ and $g_R-g_L \equiv \gdiff$.
This gives
\begin{equation}\label{eq:CrossTransVar} 
\begin{split}
\delta \int_{\mathcal{M}_{2n+1}} \VP^{\hat{R}\hat{L}}
 &=  \int_{\mathcal{M}_{2n+1}} \sqrt{-g_{2n+1}} \bigbr{  P^b_c \JHh^c \delta \adiff_b 
 + \half \mu \prn{ P_a^b u^c + u^b P_a^c} \JHh^a \delta \gdiff_{bc} } \,,\\
-\delta \int_{\mathcal{M}_{2n}}\WCS^{\hat{R}\hat{L}}
  &= -\int_{\mathcal{M}_{2n}} \sqrt{-g_{2n}} \bigbr{ P^\beta_\rho \JBZh^\rho \delta \adiff_\beta
  + \half \mu \prn{ P_\alpha^\beta u^\rho + u^\beta P_\alpha^\rho} \JBZh^\alpha \delta \gdiff_{\beta\rho} } \,.
\end{split} 
\end{equation} 
As we argue in the main text, this is exactly the contribution needed to cancel the shadow contributions to 
the conservation equations.

\section{Derivation of anomalous dynamics from effective action}
\label{sec:stensorA}

In this appendix we present a similar argument as in \cite{Jensen:2012kj}
in order to derive the current and stress tensor conservation equations from the requirement of gauge and diffeomorphism invariance
of the full action. For the sake of concreteness we will sometimes refer explicitly to the abelian action 
$S_{anom} = S_{wzI} + S_{wzII} + S_\text{CS} \equiv S_a^{(2n)} + S_\text{CS}$ in $d=2n$ as given in \S\ref{sec:GeneralResult}. 
However, we keep the notation sufficiently abstract such that the discussion will be very general and applies to every 
theory whose total effective action is a sum of any boundary action $S_a^{(2n)}$ and a Chern-Simons term $S_\text{CS}$.

Consider first the variation of the $2n$-dimensional part of the action under variations of the metric, gauge field and $\psi$:
\begin{align}
\delta S_a^{(2n)} &= \int_{\mathcal{M}_{2n}} \sqrt{-g_{2n}} 
\left(\frac{1}{2} T^{\alpha\beta}_{(a)} \delta g_{\alpha\beta} 
+ J^\alpha_{(a)} \delta A_\alpha + Z^\alpha_{(a)} \delta (\partial_\alpha\psi)\right) \, .
 \label{DeltaSa}
\end{align}
Now consider infinitesimal gauge and diffeomorphism transformations
\begin{align}
\delta_\lambda A_\alpha &= -\nabla_\alpha \Lambda + A_\beta \nabla_\alpha \xi^\beta + (\xi\cdot \nabla) A_\alpha 
   = \nabla_\alpha(A_\beta \xi^\beta - \Lambda) + \xi^\beta F_{\beta\alpha}  \,,\notag\\
\delta_\lambda g_{\alpha\beta} &= \nabla_\alpha \xi_\beta + \nabla_\beta \xi_\alpha \,,\label{InfDiff}\\
\delta_\lambda (\partial_\alpha \psi) &= \nabla_\alpha \Lambda \,.\notag
\end{align}
After an integration by parts, Eq.\ (\ref{DeltaSa}) thus yields
\begin{align}
\delta_\lambda S_a^{(2n)} &= \int_{\mathcal{M}_{2n}} \sqrt{-g_{2n}} \left[ 
\Lambda\,  \nabla_\alpha \prn{J^\alpha_{(a)}- Z^\alpha_{(a)}}
  + \xi_\alpha \left(-\nabla_\beta T^{\alpha\beta}_{(a)} + F^\alpha{}_\beta J^\beta_{(a)} - A^\alpha \nabla_\beta J^\beta_{(a)} \right)\right] \,.
\end{align}
In order to get consistent equations of motion, we need to take into account the anomaly inflow from the bulk 
action $S_\text{CS}$, as well. 
We parameterize the variation of $S_\text{CS}$ as  
\begin{align}
\delta S_\text{CS} &= \int_{\mathcal{M}_{2n+1}} \sqrt{-g_{2n+1}} \left[ \frac{1}{2} T^{ab}_{(2n+1)} \delta g_{ab} 
+ J_{(2n+1)}^a \delta A_a + Z^a_{(2n+1)}\delta(\partial_a\psi)\right]\notag\\
  &\quad + \int_{\partial \mathcal{M}_{2n+1}} \sqrt{-g_{2n}} \left[ \frac{1}{2} T_\text{CS}^{\alpha\beta} \delta g_{\alpha\beta}
    + J_\text{CS}^\alpha \delta A_\alpha + Z^\alpha_\text{CS} \delta(\partial_\alpha\psi)\right] \, ,
\end{align}
where Latin indices refer to the $(2n+1)$-dimensional bulk spacetime $\mathcal{M}_{2n+1}$ and Greek indices refer to the boundary
$\mathcal{M}_{2n} = \partial \mathcal{M}_{2n+1}$, respectively (the bulk direction is denoted by $\perp$, \emph{i.e.}\
$a=(\perp,0,1,\ldots,2n-1)$). For concreteness, if we want to work out the anomaly stemming from the Chern-Simons action
\[ S_\text{CS} = \int_{\mathcal{M}_{2n+1}} \prn{ \ICS - \hICS } \, , \]
This action would give explicitly 
\begin{equation}\begin{split}\label{eq:anInf_1}
 T^{ab}_{(2n+1)} &= -\mu  \left(P_c^a u^b + P_c^b u^a \right)\JHh^c \,,\quad 
 J^a_{(2n+1)} = \JH^a - P^a_b\JHh^b\ , \quad
 Z^a_{(2n+1)} =-u^a u_b \JHh^b\ ,\\
 T^{\alpha\beta}_\text{CS} &=-\mu \,
     \left( P_\gamma^\alpha u^\beta + P_\gamma^\beta  u^\alpha \right) \hat{J}_{_{BZ}}^\gamma\ ,\quad
 J^\alpha_\text{CS} =\JBZ^\alpha - P^\alpha_\beta \hat{J}_{_{BZ}}^\beta\ ,\quad
 Z^\alpha_\text{CS} = -u^\alpha u_\beta \hat{J}_{_{BZ}}^\beta \,.
\end{split}\end{equation}
For the abelian case, we have 
\begin{equation}\begin{split}\label{eq:anInf_2}
\JH^c &= \frac{(n+1) c_{_A}}{2^n}\,\eps^{a_1b_1\cdots a_n b_nc} \,{F}_{a_1b_1}\cdots{F}_{a_nb_n} \\
\JHh^c &= \frac{(n+1) c_{_A}}{2^n}\,\eps^{a_1b_1\cdots a_n b_nc} \,\hat{F}_{a_1b_1}\cdots\hat{F}_{a_nb_n} \\
{J}_{_{BZ}}^\beta &= \frac{n\ c_{_A}}{2^{n-1}}\, \eps^{\beta\gamma_1\cdots\beta_n\gamma_n}  {A}_{\gamma_1} 
    {F}_{\beta_2 \gamma_2}\cdots {F}_{\beta_n\gamma_n}\\
\hat{J}_{_{BZ}}^\beta &= \frac{n\ c_{_A}}{2^{n-1}}\, \eps^{\beta\gamma_1\cdots\beta_n\gamma_n}  \hat{A}_{\gamma_1} 
    \hat{F}_{\beta_2 \gamma_2}\cdots \hat{F}_{\beta_n\gamma_n}
\end{split}\end{equation}

We take $u^\perp = 0$, i.e., the extension of the fluid on the $2n$-dimensional physical space into the bulk $\mathcal{M}_{2n+1}$ is 
accomplished by working with a $(2n+1)$-dimensional fluid that doesn't move in the bulk direction. Furthermore, the bulk
is consistently connected to the 
boundary by demanding
\begin{align}
 \sqrt{-g_{2n+1}} \;\eps_{(2n+1)}^{\perp \alpha_1 \cdots \alpha_{2n}} 
 = \sqrt{-g_{2n}}\; \eps^{\alpha_1 \cdots \alpha_{2n}}_{(2n)} \,,
\end{align}
where subscripts denote dimension of the spacetime that the totally antisymmetric tensor lives in.
Using integration by parts, the variation under the infinitesimal coordinate transformation (\ref{InfDiff}) 
of $S_\text{CS}$ can then be written as
\begin{equation}\label{DeltaScs}
\begin{split}
 \delta_\lambda S_{\text{CS}} &= \int_{\mathcal{M}_{2n+1}}\sqrt{-g_{2n+1}}\ \Bigl\{
  \Lambda \nabla_a \prn{ J_{(2n+1)}^a - Z^a_{(2n+1)}}\Bigr.\\
 &\qquad\qquad\Bigl. + \xi_a \prn{
   - \nabla_b T^{ba}_{(2n+1)}+ F^a{}_b J^b_{(2n+1)} -A^a \nabla_b J^b_{(2n+1)} }\Bigr\} \\
 &\quad + \int_{\partial\mathcal{M}_{2n+1}}\sqrt{-g_{2n}}\ \Bigl\{
   \Lambda\brk{  \nabla_\alpha \left(J^\alpha_\text{CS} - Z^\alpha_\text{CS}\right)
   -\left(J^\perp_{(2n+1)} - Z_{(2n+1)}^\perp\right) } \Bigr.\\
 &\qquad\qquad+ \xi_\mu \brk{
   -\nabla_\beta T^{\alpha\beta}_\text{CS} + F^\alpha{}_\beta J^\beta_\text{CS} 
   - A^\alpha\left( \nabla_\beta J^\beta_\text{CS}
    - J^\perp_{(2n+1)}\right) + T_{(2n+1)}^{\perp\alpha}} \Bigr\}  \,.
\end{split}\end{equation}

We are now in the position to demand gauge and diffeomorphism invariance of the $2n$-dimensional theory, i.e.,
\begin{align}
\delta_\lambda S_a^{(2n)} + [\delta_\lambda S_\text{CS}]_{\partial \mathcal{M}_{2n+1}} = 0 \,.  \label{InvCondition}
\end{align}
Evaluating this equation using $T^{\alpha\beta}_{anom} = T^{\alpha\beta}_{(a)} + T^{\alpha\beta}_\text{CS}$
and $J^\alpha_{anom} = J^\alpha_{(a)} + J^\alpha_\text{CS}$ gives the general equations of motion for an anomalous theory:
\begin{align}
 \nabla_\alpha J^\alpha_{anom}  &= 
    J_{(2n+1)}^\perp  \,, \qquad
  \nabla_\beta T^{\alpha\beta}_{anom} = F^\alpha{}_\beta J^\beta_{anom} + T^{\perp\alpha}_{(2n+1)}  \,. \label{EoM}
\end{align}
where we have
\begin{equation}\begin{split}\label{AnInflow}
    J_{(2n+1)}^\perp = \JH^\perp - \JHh^\perp \,, 
   \qquad T_{(2n+1)}^{\perp\alpha} = -\mu\, u^\alpha\ \JHh^\perp,. 
\end{split}\end{equation}

For the particular case of $S_{anom}$ being the abelian anomalous action from \S\ref{sec:GeneralResult}, $J_{anom}^\alpha$ 
and $T_{anom}^{\alpha\beta}$ are given by the expressions (\ref{J_2n_sol}, \ref{Pi_2n_sol}) and the 
Chern-Simons contributions \eqref{eq:anInf_2} give the following anomaly inflow:
\begin{equation}\begin{split}
\JH^\perp &= \frac{(n+1) c_{_A}}{2^n}\,\eps^{\alpha_1\beta_1\cdots \alpha_n \beta_n} 
\,{F}_{\alpha_1\beta_1}\cdots{F}_{\alpha_n\beta_n} \\
\JHh^\perp &= \frac{(n+1) c_{_A}}{2^n}\,\eps^{\alpha_1\beta_1\cdots \alpha_n \beta_n} 
\,\hat{F}_{\alpha_1\beta_1}\cdots\hat{F}_{\alpha_n\beta_n} \\
\end{split}\end{equation}
For the general non-abelian case, we have 
\begin{equation}
 \begin{split}
 \fP &\equiv d\ICS\ \quad ,\quad
 \star_{2n+1} \fJH \equiv \frac{\partial \fP}{\partial \fF} \quad ,\quad
  \star \fJBZ \equiv \frac{\partial \ICS}{\partial \fF}. 
\end{split}
\end{equation}

For the current conservation equation, we have used that the detailed form of $S_a^{(2n)}$ has been carefully 
engineered such that gauge invariance is preserved in the full theory, which implies 
$\nabla_\alpha (Z_{(a)}^\alpha+Z^\alpha_\text{CS}) = 0$ (as one can explicitly check). 
A comment is in order concerning the invariance condition (\ref{InvCondition}) which 
should also hold true for the $(2n+1)$-dimensional theory which has been constructed 
such that it is gauge invariant and manifestly diffeomorphism invariant. Demanding 
the $(2n+1)$-dimensional integral in Eq.\ (\ref{DeltaScs}) to vanish gives a set of 
bulk equations of motion which must be satisfied for consistency. Those equations corresponding 
to gauge invariance ($\Lambda$) are almost trivially seen to be true. The equations 
corresponding to invariance under general coordinate transformations ($\xi_a$) are more involved. 
They look like stress tensor conservation equations in $(2n+1)$-dimensional fluid dynamics 
and their purpose is to constrain in a consistent manner the (a-priori undetermined) bulk components of the various fields involved.

The equations of motion (\ref{EoM}) in general and the anomaly inflow (\ref{AnInflow}) for our theory of interest in particular, are the main results of this section.

\section{Brief review of Schwinger-Keldysh formalism for hydrodynamics}
\label{sec:SK-review}

In this appendix we review some more details of the Schwinger-Keldysh (SK) technique in general, thus motivating our calculations in \S\ref{sec:Schwinger-Keldysh}. 
We will first explain the necessity of a more refined formalism for non-equilibrium quantum field theory in general and then introduce the single-time representation
which allows for a more convenient way to write SK actions because it does not require complicated integration contours. The seminal papers 
by Schwinger and Keldysh are \cite{Schwinger:1960qe, Keldysh:1964ud}. Our discussion will mainly follow the presentation in \cite{Chou:1984es, Maciejko, Kamenev:2009jj}.

Consider the fundamental problem of quantum field (or many body) theory of calculating, for example, the 2-point
Green's function of some complex field Heisenberg operator $\psi$: 
\begin{align}
 G(x,x') = -i \langle \Omega | T[\psi(x) \psi^\dagger(x')]| \Omega \rangle \,,
\end{align}
where $T$ denotes standard time ordering and $|\Omega\rangle$ is the ground state of the full interacting theory. The usual trick to construct a 
perturbative expansion is by splitting off the interactions from the Hamiltonian, i.e., $H = H_0 + H_{int}$
and switching to the interaction picture. The operator $U(t_0,t) = T\, \exp \left( -i \int_{t_0}^t
dt' \, H_{int}(t')\right)$ then defines time evolution of interaction picture states and one finds in interaction picture
\begin{align}
 G(x,x') = -i \langle 0 | S^\dagger T[\psi(x)\psi^\dagger(x') S ]| 0 \rangle 
         = -i \frac{\langle 0| T[S  \psi(x)\psi^\dagger(x')] |0\rangle}{\langle 0|S  |0\rangle} \,, \label{G-calc}
\end{align}
with the S-matrix $S \equiv U(-\infty, \infty)$ and $|0\rangle$ the (early time) ground state of the non-interacting theory defined by $H_0$.
The right hand side of this equation is the starting point for the usual perturbative expansion. 
The second step in Eq.\ (\ref{G-calc}) comes about by writing the late time groundstate in terms of the early time groundstate, i.e.,
$\langle 0 | S^\dagger = \langle 0 | e^{i\alpha}$ for some phase $\alpha$ which is
compensated by $\langle 0 | S |0 \rangle = e^{i\alpha}$. 
Physically this corresponds to the crucial assumption that the system evolves \emph{adiabatically} and 
slowly follows its ground state. The physical content of the non-interacting ground state $|0\rangle$ is therefore assumed not to be changed by 
the time evolution which means that it can change only by a phase factor $e^{i\alpha}$. 
This assumption is not justified in non-equilibrium situations. 

The SK formalism solves this problem by avoiding any reference to the evolution of the ground state at late times and by referring instead 
only to the ground state $|0\rangle$ at $t=-\infty$. This can be done by \emph{defining} the SK S-matrix $S_\mathcal{C} \equiv T_\mathcal{C} 
\exp\left( -i \int_\mathcal{C} dt' \, H_{int}(t') \right)$ with the contour $\mathcal{C}$ consisting of two antiparallel contours along the 
real axis (see Fig.~\ref{fig:SK-contour1}) and
$T_\mathcal{C}$ being the corresponding contour ordering operator which orders its arguments by their order along the contour.

The corresponding SK Green's function is defined as 
\begin{align}
 G_\mathcal{C}(x,x') = -i \langle \Omega | T_\mathcal{C} [ \psi(x) \psi^\dagger(x') ]|\Omega\rangle
                     = -i \langle 0 | T_\mathcal{C}[ S_\mathcal{C} \,\psi(x) \psi^\dagger(x')]|0\rangle \,. 
   \label{G-SK}
\end{align}
Note that there is no normalizing denominator any more since the SK S-matrix satisfies $S_\mathcal{C} | 0\rangle = |0\rangle$. 
In order to recover the analog of the causal Feynman propagator, the time components of $x$ and $x'$ in the expression (\ref{G-SK}) are both
placed on the upper branch $\mathcal{C}_R$. More generally, however, the times can be inserted on $\mathcal{C}_L$, as well. Therefore
the SK 2-point Green's function is actually a $2\times 2$ matrix containing \emph{real time} Green's functions that correspond to the $4$ combinations
of inserting $t$ and $t'$ on the branches $\mathcal{C}_R$ and $\mathcal{C}_L$:
\begin{align}
 G(x,x') = \begin{pmatrix} G_{RR} & G_{RL} \\ G_{LR} & G_{LL} 
            \end{pmatrix}
         = \begin{pmatrix} G_F & G_< \\ G_> & G_{\tilde F}
           \end{pmatrix} \,, \label{SK-CorrUnphys}
\end{align}
where $G_F(x,x')$ is the well known Feynman propagator, $G_{\tilde F}(x,x')$ is an ``anti-causal'' Feynman propagator with reversed time ordering
and there are two new cross-contour correlators:
\begin{align}
 G_F(x,x') &= -i \langle \Omega | T[\psi(x)\psi^\dagger(x')]|\Omega\rangle \,, \\
 G_{\tilde F}(x,x') &= -i \langle\Omega| \tilde T [\psi(x)\psi^\dagger(x')]|\Omega\rangle \,,\\
 G_< (x,x') &=  i\langle \Omega| \psi^\dagger(x')\psi(x)|\Omega\rangle \,,\\
 G_> (x,x') &= -i \langle \Omega| \psi(x)\psi^\dagger(x') |\Omega\rangle \,.
\end{align}

These four Green's functions above are related by $G_F + G_{\tilde F} = G_> + G_<$.
It is thus useful to perform a change of basis which results in the following \emph{physical} Green's functions as linear combinations of the above:
\begin{align}
 G^R(x,x') &\equiv -i \theta(t-t') \, \langle \Omega | \{\psi(x), \psi^\dagger(x')\}   |\Omega\rangle = G_F - G_< \,, \label{SK-CorrPhys1}\\
 G^A(x,x') &\equiv i \theta(t'-t) \, \langle \Omega |\{ \psi(x), \psi^\dagger(x') \}|\Omega \rangle = G_F- G_> \,, \label{SK-CorrPhys} \\
 G^K(x,x') &\equiv -i \langle \Omega | [\psi(x), \psi^\dagger(x') ]|\Omega \rangle = G_F + G_{\tilde F} \,.\label{SK-CorrPhys2}
\end{align}
The first two of these are the familiar retarded and advanced Green's functions, while the third one is an additional ``Keldysh'' Green's function.

We have now seen how the treatment of non-equilibrium systems forces us to consider Green's functions that involve integration along
a SK contour $\mathcal{C}$.\footnote{Let us make a few remarks concerning other integration contours. For details, we refer to \cite{Maciejko}. If the density matrix of the full 
system is known at some finite time $t_0$, then it is not actually necessary to integrate time from $-\infty$ to $\infty$ and back. It can be 
shown that it is sufficient to integrate from $t_0$ to $\text{max}(t,t')$ and back again. The resulting correlation functions are, 
however, not suited for a perturbative expansion if the system is interacting. Treating interactions by splitting the Hamiltonian 
as in the interaction picture, requires to add a third part to the integration contour which looks like half a Matsubara
contour and runs from $t_0$ to $t_0-i\beta$ with $\beta$ being the inverse temperature. 
For our purposes, this three part ``Kadanoff-Baym'' contour is more complicated than necessary. Since we will be able to ignore initial correlations at $t_0$, we can 
safely take the limit $t_0 \rightarrow -\infty$. Practically, this means that it will always be enough to work in SK formalism defined by
the integration contour $\mathcal{C}$ in Fig.~\ref{fig:SK-contour1}. 
}
We now want to put these insights about 2-point correlators in a more general context and 
discuss generating functionals in this formalism. The generating functional clearly involves integration over $\mathcal{C}$ with a source
term $\mathcal{J}(x)$ that can a-priori take different values on the two branches. If we consider for simplicity just a real bosonic field $\varphi(x)$,
the generating functional reads 
\begin{align}
 Z[J(x)] &\equiv \langle \Omega | T_\mathcal{C} \, \exp\left[ i \int_\mathcal{C} \mathcal{L}[\varphi(x)] + \mathcal{J}(x) \varphi(x)\right] |\Omega\rangle \notag \\
    &= \langle \Omega | T_\mathcal{C} \, \exp\left[ i \int_{t=-\infty}^{t=\infty}\mathcal{L}[\varphi_R(x)] + \mathcal{J}_R(x) \varphi_R(x)
       - \mathcal{L}[\varphi_L(x)] - \mathcal{J}_L(x)\varphi_L(x)\right] |\Omega \rangle \,, \label{SK-GenFc}
\end{align}
where the \emph{single time representation} in the second line contains subscripts $R$ and $L$ that indicate (generally independent) fields 
and sources on $\mathcal{C}_R$ and $\mathcal{C}_L$ which have been merged on the real time axis. The minus sign in front of the part of the 
Lagrangian and sources corresponding to $\mathcal{C}_L$ compensates for the fact that these terms should be integrated backwards in time. 
Calculating correlation functions from $Z[\mathcal{J}_R(x),\mathcal{J}_L(x)]$ as given by Eq.\
(\ref{SK-GenFc}) gives the SK correlation functions of Eq.\ (\ref{SK-CorrUnphys}). The same generating functional also encodes the physical 
correlators such as those in Eqs.\ (\ref{SK-CorrPhys1}-\ref{SK-CorrPhys2}). In order to get the latter directly from functional derivatives 
with respect to arguments of the generating functional, we switch to the ``physical'' basis for the fields and external currents:
\begin{align}
  \begin{pmatrix} \varphi_c \\ \varphi_d \end{pmatrix} = \begin{pmatrix} \tfrac{1}{2} (\varphi_R+\varphi_L) \\ \varphi_R-\varphi_L \end{pmatrix}
   \,,\qquad 
  \begin{pmatrix} {\tt J} \\ {\tt j} \end{pmatrix} = \begin{pmatrix} \tfrac{1}{2} (\mathcal{J}_R + \mathcal{J}_L) \\ \mathcal{J}_R-\mathcal{J}_L \end{pmatrix} \,. 
\end{align}
The generating functional becomes
\begin{align}
 Z[{\tt J}(x), {\tt j}(x)] = \langle\Omega| T_\mathcal{C}\, \exp\left[ i \int_{t=-\infty}^{t=\infty} 
   \mathcal{L}[\varphi_c + \tfrac{1}{2} \varphi_d] - \mathcal{L}[\varphi_c - \tfrac{1}{2}\varphi_d]
   + {\tt J}\, \varphi_d - {\tt j}\, \varphi_c \right] |\Omega\rangle \,,
\end{align}
such that the difference source ${\tt j}(x)$ generates the response as a functional of the physical common field $\varphi_c(x)$. This is 
the piece of information that is most relevant to our discussion in \S\ref{sec:Schwinger-Keldysh}: the hydrodynamical current is the one
which corresponds to a causal (retarded) response and it is thus in SK formalism identified as deriving from variations with respect to 
the difference source ${\tt j}(x)$ (the sources are in hydrodynamics of course the background gauge field or the metric). 

Note that the single time formulation is just a calculational tool. At the end of a calculation, the fields that correspond to 
the forward and backward branches $\mathcal{C}_R$ and $\mathcal{C}_L$ need to be set equal, i.e., $\varphi_d = 0$. In this sense the difference
field $\varphi_d$ describes (quantum) fluctuations around the common physical field $\varphi_c$. It is worth noting that the common source ${\tt J}(x)$ 
does not need to vanish in equilibrium; due to the simplified normalization of SK Green's functions (compare e.g.\ Eq.\ (\ref{G-calc}) against (\ref{G-SK}))
external sources $\mathcal{J}_R$ and $\mathcal{J}_L$ need to be set equal at the end of a calculation, but there is no need for them to be set to zero.  

Also note that the fields of the forward and backward integration (or $\varphi_R$ and $\varphi_L$ in the single time representation) 
are not completely independent. This becomes particularly important if the Lagrangian $\mathcal{L}[\phi]$ has some symmetry group $G$ acting
on $\varphi$ \cite{Chou:1984es}. Then $\mathcal{L}[\varphi_R]- \mathcal{L}[\varphi_L]$ seems to be symmetric under $G_R\times G_L$. However, 
since there are cross-contour correlation functions such as $G_<(x,x')$ and $G_>(x,x')$ which are generically non-zero and must respect the
symmetry of the Lagrangian, one can infer that this enlarged symmetry must always be spontaneously broken to a single $G_\text{diag} \subset
G_R\times G_L$ acting on the whole contour $\mathcal{C}$.

\section{Conventions for differential forms}
\label{appendix:conventions}

In this appendix, we briefly fix our conventions concerning Hodge duality and integration of differential forms. In order to make our results well comparable, we adopt the conventions of \cite{Loganayagam:2011mu}.

If $X_\mu \, dx^\mu$ and $Y_\mu \, dx^\mu$ are 1-forms, we define the Hodge dual $(2n-1)$-form $\star {\bf Y}$ by
\begin{align}
 {\bf X} \wedge (\star {\bf Y}) = - (\star {\bf Y}) \wedge {\bf X} = X_\mu Y^\mu \; \textbf{Vol}_{2n} \,,
\end{align}
where $\textbf{Vol}_{2n}$ is the volume form (the Hodge dual of a function $f$ is $\star f = f \;\textbf {Vol}_{2n}$).
This allows us, given a 1-form current $\form{J} = J_\mu \, dx^\mu$, to talk about the dual $(2n-1)$-form current $\star \form{J}$ instead. 

When we write integrals and switch between differential form language and expressions in coordinate charts, it is important to keep track of numerical factors. In particular, a $k$-form ${\bf X}$ is related to its components by 
\begin{align}
 {\bf X} = \frac{1}{k!} X_{\mu_1\cdots \mu_k} \, dx^{\mu_1}\wedge \cdots \wedge dx^{\mu_k} \,.
\end{align}

 We define the Hodge-dual of a $p$-form $\form{V}$ via
\begin{equation}
\label{eq:HodgeDualDef}
(\hodge \form{V})_{\mu_1 \mu_2\ldots \mu_{d-p}}
\equiv \frac{\text{Sign}[g]}{p!} V_{\nu_1\nu_2\ldots \nu_p} \varepsilon^{\nu_1\nu_2\ldots\nu_p}{}_{\mu_1 \mu_2\ldots \mu_{d-p}} \, ,
\end{equation}
or, in other words,
\begin{equation}
\hodge \form{V}
\equiv \frac{\text{Sign}[g]}{p!(d-p)!}\ V_{\nu_1\nu_2\ldots \nu_p}\ \varepsilon^{\nu_1\nu_2\ldots\nu_p}{}_{\mu_1 \mu_2\ldots \mu_{d-p}}\
dx^{\mu_1} \wedge dx^{\mu_2} \ldots \wedge dx^{\mu_{d-p}} \, .
\end{equation}
We note that acting on a $p$-form, the square of the Hodge-dual is given by
\[ {\hodge}^2 = \text{Sign}[g] (-1)^{p(d-p)}\, , \]

It is useful for various manipulations in the text to note that for a $(d-p)$-form $\form V$ , defined via its Hodge dual in terms of lower forms $\form{A}_i $ of rank $q_i$, with $\sum_{i=1}^k\,q_i= p$, i.e.,
\begin{equation}
\hodge \form{V} = \form{A}_1 \wedge \form{A}_2 \wedge\ldots \wedge \form{A}_k \,,
\end{equation}
 we have in components
\begin{equation}\label{eq:starToComponents}
\begin{split}
V^{\mu_1\mu_2\ldots\mu_{d-p}} &= \frac{ 1 }{q_1 ! q_2 ! \ldots q_k !}
 \varepsilon^{\mu_1\mu_2\ldots\mu_{d-p} \alpha_1\ldots \alpha_{q_1}\beta_1\ldots
\beta_{q_2}\ldots \lambda_1\ldots \lambda_{q_k} } \\
&\qquad  \qquad \
(A_1)_{\alpha_1\ldots \alpha_{q_1}} (A_2)_{\beta_1\ldots \beta_{q_2}} \ldots (A_k)_{\lambda_1\ldots \lambda_{q_k}}\ \, . \\
\end{split}
\end{equation}




\providecommand{\href}[2]{#2}\begingroup\raggedright\endgroup
\end{document}